\documentclass[final,3p,times]{elsarticle}
\pdfoutput=1
\usepackage{amssymb}
\usepackage{amsthm}

\usepackage{mathtools, float}
\usepackage{booktabs} %Tabelas
\usepackage{multirow}
\usepackage{makecell}
\usepackage{multicol}
\usepackage{tabularx}
\usepackage{caption}
\usepackage{subcaption}
\usepackage{xcolor}

\usepackage[colorlinks=false, citecolor=red, linkcolor=blue, pdfborder={0 0 0}]{hyperref}
\usepackage[T1]{fontenc}
\usepackage{selinput}

\makeatletter
\def\ps@pprintTitle{%
 \let\@oddhead\@empty
 \let\@evenhead\@empty
 \def\@oddfoot{}%
 \let\@evenfoot\@oddfoot}
\makeatother

\journal{Mechanical Systems and Signal Processing}

\begin{document}

\begin{frontmatter}



%\title{Complete framework to compute stable coupled state-space models for dynamic substructuring applications}

%\title{Computing stable coupled state-space models for dynamic substructuring applications: a complete framework}
\title{On the computation of stable coupled state-space models for dynamic substructuring applications}

\author[inst1]{R.S.O. Dias \texorpdfstring{\corref{cor1}}{Lg}}
\cortext[cor1]{Corresponding author.}
\ead{r.dasilva@staff.univpm.it}
\affiliation[inst1]{organization={Department of Industrial Engineering and Mathematical Sciences, Universit\`a Politecnica delle Marche},%Department and Organization
            addressline={Via Brecce Bianche}, 
            city={Ancona},
            postcode={60131}, 
            state={Marche},
            country={Italy}}

\author[inst1]{M. Martarelli}
\ead{m.martarelli@staff.univpm.it}
\author[inst2]{P. Chiariotti}
\ead{paolo.chiariotti@polimi.it}

\affiliation[inst2]{organization={Department of Mechanical Engineering, Politecnico di Milano},%Department and Organization
            addressline={Via Privata Giuseppe La Masa}, 
            city={Milan},
            postcode={20156}, 
            state={Lombardia},
            country={Italy}}
            
%\affiliation[cor1]{r.dasilva@staff.univpm.it}

\begin{abstract}
%In this article, all steps to properly compute stable coupled state-space models are described and deeply analyzed. Moreover, two novel approaches are here developed. Firstly, a procedure to impose Newtons's second law without relying on the use of undamped RCMs is presented. Secondly, a novel approach to impose stability on unstable coupled state-space models is also developed. By exploiting experimentally acquired Frequency Response Functions (FRFs), accurate state-space models respecting \textcolor{black}{Newton's second law} are computed. Then, decoupling and coupling operations are performed with the identified state-space models being obtained a reliable, yet unstable, coupled state-space model. As final analysis, stability is imposed on the computed unstable coupled model by following the methodology proposed in this paper. It is shown that the obtained stable coupled state-space model is suitable to be exploited in time-domain simulations, turning out to be an accurate representation of the original unstable coupled state-space model. 

This paper aims at introducing a methodology to compute stable coupled state-space models for dynamic substructuring applications by introducing two novel approaches targeted to accomplish this task: a) a procedure to impose \textcolor{black}{Newtons's second law} without relying on the use of undamped RCMs \textcolor{black}{(residual compensation modes)} and b) a novel approach to impose stability on unstable coupled state-space models. \textcolor{black}{The enforcement of stability is performed by dividing the unstable model into two different models, one composed by the stable poles (stable model) and the other composed by the unstable ones (unstable model). Then, the poles of the unstable state-space model are forced to be stable, leading to the computation of a stabilized state-space model. If this model is composed by real poles, it should be divided into two different ones, one composed by the pairs of complex conjugate poles and the other composed by the real poles. Afterwards, to make sure that the Frequency Response Functions (FRFs) of the stabilized model well match the FRFs of the unstable model, the Least-Squares Frequency Domain (LSFD) method is exploited to update the modal parameters of the stabilized model composed by the pairs of complex conjugate poles.} The validity of the proposed methodologies is presented and discussed by exploiting experimental data. Indeed, by exploiting the FRFs of a real system, accurate state-space models respecting \textcolor{black}{Newton's second law} are computed. Then, decoupling and coupling operations are performed with the identified state-space models, no matter the models resultant from the decoupling/coupling operations are unstable. Stability is then imposed on the computed unstable coupled model by following the approach proposed in this paper. The methodology proved to work well on these data. Moreover, the paper also shows that the coupled state-space models obtained using this methodology are suitable to be exploited in time-domain analyses and simulations. 
\end{abstract}


\begin{highlights}
\item Accurate SSMs verifying \textcolor{black}{Newton's second law} can be computed by using damped RCMs
\item Reliable stable coupled SSMs can be computed from unstable coupled SSMs
\item \textcolor{black}{Iterative} algorithms are not mandatory to compute stable coupled SSMs
\item All steps to compute accurate stable coupled SSMs are described and deeply analyzed
\item Experimental validation of the discussed approaches is provided
\end{highlights}


\end{frontmatter}


%%main text

\textcolor{black}{\section{Nomenclature and Abbreviations}}

For easier understanding of all the parameters used in this paper, the nomenclature is given in table \ref{table:Nomenclature}, \textcolor{black}{while the abbreviations used in this article are defined in table \ref{table:Abbreviations}.}

\begin{table}[ht]
\centering
\caption{Nomenclature}
\begin{tabular}{@{}llll@{}}
\toprule
$A$ & State matrix & $l$ & Modal participation factor vector \\
$B$ & Input matrix & $T$ & Transformation matrix\\
$C$ & Output matrix & $U$ & Left eigenvectors of a matrix\\
$D$ & Feed-through matrix & $UR$ & Upper residuals matrix\\ 
$H$ & FRF matrix & $u$ & Input vector\\ 
$H_{target}$ & Target FRF matrix & $V$ & Right eigenvectors of a matrix\\
$H_{ref}$ & Reference FRF matrix & $x$ & State vector\\ 
$L$ & Modal participation factors matrix & $y$ & Output vector\\ 
$LR$ & Lower residuals matrix & \\ 
\vspace{1mm} & & &\\
$\Re(\bullet)$ & Real part of a variable & $\Im(\bullet)$ & Imaginary part of a variable\\
\vspace{1mm} & & &\\
$\Lambda$ & Diagonal matrix composed by the poles of a system & $\Psi$ & Mode shapes matrix\\
$\lambda$ & Pole of a system & $\psi$ & Mode shape vector\\
$\sigma$ & Singular values of a matrix & &\\ 
\vspace{1mm} & & &\\
$\bullet_{accel}$             & Acceleration state-space model & $\bullet_{pcp}$ & Model composed by \textcolor{black}{PCPs}\\
$\bullet_{CB}$ & RCMs associated with $[C_{full}][B_{full}]$ & $\bullet_{rp}$ & Model composed by \textcolor{black}{RPs}\\
$\bullet_{df}$      & Model transformed into diagonal form & $\bullet_{st}$ & Stable state-space model\\
$\bullet_{full}$ & Complete state-space model & $\bullet_{UR}$ & RCMs associated with $[UR]$\\
$\bullet_{ib}$ & Model representative of the in-band modes & $\bullet_{ut}$ & Unstable state-space model\\
$\bullet_{LR}$ & RCMs associated with $[LR]$ & $\bullet_{vel}$ & Velocity state-space model\\
$\bullet_{or}$ & Original coupled state-space model & &\\
\vspace{1mm} & & &\\
$\bullet^{-1}$            & Inverse of a matrix & $\bullet^{\dag}$  & Pseudoinverse of a matrix\\ 
$\bullet^{INL}$ & Model forced to verify \textcolor{black}{Newton's second law} & $\bullet^{T}$  & Transpose of a vector/matrix\\
$\bullet^{op}$ & Model constructed with optimized modal parameters & $\bullet^{stbz}$ & Model forced to be stable\\
\vspace{1mm} & & &\\
$\dot{\bullet}$  & First order time derivative & $\ddot{\bullet}$  & Second order time derivative\\
$\bar{\bullet}$ & Coupled vector/matrix & $|\bullet|$ & Absolute value of a variable\\
$\bullet^{*}$ & Complex conjugate of a variable & &\\ \midrule
\label{table:Nomenclature}
\end{tabular}
\end{table}

\begin{table}[ht]
\centering
\caption{Abbreviations}
\begin{tabular}{@{}llll@{}}
\toprule
\color{black} DOF & \color{black} Degree of freedom & \color{black} RCM & \color{black} Residual compensation mode\\
\color{black} DS & \color{black} Dynamic substructuring & \color{black} RP & \color{black} Real pole\\
\color{black} FRF & \color{black} Frequency response function & \color{black} SISO & \color{black} Single input single output\\
\color{black} LM-SSS & \color{black} Lagrange multiplier state-space substructuring & \color{black} SSM & \color{black} State-space model\\ 
\color{black} LSFD & \color{black} Least-squares frequency domain & \color{black} SSS & \color{black} State-space substructuring\\ 
\color{black} MIMO & \color{black} Multiple input multiple output & \color{black} SVD & \color{black} Singular value decomposition\\
\color{black} ML-MM &  \color{black} Maximum likelihood modal parameter method & \color{black} UCF & \color{black} Unconstrained coupling form\\ 
\color{black} PCP & \color{black} Pair of complex conjugate poles & \color{black} VPT & \color{black} Virtual point transformation\\ 
\color{black} PolyMAX & \color{black} Polyreference least-squares complex frequency-domain & \color{black} VPT-SS & \color{black} State-space realization of VPT\\  \midrule
\label{table:Abbreviations}
\end{tabular}
\end{table}

\section{Introduction}\label{Introduction}

The term Dynamic Substructuring (DS) refers to that method of structural dynamics that targets the characterization of the dynamic behavior of complex structures by separately analyzing the dynamics of their constituent parts and further coupling them.
%To model the dynamic behaviour of complex structures, it is quite common to divide them into several simple substuctures. Then, by characterizing each of those simple substructures and by coupling them, a characterization of the dynamic behaviour of the initial complex structure is possible. The approach just described is commonly addressed as dynamic substructuring (DS). 
Over the last decades, several DS techniques were established, making it possible the application of the DS concept in different domains  (see \cite{DK_20081169}), e.g. in the modal domain (see, for example \cite{RM_1968}, \cite{DJ_2004}), in the frequency domain (see, for instance \cite{BJ_198855}, \cite{DK06}) and in the time domain (see, for example \cite{RD_MSSP_2022}, \cite{RD_JSV_2022}). 

In this article, we will focus our attention on the DS techniques clustered under the label State-Space Substructuring (SSS). This group of methods exploits the DS concept in time domain and describes the dynamic behaviour of the components involved in the DS analysis by means of state-space models. For this reason, the first challenge of exploiting these techniques is to construct a state-space model that accurately characterizes the components under study. A possible approach to compute the intended state-space models is to exploit data collected from an experimental modal characterization of these structures. To establish a state-space model from experimentally acquired data, it is a common practice to use a system identification algorithm. Examples of suitable algorithms for this task are the subspace methods (see, \cite{LJ_1999}, \cite{PO_1994},\cite{PO_1996}, \cite{TM_1996}), the rational fractional polynomial method presented in \cite{MR_1982}, the PolyMAX \textcolor{black}{(polyreference least-squares complex frequency-domain)} method proposed by Peeters et al. in \cite{BPet_2004395} and the Maximum Likelihood Modal Parameter method (ML-MM) (see, \cite{MEL_2013}, \cite{MEL_2014}, \cite{MEL_2015}, \cite{MEL_2015567}). Here, the state-space models will be estimated in two different steps. Firstly, both PolyMAX and ML-MM methods will be used to estimate accurate modal parameters from FRFs collected from the experimental modal characterization of the components being analyzed. Then, from the estimated parameters and by establishing tailored residual compensation modes (RCMs) to compensate the contribution of the out-of-band modes in the frequency range of interest, state-space models will be constructed in accordance with the procedures presented in \cite{ME_2022}. 

When identifying state-space models to be used in DS applications, it is not sufficient to estimate a state-space model whose FRFs match the measured ones. In fact, the identified state-space model must also be physically consistent. As reported in \cite{SJO_20072697}, the estimated state-space models must be stable, reciprocal (in case that reciprocity holds for the structure under study), obey \textcolor{black}{Newton's second law} and passive. By constructing the state-space models from the modal parameters estimated by PolyMAX and refined by ML-MM, we have the guarantee that the estimated poles are stable (i.e. their real part will be either zero or negative). Consequently, the estimated state-space models will also be stable. In case that the component under study is reciprocal, one may estimate the modal parameters by assuming a reciprocal modal model as described in \cite{MEL_2015567}. Then, by using the approach described in \cite{ME_2022} to set-up the intended state-space model, a reciprocal model can be obtained.

The computation of state-space models violating \textcolor{black}{Newton's second law} is also unacceptable, as it promotes a bias between the acceleration FRFs of the estimated modal model and the acceleration FRFs of the estimated state-space model (see \cite{RD_ISMA_2022}). Furthermore, a state-space model that does not obey \textcolor{black}{Newton's second law} cannot be properly transformed into \textcolor{black}{coupling} form \cite{SJO_20072697}, turning the elimination of the redundant states originated from coupling and decoupling operations infeasible, unless the state-space models involved in these operations are established in the physical domain \cite{RD_MSSP_2022} (which is not the common situation). \textcolor{black}{In \cite{Bylin_2018}, state-space models are forced to respect Newton's second law by exploiting an iterative approach to re-estimate their input and output matrices. This approach imposes Newton's second law by using linear least-squares constraints, thus, leading to the computation of state-space models verifying Newton's second law in a weak sense. However, as argued in \cite{MG_2020}, imposing this physical law in a weak sense may lead to the computation of state-space models that do not strictly obey Newton's second law and, hence, that are not adequate for being exploited in DS operations. Alternatively, we may exploit the method proposed by Liljerehn in \cite{AL_2016}. This approach includes tailored undamped RCMs on state-space models to force them to respect Newton's second law. This method showed to be accurate to compute state-space models, strictly, respecting Newton's second law, even when dealing with models estimated from experimentally acquired data (see, for example \cite{MG_2020},\cite{RD_ISMA_2022}). Nevertheless, as this approach relies on the use of undamped RCMs, the performance of time-domain simulations with the computed state-space models may lead to instabilities (see \cite{ME_2022}).} 

To be physically consistent, the state-space models must also be passive. The negative consequences of performing coupling/decoupling operations with state-space models violating passivity is that it leads to the computation of non-passive coupled models and, generally, it also leads to the computation of unstable state-space models (i.e. state-space models composed by poles presenting positive real part) \cite{SJO_20072697}. Therefore, it is common that stable time-domain simulations cannot be conducted with state-space models originated from coupling/decoupling operations. Hence, SSS techniques cannot, in general, be exploited to compute state-space models suitable for time-domain analysis, which limits their applications. In literature, several methods to enforce passivity on state-space models can be found. These methods are in general divided into two different classes: the optimal and the sub-optimal algorithms. \textcolor{black}{In the class of the optimal algorithms are included techniques that make the identified state-space models passive by forcing them to verify passivity constraints, such as the Positive or Bounded real lemmas (see \cite{SB_1994})}. At same time, these methods make use of least-square constraints to force the FRFs of the passive model to be as close as possible to the FRFs of the identified model. Examples of this kind of methods can be found in \textcolor{black}{\cite{CC_2004}, \cite{HC_2003}. An example of other techniques belonging to the class of the optimal algorithms can be found in \cite{Calafiore_2012}. This approach imposes passivity on state-space models by finding the perturbation matrix presenting the lowest Frobenius norm that, when summed with the output matrix of a given non-passive state-space model, makes its H-infinity norm (see, for instance, \cite{Chen_1995},\cite{Levine_1996}) lower or equal to 1 and hence, that forces this non-passive state-space model to be passive.} While accurate, \textcolor{black}{the methods belonging to the class of optimal algorithms} hold the disadvantage of demanding high computational effort. This is an important disadvantage, specially, when dealing with state-space models composed by several states, inputs and outputs. This type of models are the ones, generally, found on structural dynamics applications (which, are the target applications of this manuscript) (see \cite{MG_2020}). Conversely, the algorithms labelled as sub-optimal require less computational effort. However, by using these methods, the accuracy of the computed passive state-space models is lower than the one of the state-space models obtained with the optimal algorithms. An example of a sub-optimal method can be found in \cite{SG_2004}. This algorithm makes use of a constrained iterative scheme based on Hamiltonian eigenvalue extraction and perturbation. \textcolor{black}{For an overview of different methodologies to force state-space models to obey passivity, the readers are refereed to \cite{Grivet_2015}}.

\textcolor{black}{The methods described above to impose passivity are not completely suitable to deal with structural dynamics applications involving SSS, because they either lead to the computation of passive models with limited accuracy or they require extremely high computational effort. The computation of accurate passive models with associated high computational costs is, indeed, not acceptable for many applications, where the use of SSS methods is advantageous. For example, applications involving the characterization of the dynamics of an assembled system (composed by several substructures) presenting time-varying dynamic behaviour. To deal with this kind of applications, it is common to compute a linear-parameter varying (LPV) model (see for example, \cite{JC_2009},\cite{JC_2014},\cite{FF_2017},\cite{FF_2007}), which is set-up by using a set of state-space models representative of the dynamics of the assembled system for some pre-selected fixed operating conditions. Therefore, if we intend to set-up a LPV model from, for instance, 5 different coupled state-space models representative of the assembled system for 5 different operating conditions, we are forced to impose passivity on $5 \times n_{sub}$ state-space models, where $n_{sub}$ denotes the number of substructures to be coupled to compute a coupled state-space model. Hence, it is evident that to tackle this kind of applications, a direct method to impose passivity on state-space models that does not require extremely high computational effort and that is still capable of computing accurate passive models is fundamental.}

\textcolor{black}{To avoid the use of \textcolor{black}{optimal algorithms} and their associated high computational costs, while obtaining accurate passive state-space models, Liljerehn and Abrahamsson in \cite{AL_14} proposed a novel method to directly impose passivity on velocity state-space models. This method was developed to target structural dynamics applications involving SSS (which are also the applications that this paper is targeting). By following this approach, passivity is imposed by a minimal adjustment of the elements of both output and input matrices associated with each mode included in the model to make sure that all direct elements of their FRFs are positive real. In \cite{AL_14}, it was shown that by using this approach, an accurate passive model could be obtained from a non-passive model identified from measured FRFs. Moreover, it was also shown that by coupling this passive model with a passive state-space model computed with information extracted from a Finite Element model, an accurate passive stable coupled state-space model could be computed. However, the models involved on the coupling operation were representative of two simple structures. As argued in \cite{MEL_2019325},  forcing the direct elements of the FRFs of all the modes of a state-space model to present positive real part is a too strict requirement that can lead to important deviations between the FRFs of the passive model and the FRFs of the original model. Hence, less accurate results are expected when dealing with more complex structures modelled by high order models. Furthermore, when using this method we need to select a frequency band for which the direct elements of the FRFs of the modes will be analyzed and forced to be positive real. Therefore, we are only sure that passivity is imposed for the frequency points included in the selected frequency band. Thereby, we have no guarantee that the state-space model will be passive for frequencies in between consecutive frequency points and for frequencies outside of the selected frequency band. For this reason, the computed model might be globally non-passive. This is not acceptable, as performing coupling with globally non-passive models, generally, leads to the computation of unstable coupled models.}

\textcolor{black}{Here, we will propose a different strategy to compute stable coupled state-space models. Instead of imposing passivity on the state-space models involved in the coupling/decoupling operations, a procedure to force unstable coupled state-space models to be stable is presented. In this way, we intend to come up with a direct procedure to enable the computation of coupled state-space models that does not necessarily rely on \textcolor{black}{iterative algorithms}. At same time, we aim at mitigate the limitations pointed to the approach presented in \cite{AL_14} by proposing a strategy that is robust enough to deal with problems involving complex mechanical systems modelled by high order state-space models.}

To be suitable for being exploited in DS analysis, the identified state-space models must also present their outputs and inputs placed at the interface of the substructures under study. However, in many cases the placement of sensors and /or actuators at the interfaces of the component under analysis is not feasible. A possible approach to tackle this requirement is the use of the VPT-SS technique proposed in \cite{RD_ISMA_2022}. This method represents the extension of the well-known Virtual Point Transformation (VPT) (see \cite{MV_2013}) technique into the state-space domain. In this way, by assuming local rigid behaviour in the frequency band of interest, virtual points placed at the interface of the substructure under study can be defined. Then, by applying VPT-SS the outputs and inputs of the state-space model, originally, identified from the set of measured FRFs can be transformed into the virtual points locations. Hence, a state-space model, whose outputs and inputs are placed at the interface of the component under analysis (as required by SSS techniques) is obtained. In \cite{RD_ISMA_2022}, VPT-SS demonstrated to be a promising technique, as it makes possible the identification of a more accurate state-space model that is also composed by a lower number of states, when compared with the state-space model that could be computed by identifying a state-space model from the interface FRFs obtained from the application of the VPT approach on the measured FRFs. Moreover, it was shown that unlike the estimation of a state-space model from interface FRFs, by using VPT-SS errors on the transformation matrices involved in the VPT approach can be easily fixed as the state-space model representative of the measured FRFs is available (see, \cite{RD_ISMA_2022}).

With respect to the state-of-the-art, this paper aims at presenting the following novel elements:

\begin{itemize}
    \item develop an accurate procedure to force the estimated state-space models to obey \textcolor{black}{Newton's second law}, without relying on the use of undamped RCMs;
    \item present an approach to force unstable coupled state-space models to be stable, by keeping the quality of the fit between the FRFs of the computed unstable coupled state-space model and the FRFs of the assembled structure;
    \item describe in detail all the required steps to compute reliable stable coupled state-space models (i.e. identification of state-space models from measured FRFs, enforcement of physical constraints, coupling and imposing stability on the coupled state-space model);
    \item validate the whole approach experimentally.
\end{itemize}

The construction of state-space models from estimated modal parameters is presented in section \ref{Constructing state-space models}, while a novel procedure to impose \textcolor{black}{Newton's second law} is developed in section \ref{Imposing the second law of Newton}. Then, in section \ref{Enforcing_Stability} an approach to force unstable coupled state-space models to be stable is introduced, whereas in section \ref{Experimental Validation} the performance of the approaches discussed in this paper is evaluated by exploiting experimentally acquired data. Finally, the conclusions are presented in section \ref{Conclusion}.

\section{Constructing state-space models}\label{Constructing state-space models}

In this section, a brief insight into the construction of state-space models by using  modal parameters estimated from a given set of FRFs is given. Here, we will focus our attention on the construction of a state-space model that includes the information of both in-band and out-of-band modes. For a detailed description of the approach, the readers are forwarded to \cite{ME_2022} and \cite{RD_ISMA_2022}.

The estimation of modal parameters from a given set of measured FRFs is in general performed by using a system identification algorithm. Let us assume that both PolyMAX \cite{BPet_2004395} and Maximum Likelihood Modal Parameter method (ML-MM) \cite{MEL_2015567} are exploited. Hence, the outcome of the identification procedure is a set of poles, modal participation factors and mode shapes, which are the modal parameters associated with the in-band modes, and a lower and an upper residuals matrices, which model the contribution of the lower and upper out-of-band modes in the frequency band of interest, respectively. By using these estimated modal parameters, a modal model can be established as follows (see \cite{ME_2022}):

\begin{equation}\label{eq:modalmodel}
    [H(j\omega)]=\sum_{r=1}^{n_{m}}\left (\frac{\{\psi_{ib,r}\}\{l_{ib,r}\}}{j\omega-\lambda_{ib,r}}+\frac{\{\psi_{ib,r}\}^{*}\{l_{ib,r}\}^{*}}{j\omega-\lambda_{ib,r}^{*}}\right)+\frac{[LR]}{(j\omega)^{2}}+[UR]
\end{equation}
\textcolor{black}{where}, $[H(j\omega)] \in \mathbb{C}^{n_{o} \times n_{i}}$ is the FRF matrix of the modal model, $\lambda_{r}$ is the $r^{th}$ pole, $\{l_{r}\} \in \mathbb{C}^{1 \times n_{i}}$ is the $r^{th}$ modal participation factor, $\{\psi_{r}\} \in \mathbb{C}^{n_{o} \times 1}$ is the $r^{th}$ mode shape, $[LR] \in \mathbb{R}^{n_{o} \times n_{i}}$ and $[UR] \in \mathbb{R}^{n_{o} \times n_{i}}$ are the lower and upper residuals matrices, whose function is modelling the contribution of the lower and upper out-of-band modes in the frequency band of interest, respectively. Finally, superscript $\bullet^{*}$ denotes the complex conjugate of a variable, subscript $ib$ denotes variables associated with the in-band modes of a system, while the variables $n_{o}$, $n_{i}$ and $n_{m}$ represent the number of outputs, inputs and identified modes, respectively.

\textcolor{black}{At this point, by considering the structure of a Linear Time Invariant state-space model and by using the modal parameters associated with the in-band modes, a state-space model can be computed as follows:}

\begin{equation}\label{eq:ss_in_band_modes}
\begin{gathered}
\textcolor{black}{\{\dot{x}_{ib}(t)\}=[A_{ib}]\{x_{ib}(t)\}+[B_{ib}]\{u_{ib}(t)\}}\\
\textcolor{black}{\{y_{ib}(t)\}=[C_{ib}]\{x_{ib}(t)\}}
\end{gathered}
\end{equation}
\textcolor{black}{where}, matrices $[A_{ib}]$, $[B_{ib}]$ and $[C_{ib}]$ are given as follows:

\begin{equation}\label{eq:statespacemodelmodal}
[A_{ib}]=\left[
\begin{matrix}
\Lambda_{ib} & 0\\
0 & \Lambda_{ib}^{*}
\end{matrix}
\right],\ \ \ 
[B_{ib}]=\left[
\begin{matrix}
L_{ib}\\
L_{ib}^{*}
\end{matrix}
\right],\ \ \ 
[C_{ib}]=[
\begin{matrix}
\Psi_{ib} & \Psi_{ib}^{*}
\end{matrix}
]%,\ \ \
%\textcolor{black}{[D_{ib}]=[C_{ib}][A_{ib}][B_{ib}]}
\end{equation}
\textcolor{black}{where}, matrix $[\Lambda] \in \mathbb{C}^{n_{m} \times n_{m}}$ represents a diagonal matrix composed by the poles of the system, $[\Psi] \in \mathbb{C}^{n_{o} \times n_{m}}$ denotes the mode shape matrix and $[L] \in \mathbb{C}^{n_{m} \times n_{i}}$ is the modal participation factors matrix.

In order to compute a complete state-space model (i.e. that includes the contribution of both in-band and out-of-band modes), the contribution of the out-of-band modes must be included into the state-space model given by expression \eqref{eq:ss_in_band_modes}. To start, let us analyze the contribution of the upper out-of-band-modes which is modelled by the upper residual matrix $[UR]$. A possible approach to include the contribution of these modes is to perform the singular value decomposition (SVD) of the matrix $[UR]$ as follows

\begin{equation}\label{eq:UR_SVD}
[UR]=[U_{UR}][\sigma_{UR}][V_{UR}]^{T}=\sum_{r=1}^{n_{UR}}\sigma_{UR,rr}\{U_{UR,r}\}\{V_{UR,r}\}^{T}
\end{equation}
\textcolor{black}{where}, $[U_{UR}] \in \mathbb{R}^{n_{o} \times n_{UR}}$ is the matrix composed by the left singular vectors of $[UR]$, $[V_{UR}] \in \mathbb{R}^{n_{UR} \times n_{i}}$ is the matrix composed by the right singular vectors of $[UR]$ and $[\sigma_{UR}] \in \mathbb{R}^{n_{UR}\times n_{UR}}$ is a diagonal matrix composed by the singular values of $[UR]$ \cite{DJ_2000}. Subscript $UR$ denotes variables associated with the upper residual matrix, whereas $n_{UR}=min(n_{i},n_{o})$. From the SVD decomposition of matrix $[UR]$, we may compute tailored residual compensation modes (RCMs) that are suitable to model the contribution of the upper out-of-band modes. The modal parameters of those RCMs can be computed as follows:

\begin{equation}\label{eq:UR_poles}
\textcolor{black}{\lambda_{UR,r}=-\xi_{UR,r}\omega_{UR,r}+j\omega_{UR,r}\sqrt{1-\xi_{UR,r}^{2}}}
\end{equation}

\begin{equation}\label{eq:UR_mode_shapes}
\{\psi_{UR,r}\}=\frac{\omega_{UR,r}}{\sqrt{1-\xi_{UR,r}^2}}\sqrt{\sigma_{UR,rr}}\{U_{UR,r}\}
\end{equation}

\begin{equation}\label{eq:UR_modal_part}
\{l_{UR,r}\}=-\frac{j}{2}\sqrt{\sigma_{UR,rr}}\{V_{UR,r}\}^{T} 
\end{equation}
\textcolor{black}{where}, $\omega_{UR,r}$ and $\xi_{UR,r}$ are the selected natural frequency and damping ratio of the $r^{th}$ mode of the RCMs computed from the $[UR]$ matrix, respectively. \textcolor{black}{Note that, in general the value of the natural frequencies and damping ratios is selected to be equal for all the RCMs computed from the upper residual matrix (see \cite{RD_ISMA_2022}), hence $\omega_{UR}=\omega_{UR,r}$ and $\xi_{UR}=\xi_{UR,r}$.} From the modal parameters of the computed RCMs (equations \eqref{eq:UR_poles}, \eqref{eq:UR_mode_shapes} and \eqref{eq:UR_modal_part}), a state-space model that models the contribution of the upper out-of-band modes can be defined as follows

\begin{equation}\label{eq:ss_RCMs_UR}
\begin{gathered}
\{\dot{x}_{UR}(t)\}=[A_{UR}]\{x_{UR}(t)\}+\left[\begin{matrix}
B_{UR}
\end{matrix}
\right]\left\{\begin{matrix}
u_{UR}(t)
\end{matrix}
\right\}\\
\{y_{UR}(t)\}=\left[\begin{matrix}
C_{UR}
\end{matrix}
\right]\{x_{UR}(t)\}
\end{gathered}
\end{equation}
with,

\begin{equation}\label{eq:ss_RCMs_UR2}
[A_{UR}]=\left[
\begin{matrix}
\Lambda_{UR} & 0\\
0 & \Lambda_{UR}^{*}
\end{matrix}
\right],\ \ \ 
[B_{UR}]=\left[
\begin{matrix}
L_{UR}\\
L_{UR}^{*}
\end{matrix}
\right],\ \ \ 
[C_{UR}]=[
\begin{matrix}
\Psi_{UR} & \Psi_{UR}^{*}
\end{matrix}
]
\end{equation}
\textcolor{black}{where}, matrices $[\Lambda_{UR}]$, $[L_{UR}]$ and $[\Psi_{UR}]$ are given below.

\begin{equation}\label{eq:UR_ss_matrices}
[\Lambda_{UR}]=\left[
\begin{matrix}
\lambda_{UR,1} & & \\
& \lambda_{UR,2} &\\
& & \ddots
\end{matrix}
\right],\ \ \ 
[L_{UR}]=\left[
\begin{matrix}
l_{UR,1}\\
l_{UR,2}\\
\vdots
\end{matrix}
\right],\ \ \ 
[\Psi_{UR}]=[
\begin{matrix}
\psi_{UR,1} & \psi_{UR,2} & \hdots
\end{matrix}
]
\end{equation}

\color{black}

It can be proved that the higher the selected value for the natural frequencies of the RCMs, the more accurate the RCMs computed from $[UR]$ (see \citep{RD_ISMA_2022}). However, in case that one intends to discretize the constructed state-space models, for example for performing time-domain simulations, the selection of an extremely high value for the natural frequencies of these RCMs is not recommended. Otherwise, the sampling frequency required to properly discretize the constructed state-space model will also be very high, because the sampling frequency used in the discretization is recommended to be at least twice the frequency value of the highest natural frequency of the modes included in the model (see \cite{ME_2022}). Thereby, if very high natural frequencies are used to set-up these RCMs, even higher sampling frequencies will be needed to discretize the model, which may lead to an important increment of the computational effort required to perform calculations with the discretized model.

Furthermore, it can also be proved that as the value of the damping ratios of the RCMs is selected to be lower, more accurate will be the computed RCMs (see \citep{RD_ISMA_2022}). Nevertheless, these RCMs should not be undamped, because the introduction of undamped modes on the constructed state-space models may lead to  numerical instabilities, when performing time-domain simulations with them (see \cite{ME_2022}).

In \cite{RD_ISMA_2022}, it is proposed that the value of $\omega_{UR}$ must respect $\omega_{UR} \ge 5 \times \omega_{max}$ (where, $\omega_{max}$ represents the maximum frequency of interest). Nevertheless, this value should be seen as a bottom limit for the selection of the value of the natural frequencies used to set-up these RCMs. Indeed, in many situations, there is the need to set-up the RCMs by using a higher value for the natural frequencies. In practice, the value selected for the natural frequencies of the RCMs must guarantee that the FRFs of the state-space model computed with them (see expression \eqref{eq:ss_RCMs_UR}) well match the elements of the upper residual matrix, whose contribution for the modal model is constant over frequency (see expression \eqref{eq:modalmodel}).

The value of the damping ratio of these RCMs is suggested to be $0.4$ in \cite{ME_2022}. However, the authors have realized that in the computation of many state-space models, there is no need to use such an high value for the damping ratios. For this reason, we suggest that, as a rule of thumb the value of $\xi_{UR}$ must respect $0.1 \le \xi_{UR} < 1$. Thus, one must start by selecting a damping ratio equal to $0.1$, then if numerical instabilities are found when performing time-domain simulations with the model, the user should reconstruct the state-space model by selecting an higher damping ratio. After selecting a suitable value for the damping ratios of the RCMs, the user must re-check that the FRFs of the state-space model computed from the RCMs well-match the elements of the $[UR]$ matrix in the frequency band of interest. If it is verified that the match is not of enough quality, the value of the natural frequencies of the RCMs must be increased. Note that, the value of $\xi_{UR}$ must remain below $1$, because these RCMs are computed by using a \textcolor{black}{modal} model, which is characterized for being composed by pairs of complex conjugate poles (i.e. by poles, whose damping ratio is lower than 1) (see \cite{ME_2022}, \cite{RD_ISMA_2022}).

\color{black}

To include the contribution of the lower out-of-band modes, a similar approach must be applied. However, for these modes the RCMs must be calculated from the SVD decomposition of matrix \textcolor{black}{$\frac{[LR]}{\omega_{LR}^{2}}$ (see \cite{RD_ISMA_2022})} as follows

\begin{equation}\label{eq:LR_H_LR}
\frac{[LR]}{\omega_{LR}^2}=[U_{LR}][\sigma_{LR}][V_{LR}]^{T}=\sum_{r=1}^{n_{LR}}\sigma_{LR,rr}\{U_{LR,r}\}\{V_{LR,r}\}^{T}
\end{equation}
\textcolor{black}{where}, $[U_{LR}] \in \mathbb{R}^{n_{o} \times n_{LR}}$ and $[V_{LR}] \in \mathbb{R}^{n_{LR} \times n_{i}}$ are the matrices composed by the left and right singular vectors of matrix \textcolor{black}{$\frac{[LR]}{\omega_{LR}^2}$}, respectively, whereas $[\sigma_{LR}] \in \mathbb{R}^{n_{LR}\times n_{LR}}$ is a diagonal matrix composed by the singular values of \textcolor{black}{$\frac{[LR]}{\omega_{LR}^2}$} matrix. Subscript $LR$ denotes variables associated with the \textcolor{black}{lower out-of-band modes}, while $n_{LR}=min(n_{i},n_{o})$. From the variables computed in equation \eqref{eq:LR_H_LR}, RCMs suitable to model the contribution of the lower out-of-bad modes can be defined. The modal parameters of those modes are given as follows

\begin{equation}\label{eq:LR_poles}
\lambda_{LR,r}=-\xi_{LR,r}\omega_{LR,r}+j\omega_{LR,r}\sqrt{1-\xi_{LR,r}^{2}}
\end{equation}

\begin{equation}\label{eq:LR_mode_shapes}
%\{\psi_{LR,r}\}=\frac{\omega_{LR,r}}{\sqrt{1-\xi_{LR,r}^2}}\sqrt{\sigma_{UR,r}}\{U_{UR,r}\} 
\{\psi_{LR,r}\}=\frac{\omega_{LR,r}}{\sqrt{1-\xi_{LR,r}^2}}\sqrt{\sigma_{LR,rr}}\{U_{LR,r}\} 
\end{equation}

\begin{equation}\label{eq:LR_modal_part}
\{l_{LR,r}\}=-\frac{j}{2}\sqrt{\sigma_{LR,rr}}\{V_{LR,r}\}^{T}
\end{equation}
\textcolor{black}{where}, $\omega_{LR,r}$ and $\xi_{LR,r}$ are the selected natural frequency and the damping ratio of the $r^{th}$ mode of the RCMs \textcolor{black}{responsible for modelling the contribution of the lower out-of-band modes}, respectively. \textcolor{black}{Note that, in general the value of the natural frequencies and damping ratios is selected to be equal for all the RCMs responsible for modelling the contribution of the lower out-of-band modes (see \cite{RD_ISMA_2022}), hence $\omega_{LR}=\omega_{LR,r}$ and $\xi_{LR}=\xi_{LR,r}$.} From the modal parameters of the computed RCMs (equations \eqref{eq:LR_poles}, \eqref{eq:LR_mode_shapes} and \eqref{eq:LR_modal_part}) and by following expressions \eqref{eq:ss_RCMs_UR} and \eqref{eq:UR_ss_matrices}, a state-space model that models the contribution of the lower out-of-band modes can be established. It can be proved that the lower the selected natural frequencies, the more accurate the RCMs computed from \textcolor{black}{the lower out-of-band modes.} As rule of thumb, \textcolor{black}{$\omega_{LR} \leq \frac{1}{5}\times \omega_{min}$} (where, $\omega_{min}$ represents the minimum frequency of interest) must be verified (see \cite{RD_ISMA_2022}). Moreover, the lower the damping ratio the more accurate the computed RCMs (see \citep{RD_ISMA_2022}).

\color{black}

The value of the natural frequencies of the RCMs responsible for modeling the lower out-of-band modes can be chosen with more freedom, because decreasing it does not lead to any negative consequence. The selection of the value of the damping ratios of these RCMs must follow the procedure described above to select the value of the damping ratios of the RCMs computed from $[UR]$. To verify the quality of the RCMs responsible for modelling the lower out-of-band modes, one should compare the FRFs of the state-space model constructed from these RCMs with the contribution of the lower out-of-band modes, i.e. $\frac{[LR]}{(j\omega)^{2}}$, in the frequency band of interest (see expression \eqref{eq:modalmodel}). If the match turns out to be good, it means that these RCMs were correctly set-up. Otherwise, one should reconstruct these RCMs by selecting a lower value for their natural frequencies.

\color{black}

After having constructed state-space models from the modal parameters of the RCMs responsible for modelling both lower and upper out-of-band modes, we may establish a complete state-space model as follows

\begin{equation}\label{eq:ss_Full}
\begin{gathered}
\{\dot{x}_{full}(t)\}=[A_{full}]\{x_{full}(t)\}+\left[\begin{matrix}
B_{full}
\end{matrix}
\right]\left\{\begin{matrix}
u_{full}(t)
\end{matrix}
\right\}\\
\{y_{full}(t)\}=\left[\begin{matrix}
C_{full}
\end{matrix}
\right]\{x_{full}(t)\}
\end{gathered}
\end{equation}
\textcolor{black}{where}, the value of the state-space matrices of the state-space model given by expression \eqref{eq:ss_Full} is presented below,

\begin{equation}\label{eq:Full_ss_matrices}
[A_{full}]=\left[
\begin{matrix}
\Lambda_{full} & 0\\
0 & \Lambda_{full}^{*}
\end{matrix}
\right],\ \ \ 
[B_{full}]=\left[
\begin{matrix}
L_{full}\\
L_{full}^{*}
\end{matrix}
\right],\ \ \ 
[C_{full}]=[
\begin{matrix}
\Psi_{full} & \Psi_{full}^{*}
\end{matrix}
]
\end{equation}
\textcolor{black}{where}, matrices $[\Lambda_{full}]$, $[L_{full}]$ and $[\Psi_{full}]$ are given as follows.

\begin{equation}\label{eq:Full_ss_matrices_1}
[\Lambda_{full}]=\left[
\begin{matrix}
\Lambda_{i_b} & 0 & 0\\
0 & \Lambda_{LR} & 0\\
0 & 0 &\Lambda_{UR}
\end{matrix}
\right],\ \ \ 
[L_{full}]=\left[
\begin{matrix}
L_{i_b}\\
L_{LR}\\
L_{UR}
\end{matrix}
\right],\ \ \ 
[\Psi_{full}]=[
\begin{matrix}
\Psi_{i_b} & \Psi_{LR} & \Psi_{UR}
\end{matrix}
]
\end{equation}

As final note, it is important to mention that a similarity transform can be applied to the state-space model given by expression \eqref{eq:ss_Full} in order to obtain a model composed by real-valued matrices \cite{AMRDvMTPATMR_2020},\cite{ME_2022}. In this way, the user has the possibility of substantially decreasing the numerical complexity of the calculations that might be done with the computed complete state-space model.

\section{Imposing \textcolor{black}{Newton's second law}}\label{Imposing the second law of Newton}

To be physically consistent, the estimated state-space models must obey \textcolor{black}{Newton's second law}. \textcolor{black}{Newton's second law} refers that there is a direct relation between force and acceleration, however there are no direct relations between force and displacement or velocity. Hence, the feed-through matrix of both the constructed displacement and velocity state-space models (i.e. the $[D]$ and the $[D^{vel}]$ matrices) must be null. While by definition a displacement state-space model is always composed by a null feed-through matrix, the same does not hold for a velocity model, as its feed-trough matrix is given by the product of the output and input matrices (i.e. $[C][B]$). Thus, we do not have any guarantee that the constructed velocity state-space model will respect \textcolor{black}{Newton's second law}. As mentioned and proved in \cite{RD_ISMA_2022}, if a constructed state-space model does not obey \textcolor{black}{Newton's second law}, its acceleration FRFs will not match the accelerance FRFs of the estimated modal model (see equation \eqref{eq:modalmodel}). Furthermore, in these situations a proper transformation of the state-space model into coupling form (see \cite{SJO_20072697},\cite{RD_MSSP_2022}) is not possible, as the velocity states will not represent the time differentiation of the displacement states. Consequently, one does not have the possibility of computing  minimal-order coupled state-space models, when coupling state-space models that are not established in the physical domain (see \cite{RD_MSSP_2022}).

To identify a solution to enforce \textcolor{black}{Newton's second law} on the constructed state-space model (see equation \eqref{eq:Full_ss_matrices_1}), let us start by computing the FRFs of this model (see \cite{FL_1988}) as given below.

\begin{equation}\label{eq:genericTFofFRF_displacement}
[H(j\omega)]=[C_{full}](j\omega[I]-[A_{full}])^{-1}[B_{full}]
\end{equation}

By differentiating expression \eqref{eq:genericTFofFRF_displacement}, we may compute the FRFs of the correspondent velocity model as follows.

\begin{equation}\label{eq:genericTFofFRF_velocity}
j\omega[H(j\omega)]=[C^{vel}_{full}](j\omega[I]-[A^{vel}_{full}])^{-1}[B^{vel}_{full}]+[D^{vel}_{full}]
\end{equation}
\textcolor{black}{where}, matrices $[A^{vel}_{full}]$, $[B^{vel}_{full}]$, $[C^{vel}_{full}]$ and $[D^{vel}_{full}]$ are given below.

\begin{equation}\label{eq:Full_ss_matrices_0}
[A^{vel}_{full}]=[A_{full}],\ \ \ 
[B^{vel}_{full}]=[B_{full}],\ \ \ 
[C^{vel}_{full}]=[C_{full}][A_{full}],\ \ \
[D^{vel}_{full}]=[C_{full}][B_{full}]
\end{equation}

By developing the product $[C_{full}][B_{full}]$, we obtain the expression given below.

\begin{equation}\label{eq:CB_development}
[C_{full}][B_{full}]=[
\begin{matrix}
\Psi_{ib} L_{ib} + \Psi_{LR} L_{LR} + \Psi_{UR} L_{UR} + \Psi_{ib}^{*}L_{ib}^{*} + \Psi_{LR}^{*}L_{LR}^{*} + \Psi_{UR}^{*}L_{UR}^{*}
\end{matrix}
]
\end{equation}

By computing the column-row expansion (see \citep{HA_2013}) of each of the matrix product involved in equation \eqref{eq:CB_development}, we may rewrite this equation as follows.

\begin{equation}\label{eq:CB_development_column_row_expansion}
\begin{split}
\color{black}[C_{full}][B_{full}]=\Biggl[
& \color{black} \sum_{r=1}^{n_{ib}}\left(\psi_{ib,r} l_{ib,r}\right) + \sum_{r=1}^{n_{LR}}\left(\psi_{LR,r} l_{LR,r}\right) + \sum_{r=1}^{n_{UR}}\left(\psi_{UR,r} l_{UR,r}\right)+\\
& \color{black} \sum_{r=1}^{n_{ib}}\left(\psi_{ib,r}^{*} l_{ib,r}^{*}\right) + \sum_{r=1}^{n_{LR}}\left(\psi_{LR,r}^{*} l_{LR,r}^{*}\right) + \sum_{r=1}^{n_{UR}}\left(\psi_{UR,r}^{*} l_{UR,r}^{*}\right)\Biggr]
\end{split}
\end{equation}

By analyzing equation \eqref{eq:CB_development_column_row_expansion}, it is straightforward to conclude that if the in-band modal parameters are estimated by assuming a proportionally damped modal model $[C_{ib}][B_{ib}]=[0]$ is verified, because the residue matrices \textcolor{black}{$[\psi_{ib,r}][l_{ib,r}]$} will be pure imaginary (see \textcolor{black}{\cite{EB_1996}}). For this reason, $[C_{LR}][B_{LR}]=[0]$ and $[C_{UR}][B_{UR}]=[0]$ will always be verified, as the established RCMs to model the contribution of both lower and upper out-of-band modes are defined by assuming a proportionally damped modal model (see section \ref{Constructing state-space models}). Therefore, \textcolor{black}{as $[C_{full}][B_{full}]=[C_{ib}][B_{ib}]$,} from now on to enforce \textcolor{black}{Newton's second law} we will focus our attention on matrix $[C_{ib}][B_{ib}]$. Before moving on, it is important to mention that matrix $[C_{ib}][B_{ib}]$ will be real, even if the modal parameters are estimated without assuming a proportionally damped modal model.

Even though a state-space model composed by modal parameters estimated by assuming a proportionally damped modal model will always obey \textcolor{black}{Newton's second law}, real mechanical systems, generally, do not present proportional damping. Thus, it is expected that the match quality between the measured FRFs and the estimated modal model by assuming a proportionally damped modal model be lower when compared with the match that would be obtained if the modal model was estimated without performing the mentioned assumption. For this reason, the enforcement of \textcolor{black}{Newton's second law} on the constructed state-space models continues to be worth of study.

\textcolor{black}{To properly force a constructed complete displacement state-space model (see expression \eqref{eq:ss_Full}) to obey \textcolor{black}{Newton's second law}, we must guarantee that the product of its output and input matrices is a null matrix and that the contribution of $[C_{ib}][B_{ib}]$ is properly included on the correspondent complete velocity state-space model (see expression \eqref{eq:genericTFofFRF_velocity}).} A straightforward approach is to follow a similar procedure to the one used to set up the RCMs responsible for modelling the upper out-of-band modes. This is possible, because the contribution of the upper out-of-band modes for the displacement modal model is given by the upper residuals matrix (see equation \eqref{eq:modalmodel}), while to properly impose \textcolor{black}{Newton's second law} we must include the contribution of $[C_{ib}][B_{ib}]$ on the velocity state-space model. In this way, we may perform the SVD decomposition of the $[C_{ib}][B_{ib}]$ matrix as follows 

\begin{equation}\label{eq:CB_SVD}
[C_{ib}][B_{ib}]=[U_{CB}][\sigma_{CB}][V_{CB}]^{T}=\sum_{r=1}^{n_{CB}}\sigma_{CB,rr}\{U_{CB,r}\}\{V_{CB,r}\}^{T}
\end{equation}
\textcolor{black}{where}, $[U_{CB}] \in \mathbb{R}^{n_{o} \times n_{CB}}$ and $[V_{CB}] \in \mathbb{R}^{n_{CB} \times n_{i}}$ are the matrices composed by the left and right singular vectors of matrix $[C_{ib}][B_{ib}]$, respectively, whereas $[\sigma_{CB}] \in \mathbb{R}^{n_{CB}\times n_{CB}}$ is a diagonal matrix composed by the singular values of $[C_{ib}][B_{ib}]$. Subscript $CB$ denotes vectors/matrices associated to the $[C_{ib}][B_{ib}]$ matrix, while the following relation holds for $n_{CB}$, i.e. $n_{CB}=min(n_{i},n_{o})$. From the variables computed in expression \eqref{eq:CB_SVD}, we may set up RCMs, whose modal parameters are given below. 

\begin{equation}\label{eq:CB_poles}
\lambda_{CB,r}=-\xi_{CB,r}\omega_{CB,r}+j\omega_{CB,r}\sqrt{1-\xi_{CB,r}^{2}}\\
\end{equation}

\begin{equation}\label{eq:CB_mode_shapes}
\{\psi_{CB,r}\}=\frac{\omega_{CB,r}}{\sqrt{1-\xi_{CB,r}^2}}\sqrt{\sigma_{CB,rr}}\{U_{CB,r}\}
\end{equation}

\begin{equation}\label{eq:CB_modal_part}
\{l_{CB,r}\}=-\frac{j}{2}\sqrt{\sigma_{CB,rr}}\{V_{CB,r}\}^{T} 
\end{equation}
\textcolor{black}{where}, $\omega_{CB,r}$ and $\xi_{CB,r}$ are the selected natural frequency and the damping ratio of the $r^{th}$ mode of the RCMs responsible for imposing \textcolor{black}{Newton's second law}, respectively. Note that, the value of the natural frequencies and damping ratios can be selected to be equal for all the RCMs responsible for imposing \textcolor{black}{Newton's second law}, hence $\omega_{CB}=\omega_{CB,r}$ and $\xi_{CB}=\xi_{CB,r}$. By using the modal parameters of the computed RCMs (equations \eqref{eq:CB_poles}, \eqref{eq:CB_mode_shapes} and \eqref{eq:CB_modal_part}) and considering that these RCMs were established from the contribution of $[C_{ib}][B_{ib}]$ for the complete velocity model, we may establish the following state-space model

\begin{equation}\label{eq:ss_RCMs_CB}
\begin{gathered}
\{\dot{x}_{CB}(t)\}=[A^{vel}_{CB}]\{x_{CB}(t)\}+[B^{vel}_{CB}]\{u_{CB}(t)\}\\
\{\dot{y}_{CB}(t)\}=[C^{vel}_{CB}]\{x_{CB}(t)\}+[D^{vel}_{CB}]\{u_{CB}(t)\}
\end{gathered}
\end{equation}
\textcolor{black}{where} matrices $[A^{vel}_{CB}]$, $[B^{vel}_{CB}]$, $[C^{vel}_{CB}]$ and $[D^{vel}_{CB}]$ are given as follows

\begin{equation}\label{eq:CB_ss_matrices_4}
[A^{vel}_{CB}]=\left[
\begin{matrix}
\Lambda_{CB} & 0\\
0 & \Lambda_{CB}^{*}
\end{matrix}
\right],\ \ \ 
[B^{vel}_{CB}]=\left[
\begin{matrix}
L_{CB}\\
L_{CB}^{*}
\end{matrix}
\right],\ \ \ 
[C^{vel}_{CB}]=\left[\begin{matrix}
\Psi_{CB} & \Psi^{*}_{CB}
\end{matrix}
\right],\ \ \
[D^{vel}_{CB}]=[C^{vel}_{CB}][A^{vel}_{CB}]^{-1}[B^{vel}_{CB}]
\end{equation}
\textcolor{black}{where}, matrices $[\Lambda_{CB}]$, $[L_{CB}]$ and $[\Psi_{CB}]$ are given below.

\begin{equation}\label{eq:CB_ss_matrices_value_0}
[\Lambda_{CB}]=\left[
\begin{matrix}
\lambda_{CB,1} & & \\
& \lambda_{CB,2} &\\
& & \ddots
\end{matrix}
\right],\ \ \ 
[L_{CB}]=\left[
\begin{matrix}
l_{CB,1}\\
l_{CB,2}\\
\vdots
\end{matrix}
\right],\ \ \ 
[\Psi_{CB}]=[
\begin{matrix}
\psi_{CB,1} & \psi_{CB,2} & \hdots
\end{matrix}
]
\end{equation}

The state-space model established in expression \eqref{eq:ss_RCMs_CB} models the contribution of the computed RCMs for the complete velocity state-space model. However, we aim at directly enforce \textcolor{black}{Newton's second law} on the complete displacement state-space model (see expression \eqref{eq:ss_Full}), because it is simple to calculate the complete velocity state-space model from it, whereas going from the velocity model to the displacement one involves the inversion of \textcolor{black}{the state matrix} (see expression \eqref{eq:Full_ss_matrices_0}). Hence, we must compute the displacement state-space model associated with the velocity state-space model established in expression \eqref{eq:ss_RCMs_CB}. To compute the intended state-space model, we must follow the expressions given below.

\begin{equation}\label{eq:CB_disp_ss_matrices}
[A_{CB}]=[A^{vel}_{CB}],\ \ \ 
[B_{CB}]=[B^{vel}_{CB}],\ \ \ 
[C_{CB}]=[C^{vel}_{CB}][A_{CB}]^{-1},\ \ \
[D_{CB}]=[0]
\end{equation}

By observing expression \eqref{eq:CB_disp_ss_matrices}, we realize that the computation of the inverse of matrix $[A_{CB}]$ is required to set up the output matrix of the displacement state-space model associated with the model given by expression \eqref{eq:ss_RCMs_CB}. However, $[A_{CB}]$ is a diagonal $2n_{CB}\times 2n_{CB}$ matrix composed by the poles of the computed RCMs. For these reasons, during the computation of $[A_{CB}]^{-1}$ it is not expected the introduction of numerical problems associated with ill-condition matrix inversions neither a substantial computational effort to calculate the intended matrix inversion. 

After computing the intended state-space model by following expression \eqref{eq:CB_disp_ss_matrices}, we may define a complete displacement state-space model that obeys \textcolor{black}{Newton's second law} as follows 

\begin{equation}\label{eq:ss_Full_INL}
\begin{gathered}
\{\dot{x}^{INL}_{full}(t)\}=[A^{INL}_{full}]\{x^{INL}_{full}(t)\}+[B^{INL}_{full}]\left\{\begin{matrix}
u^{INL}_{full}(t)
\end{matrix}
\right\}\\
\{y^{INL}_{full}(t)\}=[C^{INL}_{full}]\{x^{INL}_{full}(t)\}
\end{gathered}
\end{equation}
\textcolor{black}{where}, superscript $INL$ denotes variables associated with a state-space model that was forced to verify \textcolor{black}{Newton's second law} and matrices $[A^{INL}]$, $[B^{INL}]$ and $[C^{INL}]$ are given as follows.

\begin{equation}\label{eq:Full_ss_matrices_INL}
[A^{INL}_{full}]=\left[
\begin{matrix}
A_{full} & 0\\
0 & A_{CB}
\end{matrix}
\right],\ \ \ 
[B^{INL}_{full}]=\left[\begin{matrix}
B_{full}\\
B_{CB}
\end{matrix}
\right],\ \ \ 
[C^{INL}_{full}]=\left[\begin{matrix}
C_{full} & C_{CB}
\end{matrix}
\right]
\end{equation}

\color{black}

In \ref{AppendixA}, it is mathematically proven that the RCMs proposed in this section are indeed suitable to impose \textcolor{black}{Newton's second law} on estimated state-space models. To better understand how the accuracy of these RCMs is affected by the selected values for their natural frequencies and damping ratios, we should compute and analyze the expression to obtain the FRFs of the velocity state-space model given by expression \eqref{eq:ss_RCMs_CB}. The steps to compute the mentioned expression are shown in \ref{AppendixA}, here we will just resume the final expression (equation \eqref{eq:Velo_FRF_CB_final}) as follows

\begin{equation}\label{eq:Velo_FRF_CB_final_text}
    [H^{vel}_{CB}(j\omega)]=\frac{\omega_{CB}^{2}[C_{ib}][B_{ib}]}{-\omega^{2}+2j\omega \xi_{CB}\omega_{CB}+\omega_{CB}^{2}}-[C_{ib}][B_{ib}]
\end{equation}
\textcolor{black}{by} observing equation \eqref{eq:Velo_FRF_CB_final_text}, we may realize that its right-hand side presents two different terms. The first one is responsible for making sure that the contribution of $[C_{ib}][B_{ib}]$ is included in the complete velocity state-space model, while the second one is responsible for making sure that $[C^{INL}_{full}][B^{INL}_{full}]=[0]$. Thus, we aim the FRFs of the velocity state-space model computed from  the RCMs responsible for imposing \textcolor{black}{Newton's second law} (see expression \eqref{eq:ss_RCMs_CB}) to be as close as possible to be null. The accuracy of the second term does not depend on the selected values for the natural frequencies and damping ratios, this means that $[C^{INL}_{full}][B^{INL}_{full}]$ will always be close to a null matrix (it will never be a null matrix due to small numerical errors) regardless of the values selected for the natural frequencies and damping ratios. However, the accuracy of the first term is influenced by the values selected for the natural frequencies and damping ratios. By analyzing this term, we can conclude that by increasing the selected value for the natural frequencies of the RCMs and by decreasing the selected value for the damping ratios of the RCMs, the value of this term will tend to $[C_{ib}][B_{ib}]$. Thereby, more accurate
will be the inclusion of the contribution of $[C_{ib}][B_{ib}]$ into the complete velocity state-space model and,
consequently, more accurate will be the RCMs to impose \textcolor{black}{Newton’s second law}. Even though the computed RCMs would be more accurate if they were assumed to be undamped, we must not include undamped modes on the constructed state-space model, otherwise numerical instabilities might be found when performing time-domain simulations with the constructed state-space model (see \cite{ME_2022}).

As a rule of thumb, we suggest that the value selected for $\omega_{CB}$ must respect $10 \times \omega_{max} \le \omega_{CB}$. However, in practice, the value selected for the natural frequencies of the RCMs must guarantee that the first term of the right-hand side of equation \eqref{eq:Velo_FRF_CB_final_text} well matches the elements of $[C_{ib}][B_{ib}]$, whose contribution to the complete velocity state-space model is constant over frequency (see expressions \eqref{eq:genericTFofFRF_velocity} and \eqref{eq:Full_ss_matrices_0}). Hence, when needed, the value selected for the natural frequencies of these RCMs must be increased.

The value selected for the damping ratios of the RCMs must guarantee that no instabilities are found due to the presence of undamped or slightly damped poles when performing time-domain simulations with the constructed state-space model. At same time, the value of the damping ratios must be as small as possible, because the accuracy of the RCMs is deteriorated as the selected value for the damping ratios increases. As a rule of thumb, we propose that the selected value for $\xi_{CB}$ must respect $0.1 \le \xi_{CB} < 1$. Thus, one might start by selecting a damping ratio equal to $0.1$, then if numerical instabilities are found when performing time-domain simulations with the model, the user should reconstruct the state-space model by selecting an higher value for the damping ratios of the RCMs. After selecting a suitable value for the damping ratios, the user must re-check that the first term of the right-hand side of equation \eqref{eq:Velo_FRF_CB_final_text} well-matches $[C_{ib}][B_{ib}]$ in the frequency band of interest. If it is verified that the match is not of enough quality, the natural frequencies of the RCMs must be increased. Note that, the selected value for the damping ratios must remain below 1, because these RCMs are computed by using a \textcolor{black}{modal} model, which is characterized for being composed by pairs of complex conjugate poles (i.e. by poles, whose damping ratio is lower than 1).

\color{black}

As final note, it is worth comparing the approach here presented to force the constructed state-space models to obey \textcolor{black}{Newton's second law} with the one presented in \cite{AL_2016}. In that work, the used RCMs are not complex conjugate and are considered undamped. Hence, we may claim that the approach presented in \cite{AL_2016} has the advantage of using less RCMs. However, besides the advantage of providing the possibility of using damped RCMs instead of undamped ones to force the state-space model to fulfill \textcolor{black}{Newton's second law}, we may outline another important advantage of the approach here proposed over the one presented in \cite{AL_2016}. Let us assume that the state-space model under analysis is a SISO model. By following the approach reported in \cite{AL_2016} and having in mind that $[C_{ib}][B_{ib}]$ is real, the state-space model of the RCMs must be computed as follows

\begin{equation}\label{eq:ss_CB_AL}
[A_{CB,AL}]=\left[
j \omega_{CB,AL}
\right],\ \ \ 
[B_{CB,AL}]=\left[
-\sqrt{\sigma_{CB}}V_{CB}^{T}
\right],\ \ \ 
[C_{CB,AL}]=\left[
U_{CB}\sqrt{\sigma_{CB}}
\right]
\end{equation}
\textcolor{black}{where}, subscript $AL$ denotes the matrices of the state-space model representative of the RCMs used to impose \textcolor{black}{Newton's second law} by following the approach proposed by Liljerehn in \cite{AL_2016}. Hence, by computing the FRFs of the correspondent velocity state-space model, we obtain

\begin{equation}\label{eq:CB_AL}
[H^{vel}_{CB,AL}(j\omega)]=\frac{\omega_{CB,AL}U_{CB}\sigma_{CB} V_{CB}^{T}}{-\omega+\omega_{CB,AL}}-U_{CB}\sigma_{CB} V_{CB}^{T}
\end{equation}

If the same analysis is performed following the approach here presented and assuming \textcolor{black}{$\xi_{CB}=0$}, we obtain the following velocity FRFs.

\begin{equation}\label{eq:CB_RD}
    [H^{vel}_{CB}(j\omega)]=\frac{\omega_{CB}^{2}U_{CB}\sigma_{CB} V_{CB}^{T}}{-\omega^{2}+\omega_{CB}^{2}}-U_{CB}\sigma_{CB} V_{CB}^{T}
\end{equation}

It is worth mentioning that, once again, the first term on the right-hand side of equations \eqref{eq:CB_AL} and \eqref{eq:CB_RD} has the function of including the contribution of $[C_{ib}][B_{ib}]$ into the \textcolor{black}{complete velocity state-space model}, while the second term of the same equations imposes that $[C^{INL}_{full}][B^{INL}_{full}]=[0]$. By observing equations \eqref{eq:CB_AL} and \eqref{eq:CB_RD} and having in mind that $\omega_{CB}>\omega_{max}$, it is straightforward to conclude that it is necessary a smaller value of $\omega_{CB}$ to force the first term on the right-hand side of equation \eqref{eq:CB_RD} to be, approximately, equal to  $[C_{ib}][B_{ib}]$ (as intended) than to force the first term on the right-hand side of equation \eqref{eq:CB_AL} to present the same value. This is an important advantage in case that a discretized state-space model is intended to be obtained from the constructed state-space model, because the sampling frequency used in the discretization is recommended to be at least twice the frequency value of the highest natural frequency of the modes included in the state-space model (see \cite{ME_2022}). Hence, the use of the approach here presented will in general require the use of lower values for \textcolor{black}{$\omega_{CB,r}$}, leading to the possibility of use a lower sampling frequency to discretize the constructed state-space model and, thus decreasing the computational effort required to perform calculations with the discretized model.

Note that, even though we have assumed SISO state-space models to perform the proofs given in this section, the proofs are still valid for Multiple Input Multiple Output (MIMO) state-space models.

\section{Imposing Stability}\label{Enforcing_Stability}

When performing coupling by using SSS techniques, we may end-up with an unstable coupled state-space model (i.e. state-space model presenting real positive poles), specially if the models involved in the coupling operation are not passive. The computation of unstable coupled state-space models reduces the applications of SSS methods, as the performance of time-domain simulations with these models is infeasible. In an attempt to impose stability on the coupled state-space models, we will start by transforming a given displacement coupled model into diagonal form as follows (see \cite{AMRDvMTPATMR_2020}) 

\begin{equation}\label{eq:ss_Full_1}
\begin{gathered}
\{\dot{\bar{x}}_{or,df}(t)\}=[\bar{A}_{or,df}]\{\bar{x}_{or,df}(t)\}+[\begin{matrix}
\bar{B}_{or,df}
\end{matrix}]\{\begin{matrix}
\bar{u}_{or,df}(t)
\end{matrix}\}\\
\{\bar{y}_{or,df}(t)\}=[\begin{matrix}
\bar{C}_{or,df}
\end{matrix}]\{\bar{x}_{or,df}(t)\}
\end{gathered}
\end{equation}
\textcolor{black}{where}, subscript $or$ denotes matrices/vectors related to the unstable coupled state-space model directly obtained from the coupling operation (from now on tagged as original unstable coupled state-space model), subscript $df$ denotes matrices/vectors related to a state-space model transformed into diagonal form, while matrices $[\bar{A}_{or,df}]$, $[\bar{B}_{or,df}]$ and $[\bar{C}_{or,df}]$ are given as follows 

\begin{equation}\label{eq:Full_ss_matrices_value}
[A_{or,df}]=[T_{or}]^{-1}[\bar{A}_{or}][T_{or}],\ \ \ 
[B_{or,df}]=[T_{or}]^{-1}[\begin{matrix}
\bar{B}_{or}
\end{matrix}],\ \ \ 
[C_{or,df}]=[\begin{matrix}
\bar{C}_{or}
\end{matrix}][T_{or}].
\end{equation}

In \textcolor{black}{expression \eqref{eq:Full_ss_matrices_value}}, $[T_{or}]$ is a modal matrix containing all system eigenvectors as columns and can be computed by solving the following eigenvalue problem

\begin{equation}\label{eq:eigemvalue_problem}
[\bar{A}_{or}][T_{or}]=[T_{or}][\Lambda_{or}]
\end{equation}
\textcolor{black}{where}, $[\Lambda_{or}]$ is a diagonal matrix containing the eigenvalues of matrix $[\bar{A}_{or}]$ (which are also the poles of the system). Let us now observe the generic expression to calculate a pole,

\begin{equation}\label{eq:generic_pole_expression}
\lambda,\lambda^{*}=-\xi_{n}\omega_{n} \pm j\omega_{n}\sqrt{1-\xi_{n}^{2}}
\end{equation}
\textcolor{black}{where}, $\omega_{n}$ and $\xi_{n}$ are the natural frequency and the damping ratio associated to a given pole. By observing equation \eqref{eq:generic_pole_expression}, it is straightforward to conclude that having a pole with positive real part can be seen as a pole that presents a negative damping ratio. Obviously, the damping ratio of a pole must always be a positive quantity. Hence, to force the coupled state-space model to be stable, we will start by dividing our original state-space model into two state-space models, a stable one containing the stable poles and an unstable one containing the unstable poles.

\color{black}

As the original coupled state-space model was previously transformed into diagonal form, the state equations of this model are fully decoupled (see \citep{AMRDvMTPATMR_2020}), hence the division of the state-space model into a stable and an unstable model becomes a simple task. To construct the stable state-space model, we must firstly search in the diagonal of $[A_{or,df}]$ matrix the stable poles. Then, by constructing a diagonal matrix with those stable poles, we obtain the state matrix, i.e. $[A_{st}]$, of the stable model. Afterwards, the $[B_{st}]$ matrix is constructed by including the rows of the $[B_{or,df}]$ matrix associated with the stable poles, while matrix $[C_{st}]$ is obtained by including the columns of the $[C_{or,df}]$ matrix associated with the same poles. To set-up the unstable state-space model, the same procedure must be applied, but instead of searching on the diagonal of $[A_{or,df}]$ for the stable poles, we must search for the unstable ones.

\color{black}

To force the unstable state-space model to be stable, we will start by calculating the natural frequency and the damping ratio of each unstable pole by solving the system of equations given as follows

\begin{equation}\label{eq:finding_w_n_xi_n}
\begin{cases}
-\xi_{ut,r}\omega_{ut,r}=\Re(\lambda_{ut,r})\\
\omega_{ut,r}\sqrt{1-\xi_{ut,r}^{2}}=\left|\Im(\lambda_{ut,r})\right|\\
\omega_{ut,r}\ge 0
\end{cases}
\end{equation}
\textcolor{black}{where}, subscript $ut$ denotes variables associated with the unstable state-space model, $|\bullet|$ represents the absolute value of a variable, while $\Re$ and $\Im$ denotes the real and imaginary parts of a variable, respectively.

By observing the system of equations \eqref{eq:finding_w_n_xi_n} and having in mind that the poles are unstable, all the calculated damping ratios will be negative. Thus, to force the unstable model to be stable, we will multiply all the calculated damping ratios by $-1$ and, then we will re-calculate the poles of the state-space model in accordance with equation \eqref{eq:generic_pole_expression}. In this way, a stabilized version of the unstable model is obtained. However, by multiplying the damping ratio of the unstable poles by $-1$, we are modifying the FRFs of the unstable state-space model. Hence, the FRFs of the stabilized model will not properly represent the FRFs of the unstable state-space model. To improve the quality of the FRFs of the stabilized model, one possible approach is to exploit the Least-Squares Frequency Domain (LSFD) estimator (see, for instance \cite{PG_03}). 

The LSFD estimator enables the estimation of the mode shapes and of the lower and upper \textcolor{black}{residual matrices} in a linear least-squares sense, provided that the reference FRFs, the poles and the modal participation factors are known in advance. Hence, to follow this approach, we will assume that the mode shapes of the stabilized state-space model are unknown. To implement LSFD, we are required to rewrite the \textcolor{black}{modal} model (see equation \eqref{eq:modalmodel}) as follows

\begin{equation}\label{eq:LSFD_MM}
\left[\tilde{H}(j\omega)\right]=\left[\begin{matrix}
[\Upsilon] & [LR] & [UR]
\end{matrix}\right]\left[\tilde{A}(L,\lambda,j\omega)\right]
\end{equation}
\textcolor{black}{where}, matrices $\left[\tilde{H}(j\omega)\right] \in \mathbb{R}^{n_{o}\times 2n_{i}n_{f}}$, $[\Upsilon] \in \mathbb{R}^{n_{o}\times 2n_{m}}$ and $\left[\tilde{A}(L,\lambda,j\omega)\right] \in \mathbb{R}^{(2n_{m}+2n_{i})\times 2n_{i}n_{f}}$ (being $n_{f}$ the number of frequency lines) are given as follows

\begin{equation}\label{eq:H_tilde}
\left[\tilde{H}(j\omega)\right]=\left[\begin{matrix}
\Re([H_{target}(j\omega)]_{1} \ \hdots \ [H_{target}(j\omega)]_{n_{f}}) & \Im([H_{target}(j\omega)]_{1} \ \hdots \ [H_{target}(j\omega)]_{n_{f}}) 
\end{matrix}\right]
\end{equation}

\begin{equation}\label{eq:Upsilon}
[\Upsilon]=\left[\Re([\Psi]) \ \ \ \Im([\Psi]) \right]
\end{equation}

\begin{equation}\label{eq:A_tilde}
\left[\tilde{A}(L,\lambda,j\omega)\right]=\left[\begin{matrix}
\left[\tilde{a}_{\Re}(L,\lambda,j\omega)\right]_{1} & \hdots & \left[\tilde{a}_{\Re}(L,\lambda,j\omega)\right]_{n_{f}} &\left[\tilde{a}_{\Im}(L,\lambda,j\omega)\right]_{1} & \hdots & \left[\tilde{a}_{\Im}(L,\lambda,j\omega)\right]_{n_{f}}\\
\left[\tilde{b}_{\Re}(\omega)\right]_{1} & \hdots & \left[\tilde{b}_{\Re}(\omega)\right]_{n_{f}} & [0] & \hdots & [0]
\end{matrix}
\right].
\end{equation}

In \textcolor{black}{expression} \eqref{eq:Upsilon}, matrix $[H_{target}(j\omega)]$ represents the displacement FRF matrix that the stabilized state-space model must present, while matrices $\left[\tilde{a}(L,\lambda,j\omega)\right]$ and $\left[\tilde{b}(\omega)\right]$ must be computed as follows

\begin{equation}\label{eq:Full_a_Re_Im_matrices}
\left[\tilde{a}_{\Re}(L,\lambda,j\omega)\right]_{1}=\left[\begin{matrix}
\Re(\frac{l_{1}}{j\omega_{1}-\lambda_{1}})+\Re(\frac{l^{*}_{1}}{j\omega_{1}-\lambda_{1}^{*}})\\
\vdots\\
\Re(\frac{l_{n_{m}}}{j\omega_{1}-\lambda_{n_{m}}})+\Re(\frac{l^{*}_{n_{m}}}{j\omega_{1}-\lambda_{n_{m}}^{*}})\\
-\Im(\frac{l_{1}}{j\omega_{1}-\lambda_{1}})+\Im(\frac{l^{*}_{1}}{j\omega_{1}-\lambda^{*}_{1}})\\
\vdots\\
-\Im(\frac{l_{n_{m}}}{j\omega_{1}-\lambda_{n_{m}}})+\Im(\frac{l^{*}_{n_{m}}}{j\omega_{1}-\lambda^{*}_{n_{m}}})\\
\end{matrix}
\right],\ \ \ 
\left[\tilde{a}_{\Im}(L,\lambda,j\omega)\right]_{1}=\left[\begin{matrix}
\Im(\frac{l_{1}}{j\omega_{1}-\lambda_{1}})+\Im(\frac{l^{*}_{1}}{j\omega_{1}-\lambda^{*}_{1}})\\
\vdots\\
\Im(\frac{l_{n_{m}}}{j\omega_{1}-\lambda_{n_{m}}})+\Im(\frac{l^{*}_{n_{m}}}{j\omega_{1}-\lambda^{*}_{n_{m}}})\\
\Re(\frac{l_{1}}{j\omega_{1}-\lambda_{1}})-\Re(\frac{l^{*}_{1}}{j\omega_{1}-\lambda^{*}_{1}})\\
\vdots\\
\Re(\frac{l_{n_{m}}}{j\omega_{1}-\lambda_{n_{m}}})-\Re(\frac{l^{*}_{n_{m}}}{j\omega_{1}-\lambda^{*}_{n_{m}}})\\
\end{matrix}
\right]
\end{equation}

\begin{equation}
\left[\tilde{b}_{\Re}(\omega)\right]_{1}=\left[\begin{matrix}
\frac{[I]}{-\omega^{2}_{1}}\\
[I]
\end{matrix}
\right]
\end{equation}
\textcolor{black}{where} $[I]$ represents an identity matrix of dimension $n_{i}\times n_{i}$. Proof for the computation of matrix $\left[\tilde{A}(L,\lambda,j\omega)\right]$ is given in \ref{AppendixB}. Moreover, in the same appendix it is also shown how to construct $\left[\tilde{A}(L,\lambda,j\omega)\right]$, when exploiting LSFD by using velocity or acceleration reference FRFs.

%,\ \ \
%\left[\tilde{b}_{\Re}(\omega)\right]_{1}=\left[\begin{matrix}
%\frac{[I]}{-\omega^{2}_{1}}\\
%[I]
%\end{matrix}
%\right],\ \ \ 
%\left[\tilde{b}_{\Im}(\omega)\right]_{1}=\left[\begin{matrix}
%[0]\\
%[0]
%\end{matrix}
%\right]

By observing expression \eqref{eq:LSFD_MM}, we conclude that the mode shapes, the lower and the upper \textcolor{black}{residual matrices} can be computed in a linear least-squares sense as follows

\begin{equation}\label{eq:LSQ_solution}
\left[\begin{matrix}
[\Upsilon] & [LR] & [UR]
\end{matrix}\right]=\left[\tilde{H}(j\omega)\right]\left[\tilde{A}(L,\lambda,j\omega)\right]^{\dag}
\end{equation}
\textcolor{black}{where}, superscript $\dag$ represents the Moore–Penrose pseudoinverse of a matrix. 

The poles can either be real (in case that \textcolor{black}{$\xi_{n} \ge 1$}) or appear as complex conjugate pairs (see \cite{AMRDvMTPATMR_2020}). As the LSFD estimator \textcolor{black}{makes use of the modal} model, the re-estimation of the mode shapes associated with the real poles is not possible. Hence before exploiting LSFD, we must construct two models from the stabilized state-space model, one containing the real poles and the other one containing the pairs of complex conjugate poles. The partition of the stabilized state-space model into the two mentioned state-space models follows the same procedures described to set-up the stable and unstable state-space models from the original unstable coupled model. 

After dividing the stabilized model, we must identify the modal parameters from the state-space model composed by the pairs of complex conjugate poles. To perform this task, we must identify from its state matrix one pole of each pair of complex conjugate pairs, the correspondent row of its input matrix representative of the associated modal participation factor and the associated column
of its output matrix, which represents the associated mode shape. Before applying LSFD, we must define the FRFs that our state-space model must present (i.e. the target FRFs). Obviously, our goal is that the FRFs of the stabilized model be completely equal to the FRFs of the unstable state-space model, however we should not define those FRFs as the target ones, because the modal parameters of the state-space model composed by the real poles will not be optimized by LSFD. Hence, our target FRFs must exclude the contribution of those poles. Thus, the target FRFs must be computed as follows

\begin{equation}\label{eq:FRF_target_approach_1}
[H_{target}(j\omega)]=[H_{ut}(j\omega)]-[H^{stbz}_{rp}(j\omega)]
\end{equation}
\textcolor{black}{where}, $[H_{ut}]$ and $[H^{stbz}_{rp}]$ are the FRFs of the unstable state-space model and of the model composed by the real poles identified from the stabilized state-space model, respectively. 

After applying LSFD, we will end-up with a set of modal parameters and a lower and a upper \textcolor{black}{residual matrices}, which can be used to set up an optimized state-space model composed by the pairs of complex conjugate poles. Before starting to construct the intended state-space model, we must evaluate the match quality between the FRFs of the modal model that can be computed with the parameters optimized with LSFD and the target FRFs (see equation \eqref{eq:FRF_target_approach_1}). If the match quality is not satisfactory, we must further refine the estimated modal parameters, for example by exploiting the ML-MM approach (see \cite{MEL_2015567}). Once a sufficient accurate modal model is found the desired stable state-space model representative of the original unstable coupled state-space model can be derived using the optimized modal parameters as outlined in \textcolor{black}{section} \ref{Constructing state-space models}. Then, we must concatenate the constructed model, the stabilized model composed by the real poles and the stable model as follows

\begin{equation}\label{eq:stabilized_original_coupled_ss}
\begin{gathered}
\{\dot{\bar{x}}^{stbz}_{or}(t)\}=\left[\begin{matrix}
\bar{A}_{st} & \\
& \bar{A}^{op,stbz}_{pcp} & \\
& & \bar{A}^{stbz}_{rp}
\end{matrix}\right]\{\bar{x}^{stbz}_{or}(t)\}+\left[\begin{matrix}
\bar{B}_{st}\\
\bar{B}^{op,stbz}_{pcp}\\
\bar{B}^{stbz}_{rp}
\end{matrix}
\right]\left\{\begin{matrix}
\bar{u}^{stbz}_{or}(t)
\end{matrix}
\right\}\\
\{\bar{y}^{stbz}_{or}(t)\}=\left[\begin{matrix}
\bar{C}_{st} & \bar{C}^{op,stbz}_{pcp} & \bar{C}^{stbz}_{rp}
\end{matrix}
\right]\{\bar{x}^{stbz}_{or}(t)\}
\end{gathered}
\end{equation}
\textcolor{black}{where}, subscripts $st$ denote matrices/vectors associated to the stable state-space models computed from the original unstable coupled model, while subscripts $pcp$ and $rp$ denote matrices/vectors associated to the state-space models composed by the pairs of complex conjugate poles and by the real poles identified from the stabilized model, respectively. Superscript $stbz$ denotes matrices/vectors related to state-space models that were forced to be stable and superscript $op$ denotes matrices/vectors related to a state-space model constructed with the optimized modal parameters.

Finally, if the state-space model given by expression \eqref{eq:stabilized_original_coupled_ss} does not respect \textcolor{black}{Newton's second law}, it should be forced in agreement with  the approach developed in section \ref{Imposing the second law of Newton}. At the end of this step, a stable coupled state-space model representative of the original unstable coupled model is obtained.

Note that, by following the approach here proposed, the computed stable coupled state-space model will present a higher order than the original unstable coupled state-space model. This happens due to the use of RCMs to include the contribution of both lower and upper residuals matrices computed by the LSFD estimator and, if required, due to the inclusion of another set of RCMs to impose \textcolor{black}{Newton's second law}. Hence, the inclusion of these RCMs might represent an increment of $6 \times min(n_{o},n_{i})$ states. However, as for the state-space models estimated from experimentally acquired data the relation \textcolor{black}{$n_{m}>>min(n_{o},n_{i})$} is usually verified \cite{mg_2013}, the number of added states is, in general, not substantial. 

%However, it is also important to mention that to compute stable coupled state-space models, the approach here developed holds some advantages over imposing passivity on all the state-space models involved on the DS analysis. Firstly, by exploiting the approach here developed we make sure that a stable coupled state-space model is computed. Secondly, we only need to take action over the state-space model resultant from all the coupling/decoupling operations performed. This advantage is more relevant as more models are involved in the DS operations. Lastly, the approach here proposed does not necessarily rely on the use of iterative algorithms. Hence, the approach is simpler to implement (in case that LSFD estimator is exploited) and issues like convergence problems can be avoided.

As final note, we must highlight that even though the approach here discussed was presented as a procedure to force a computed unstable coupled state-space model to be stable, it continues to be valid to impose stability on unstable state-space models resulting from decoupling operations.

\section{Experimental Validation}\label{Experimental Validation}

\subsection{Testing Campaign}\label{Testing Campaign}

To experimentally validate the approaches discussed in this paper, the following components/assemblies were tested:

\begin{itemize}
    \item Two aluminum crosses (from now on labelled as cross aluminum A and B);
    \item Two steel crosses (from now on labelled as cross steel A and B);
    \item Assembly A composed by the two aluminum crosses connected by a rubber mount;
    \item Assembly B composed by the two steel crosses connected by a rubber mount.
\end{itemize}

\begin{figure}
\centering
\begin{subfigure}{0.5\textwidth}
  \centering
  \includegraphics[width=0.6\linewidth]{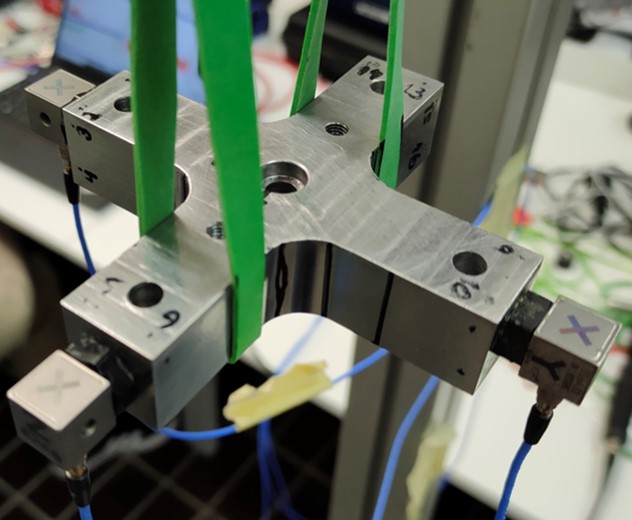}
  \caption{Crosses.}
  \label{fig:cross}
\end{subfigure}%
\begin{subfigure}{0.5\textwidth}
  \centering
  \includegraphics[width=0.6\linewidth]{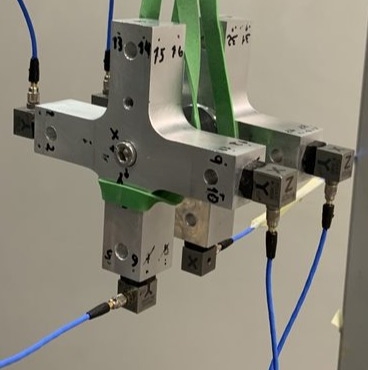}
  \caption{Assemblies.}
  \label{fig:Assembly}
\end{subfigure}
\caption{Test set up used to experimentally characterize the isolated crosses and assemblies \citep{RD_MSSP_2022}.}
\label{fig:test}
\end{figure}

Figures \ref{fig:cross} and \ref{fig:Assembly} show the test set up used to experimentally characterize the crosses and the assemblies, respectively. As we intend to perform decoupling and coupling operations with the identified state-space models, their outputs and inputs must be placed at the interfaces of the components under analysis. However, it is infeasible to excite and place sensors at the interface of the crosses. For this reason, the crosses were design to behave as rigid bodies in the frequency range of interest (i.e. from 20 Hz up to 500 Hz). In this way, we are able to excite and place sensors at accessible locations, then, by exploiting the VPT-SS technique (see \cite{RD_ISMA_2022}) state-space models, whose outputs and inputs are placed at the interfaces of the experimentally characterized components, can be computed.  

\begin{figure}[ht]
\centering
    \includegraphics[width=0.7\textwidth]{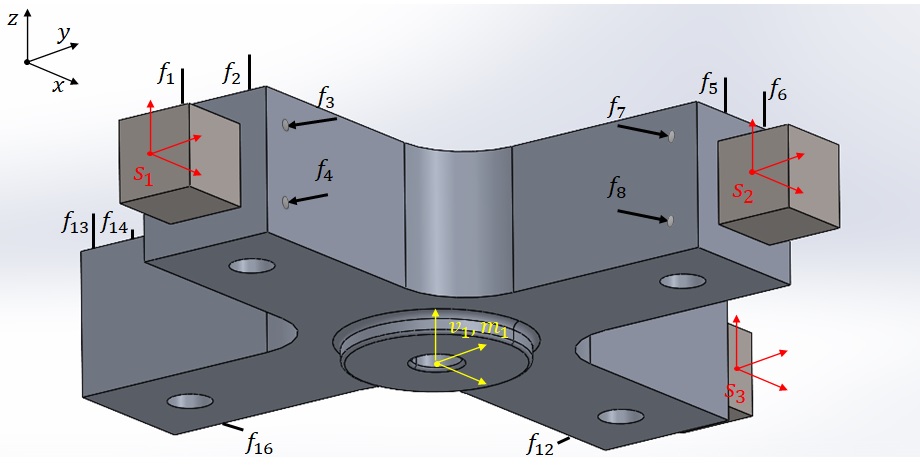}
    \caption{Locations of measurement accelerometers (red), hammer impact directions (black arrows) and virtual point (yellow) \citep{RD_MSSP_2022}.}
     \label{fig:VPT_Cross}
\end{figure}

To perform the experimental modal characterization of the mentioned structures, we exploited the roving hammer testing approach. For each cross, either isolated or included in assemblies $A$ or $B$, we provided excitation at sixteen input locations with an instrumented hammer (PCB Model 086C03). The response of the structure was measured using  three accelerometers (PCB Model TLD356A32) as depicted in Figure \ref{fig:VPT_Cross}. As final note, we must highlight that the use of fixtures with cross shape has the aim of enable an effective excitation of the rotational DOFs, which leads to a reliable six DOFs characterization of each isolated cross and a reliable twelve DOFs characterization of each of the analyzed assemblies (see \cite{MH_2020},\cite{MH_18}).  

\subsection{State-space models identification}\label{State-space models identification}

In this section, state-space models representative of the dynamic behavior of the aluminum and steel crosses and of the assembly A will be identified. No state-space model representative of assembly B will be identified \textcolor{black}{in this section}, because on section \ref{Coupling results} we aim at obtaining a coupled state-space model representative of this assembly (which will be used in section \ref{Stabilization of the coupled state-space model} to validate the approaches discussed in section \ref{Enforcing_Stability}) by using the state-space models identified in this section. 

To identify the intended state-space models, the procedures outlined in sections \ref{Constructing state-space models} and \ref{Imposing the second law of Newton} were followed and the VPT-SS approach presented in \cite{RD_ISMA_2022} was exploited. We started by estimating modal parameters from the collected sets of FRFs. To perform the estimation of the intended parameters, the Simcenter Testlab\textsuperscript{\textregistered} implementation of both PolyMAX and ML-MM methods was used. Moreover, to estimate the modal parameters a proportional damped modal model was not assumed (hence, $[C_{ib}][B_{ib}]\ne[0]$). In this way, we avoid making an assumption that is very likely not suited to model components/assembly involving rubber mounts. Moreover, not assuming proportional damping gives us the possibility to prove the accuracy of the approach proposed \textcolor{black}{in section \ref{Imposing the second law of Newton} to impose \textcolor{black}{Newton's second law} on identified state-space models}.
%In this way, we avoid the performance of an assumption that is probably not suitable to model the components under study, specially both assemblies that are composed by a rubber mount, which could lead to the estimation of state-space models, whose FRFs present a lower quality match with the measured FRFs. Furthermore, by not making the mentioned assumption we have the possibility of evaluate the accuracy of the approach proposed in section \ref{Imposing the second law of Newton} to impose \textcolor{black}{Newton's second law} on the estimated state-space models.

After having estimated the modal parameters from the collected sets of FRFs, state-space models were constructed as discussed in section \ref{Constructing state-space models}. Then, tailored RCMs were established and included into the constructed state-space model in order to include the contribution of the out-of-band modes in the frequency range of interest (see section \ref{Constructing state-space models}) and to impose \textcolor{black}{Newton's second law} as described in section \ref{Imposing the second law of Newton}. To construct all the intended state-space models, the RCMs responsible for compensating the contribution of the out-of-band modes were constructed by assuming $\omega_{LR}=$0.1 Hz, $\omega_{UR}=1.5 \times 10^{4}$ Hz and $\xi_{LR}=\xi_{UR}=$0.1, while the RCMs responsible for imposing \textcolor{black}{Newton's second law} were set up by assuming $\omega_{CB}=1.5 \times 10^{4}$ Hz and $\xi_{CB}=$0.1. At this point, a complete state-space model representative of the tested structures is obtained. However, as we intend to use the identified state-space models for DS applications, the identified models must present outputs and inputs placed at the interface of the components. To transform the original outputs and inputs locations into the intended ones, VPT-SS was exploited (see \cite{RD_ISMA_2022}). \textcolor{black}{From the state-space models obtained by exploiting VPT-SS, passive models were computed by using the approach presented in \cite{AL_14}. By following this methodology, we must select a frequency band for which the FRFs of the modes will be analyzed and forced to be positive real. No hints on how to select this band are provided in \cite{AL_14}. Therefore, to well include the dynamics of the FRFs of all modes included in the identified state-space models, we decided to select a frequency band between 0.01 Hz and 18000 Hz. The frequency interval between consecutive frequency points was chosen to be 0.01 Hz in order to try to avoid possible passivity violations for frequencies in between consecutive frequency points}. 

\begin{figure}
  \begin{subfigure}[t]{1\textwidth}
    \centering
    \includegraphics[width=.8\textwidth]{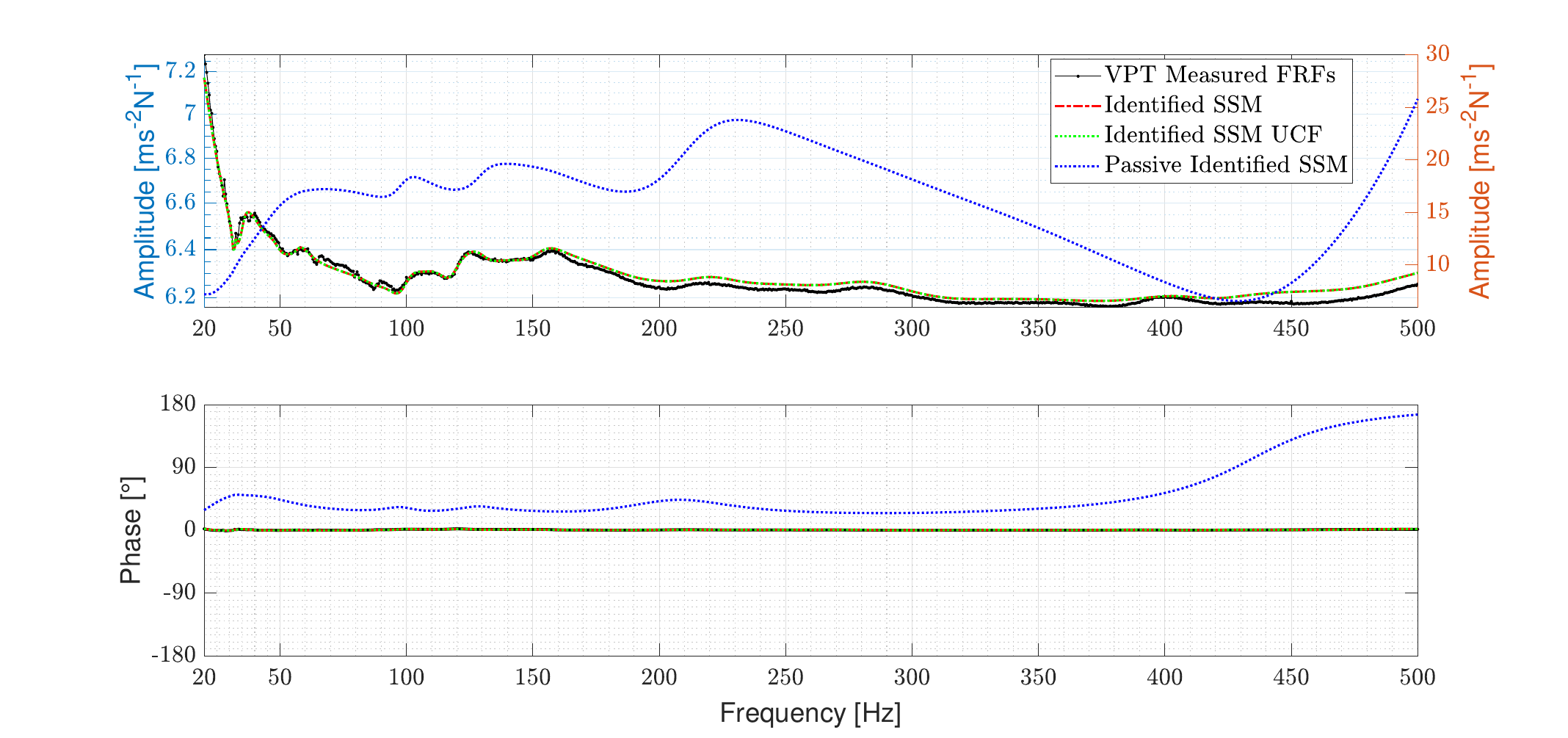}
    \caption{}
     \label{fig:Identified_VPT_Cross_Aluminum_A}
  \end{subfigure}
  %\hfill
  \\
  \medskip
  \begin{subfigure}[t]{1\textwidth}
    \centering
    \includegraphics[width=.8\textwidth]{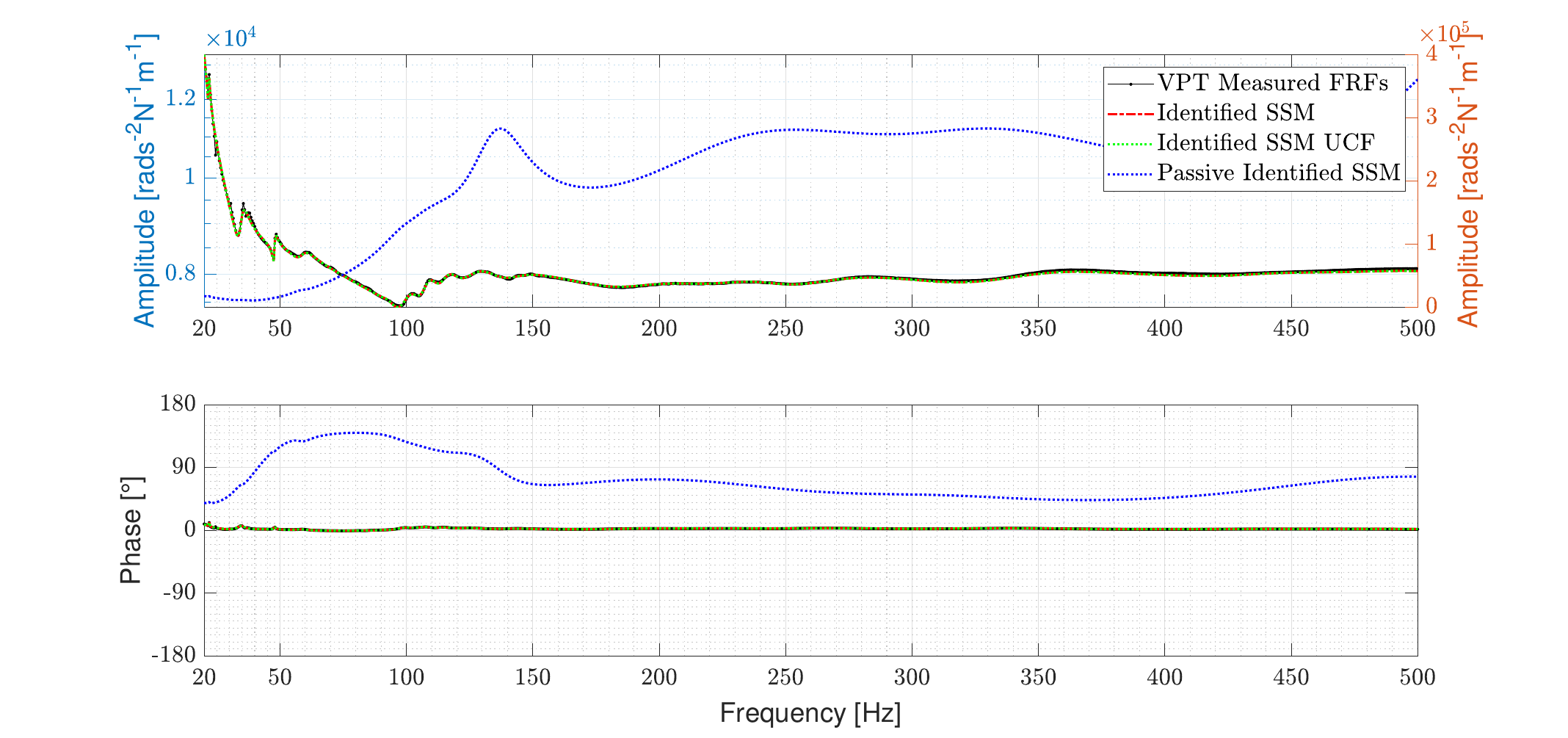}
    \caption{}
     \label{fig:Identified_VPT_Cross_Aluminum_B}
  \end{subfigure}
\caption{\textcolor{black}{Comparison of interface FRFs: i) obtained by applying VPT on the measured FRFs (black solid line - colour version only) (the amplitudes of these FRFs must be evaluated by using the left y axis); ii) from the identified SSM (red dashed line - colour version only) (the amplitudes of these FRFs must be evaluated by using the left y axis); iii) from the identified SSM transformed into UCF (\textcolor{black}{green} dotted line - colour version only) (the amplitudes of these FRFs must be evaluated by using the left y axis); iv) from the passive model computed according to the approach presented in \cite{AL_14} (blue dotted line - colour version only) (the amplitudes of these FRFs must be evaluated by using the right y axis):} a) FRF of the aluminum cross A, whose output is $v_{1}^{y}$ and the input is $m_{1}^{y}$; b) FRF of the aluminum cross B, whose output is $v_{2}^{R_{x}}$ and the input is $m_{2}^{R_{x}}$.}
  \label{fig:Identified_VPT_State_Space_Models_1}
\end{figure}

\begin{figure}
  \begin{subfigure}[t]{1\textwidth}
    \centering
    \includegraphics[width=.8\textwidth]{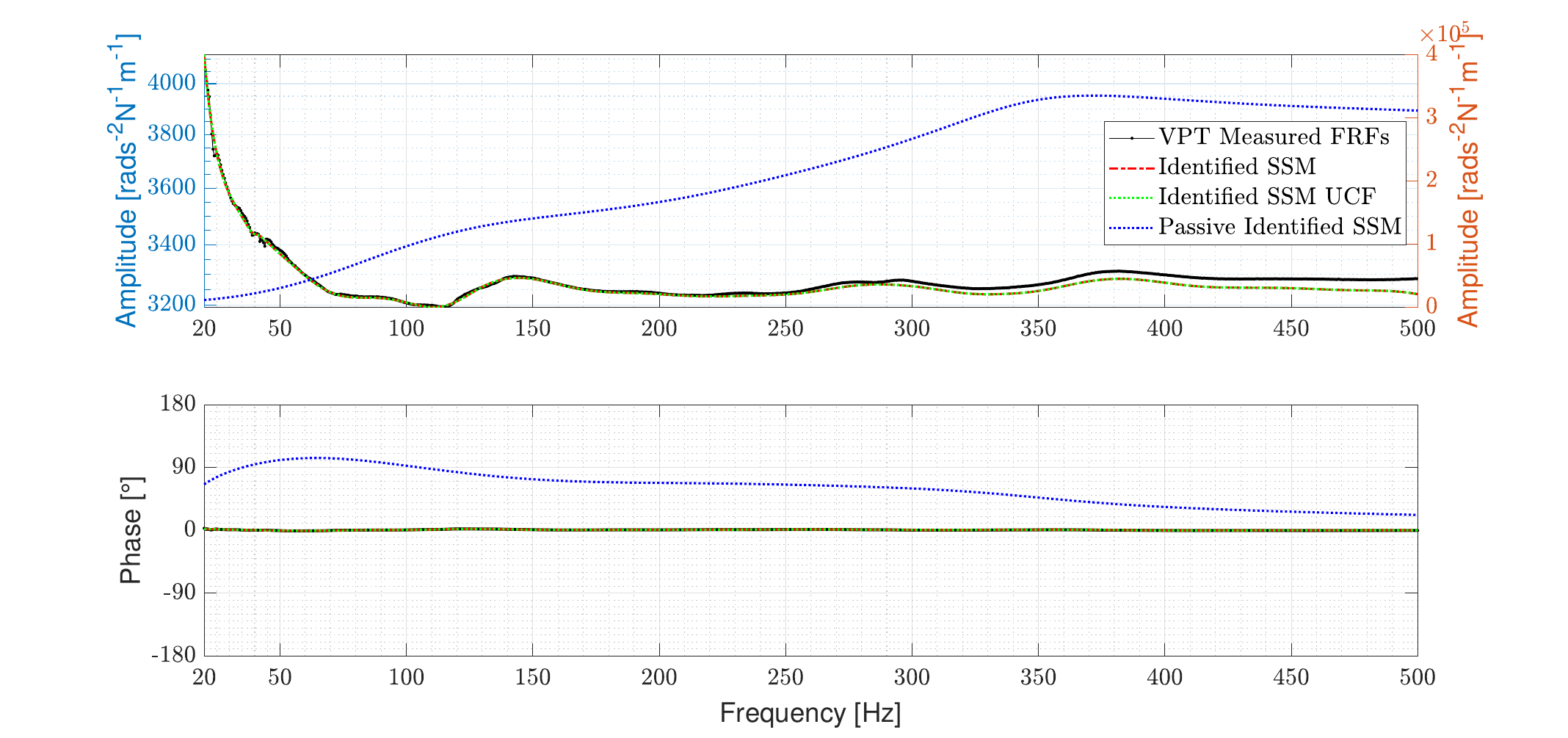}
    \caption{}
     \label{fig:Identified_VPT_Cross_Steel_A}
  \end{subfigure}
  %\hfill
  \\
  \medskip
  \begin{subfigure}[t]{1\textwidth}
    \centering
    \includegraphics[width=.8\textwidth]{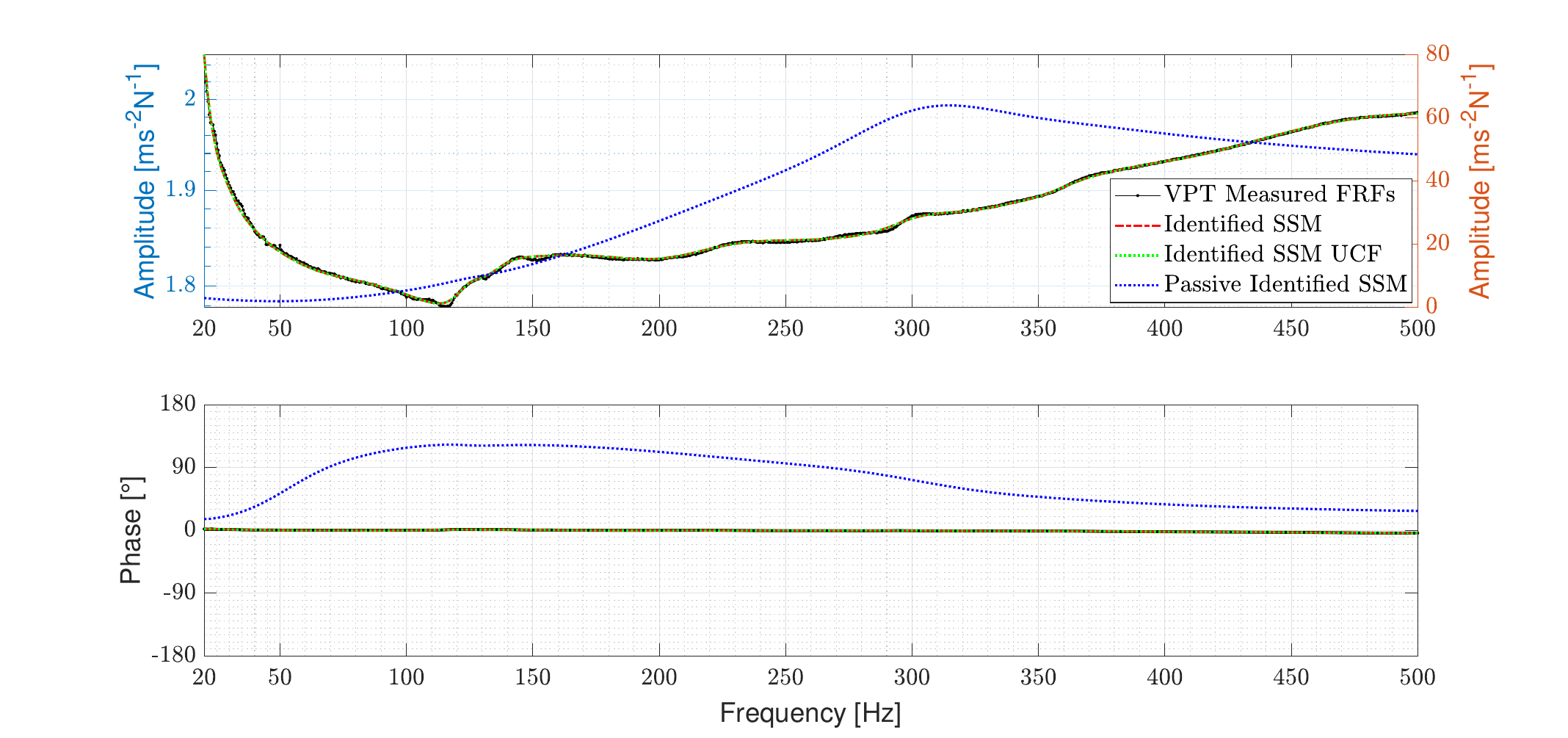}
    \caption{}
     \label{fig:Identified_VPT_Cross_Steel_B}
  \end{subfigure}
\caption{\textcolor{black}{Comparison of interface FRFs: i) obtained by applying VPT on the measured FRFs (black solid line - colour version only) (the amplitudes of these FRFs must be evaluated by using the left y axis); ii) from the identified SSM (red dashed line - colour version only) (the amplitudes of these FRFs must be evaluated by using the left y axis); iii) from the identified SSM transformed into UCF (\textcolor{black}{green} dotted line - colour version only) (the amplitudes of these FRFs must be evaluated by using the left y axis); iv) from the passive model computed according to the approach presented in \cite{AL_14} (blue dotted line - colour version only) (the amplitudes of these FRFs must be evaluated by using the right y axis):} a) FRF of the steel cross A, whose output is $v_{1}^{R_{x}}$ and the input is $m_{1}^{R_{x}}$; b) FRF of the steel cross B, whose output is $v_{1}^{z}$ and the input is $m_{1}^{z}$.}
  \label{fig:Identified_VPT_State_Space_Models_2}
\end{figure}

\begin{figure}
  \begin{subfigure}[t]{1\textwidth}
    \centering
    \includegraphics[width=.8\textwidth]{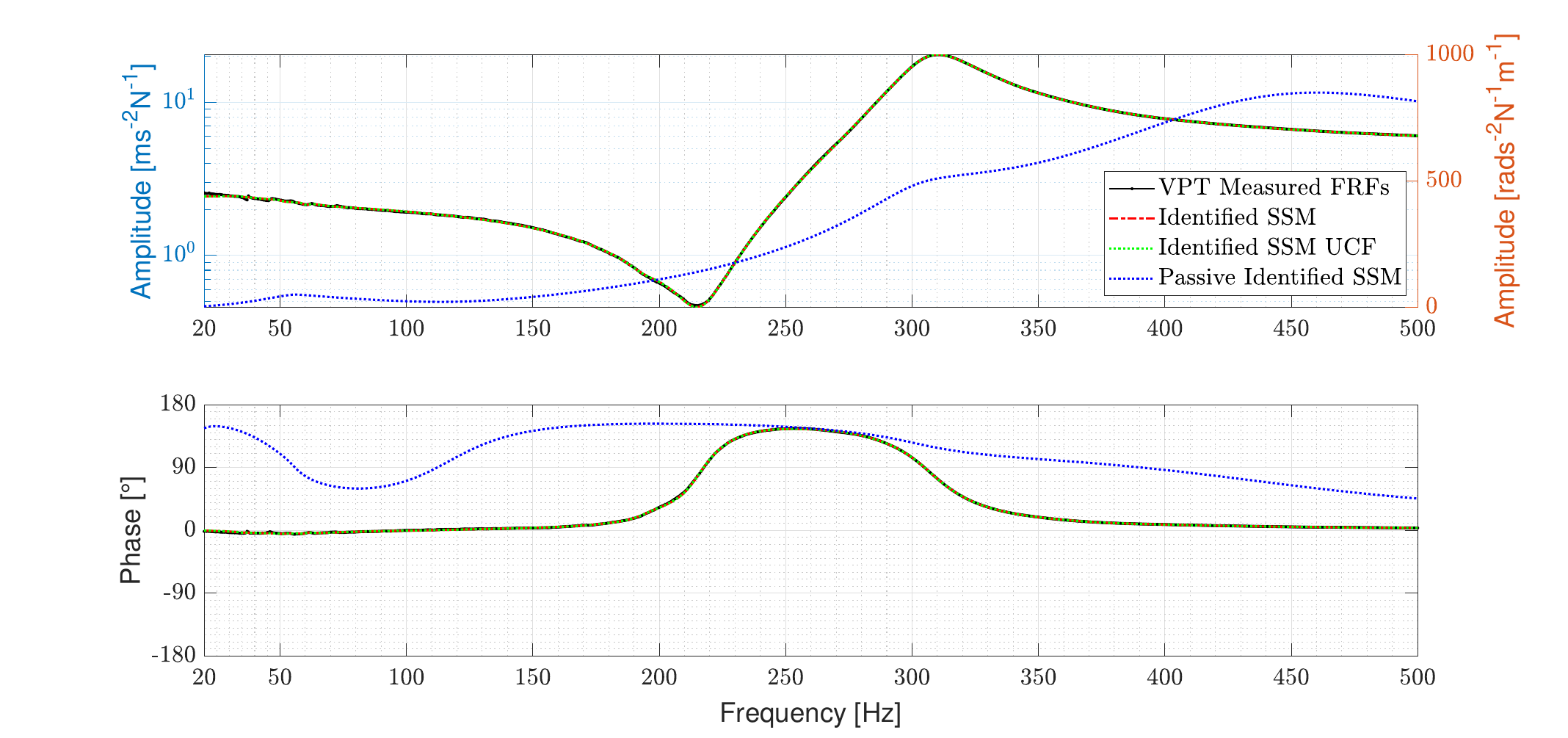}
    \caption{}
\label{fig:Identified_VPT_Assembly_Aluminum_driving_point_FRF}
  \end{subfigure}
  %\hfill
  \\
  \medskip
  \begin{subfigure}[t]{1\textwidth}
    \centering
    \includegraphics[width=.8\textwidth]{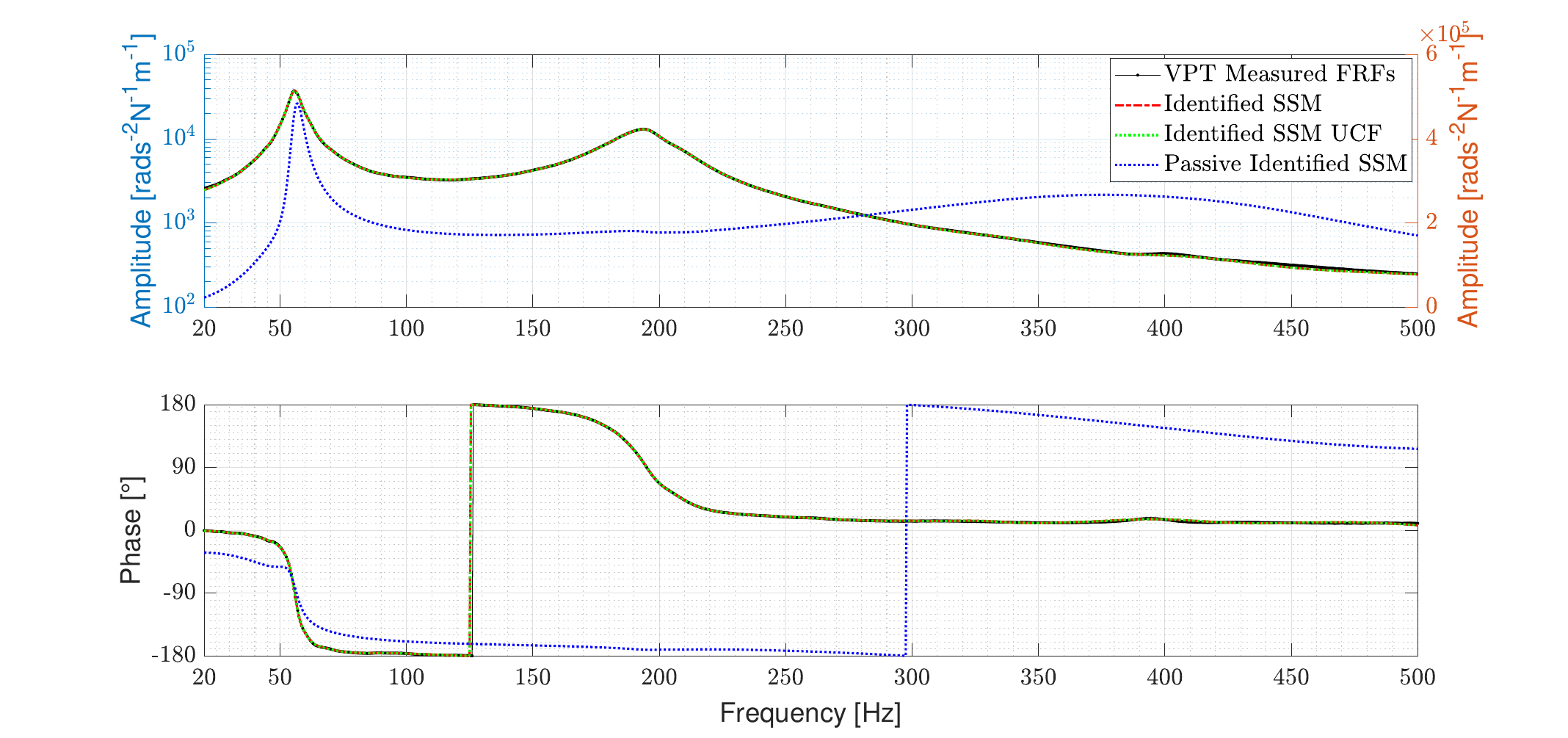}
    \caption{}
     \label{fig:Identified_VPT_Assembly_Aluminum}
  \end{subfigure}
\caption{\textcolor{black}{Comparison of interface FRFs: i) obtained by applying VPT on the measured FRFs (black solid line - colour version only)  (the amplitudes of these FRFs must be evaluated by using the left y axis); ii) from the identified SSM (red dashed line - colour version only) (the amplitudes of these FRFs must be evaluated by using the left y axis); iii) from the identified SSM transformed into UCF (\textcolor{black}{green} dotted line - colour version only) (the amplitudes of these FRFs must be evaluated by using the left y axis); iv) from the passive model computed according to the approach presented in \cite{AL_14} (blue dotted line - colour version only) (the amplitudes of these FRFs must be evaluated by using the right y axis):} a) FRF of the assembly A, whose output is $v_{2}^{z}$ and the input is $m_{2}^{z}$; b) FRF of the assembly A, whose output is $v_{1}^{R_{y}}$ and the input is $m_{2}^{R_{y}}$.}
  \label{fig:Identified_VPT_State_Space_Models_3}
\end{figure}

\textcolor{black}{Figures \ref{fig:Identified_VPT_State_Space_Models_1}, \ref{fig:Identified_VPT_State_Space_Models_2} and \ref{fig:Identified_VPT_State_Space_Models_3} show, for each identified state-space model, the comparison of the FRFs obtained by applying VPT on the measured FRFs with the same FRFs of the identified state-space model, with the same FRFs of the identified model transformed into unconstrained coupling form (UCF) (see \cite{RD_MSSP_2022}) and with the same FRFs of the passive model estimated by exploiting the approach proposed in \cite{AL_14}. Indeed, the transformation of the identified state-space models into UCF is interesting, because DS operations are to be conducted with these models. In this way, we have the possibility to exploit the post-processing procedures presented in \cite{RD_2021} to eliminate the redundant states originated from the DS operations (from either decoupling or/and coupling operations).} Note that, in the caption of Figures \ref{fig:Identified_VPT_State_Space_Models_1}, \ref{fig:Identified_VPT_State_Space_Models_2}, \ref{fig:Identified_VPT_State_Space_Models_3} and through all the paper, the virtual point loads of the crosses aluminum and steel A will be tagged as $m_{1}$, while the virtual point loads of the crosses aluminum and steel B will the denoted as $m_{2}$. The virtual point responses of the crosses aluminum and steel A will be labelled as $v_{1}$, whereas the virtual point responses of the crosses aluminum and steel B will be referred as $v_{2}$ .

By observing Figures \ref{fig:Identified_VPT_State_Space_Models_1}, \ref{fig:Identified_VPT_State_Space_Models_2} and \ref{fig:Identified_VPT_State_Space_Models_3}, it is straightforward that all the state-space models were successfully identified. Moreover, it was verified that the matrix \textcolor{black}{$[C^{INL}_{full}][B^{INL}_{full}]$} of the state-space model representative of assembly A presented the element with the maximum absolute value, whose value was, approximately, $1 \times 10^{-12}$. Hence, we may claim that the approach proposed in section \ref{Imposing the second law of Newton} is accurate to properly force the estimated state-space models to verify \textcolor{black}{Newton's second law}. \textcolor{black}{Conversely, the passive state-space models computed with the approach proposed in \cite{AL_14} showed to present a poor accuracy. The poor results are not a complete surprise and are a direct consequence of the small adjustments that are performed in the elements of the output and input matrices to ensure passivity. As the identified state-space models are composed by an high number of modes, these small adjustments sum up leading to the observed important deviations between the FRFs of the identified and passive models. Besides their poor accuracy, the passive models revealed to not be globally passive, as passivity violations were verified at frequencies outside of the selected frequency band. This is a direct consequence of enforcing passivity by taking into account a specific frequency band. When following an approach like this one, it is common to end-up with a state-space model that verifies passivity only at the frequency points included on the selected frequency band (see, \cite{CC_2004}). Moreover, it was also verified that the passive model did not respect \textcolor{black}{Newton's second law}. This was indeed expected, because the passive state-space model is obtained by adjusting the elements of both output and input matrices without using any constraint to make sure that the model verifies \textcolor{black}{Newton's second law}. Due to the lack of accuracy and global passivity of the state-space models computed with the technique proposed in \cite{AL_14}, these models will not be used in the DS operations that will be performed in section \ref{Coupling results}.}

It is worth mentioning that the computed state-space models representative of the aluminium crosses A and B presented 204 and 210 states, respectively. The state-space models representative of the steel crosses A and B presented 206 and 192 states, respectively. The state-space model of assembly A presented 260 states. Obviously, the structures under analysis do not present so many modes in the frequency band of interest. Nevertheless, as we intended to compute state-space models, whose FRFs match as perfectly as possible the FRFs obtained by applying VPT on the correspondent measured FRFs, the computation of the identified state-space models was performed by including a large number of modes. Moreover, it was expected that the estimated state-space models representative of the aluminum crosses  A and B presented the same number of states (the same was expected for the crosses steel A and B). However, as the FRFs of these components were experimentally acquired, they did not perfectly match. This is possibly due to small differences between either the mechanical properties of both crosses (e.g. different mass) or \textcolor{black}{due to small differences in the boundary conditions during the performance of each experimental modal characterization tests}. Hence, the number of identified in-band modes was different, leading to the computation of state-space models composed by a different number of states. 

Comparing the method developed in this paper to impose \textcolor{black}{Newton's second law} with the one proposed in \cite{AL_2016}, it is evident that the method here suggested holds the advantage of not relying on the use of undamped RCMs. Besides this advantage, in section \ref{Imposing the second law of Newton} we mentioned that the proposed approach enables the computation of accurate RCMs by selecting a lower natural frequency than the one required by the method proposed by Liljerehn in \cite{AL_2016}. To demonstrate the last mentioned advantage in a real context, we computed two additional state-space models representative of assembly A. Both of these models were constructed by following the procedures outlined in section \ref{Constructing state-space models}. However, to impose \textcolor{black}{Newton's second law} on the first state-space model we exploited the approach suggested in \cite{AL_2016} (state-space model A), while to impose the same law on the second model we have used the technique here proposed (state-space model B). During the computation of both models, the RCMs responsible for modelling the contribution of the out-of-band modes were computed by assuming $\omega_{LR}=$0.1 Hz, $\omega_{UR}=1.5 \times 10^{4}$ Hz and $\xi_{LR}=\xi_{UR}=$0.1. While, the RCMs responsible for imposing \textcolor{black}{Newton's second law} were constructed by assuming $\omega_{CB}=5 \times 10^{3}$ Hz and $\xi_{CB}=$0. 

\begin{figure}
  \begin{subfigure}[t]{1\textwidth}
    \centering
    \includegraphics[width=.8\textwidth]{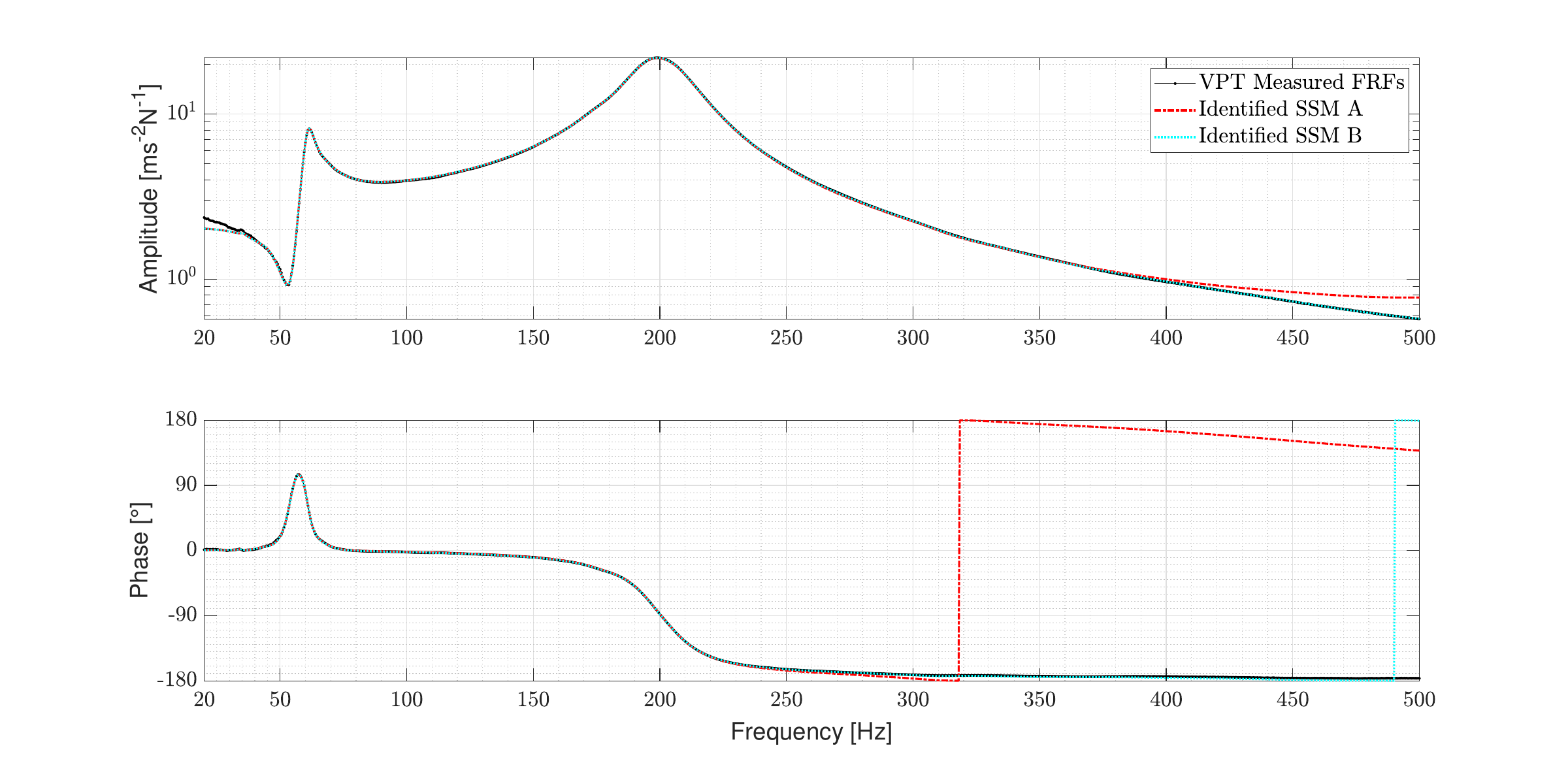}
    \caption{}
     \label{fig:Identified_VPT_Cross_Aluminum_A_comparison_AL_RD_8_2}
  \end{subfigure}
  %\hfill
  \\
  \medskip
  \begin{subfigure}[t]{1\textwidth}
    \centering
    \includegraphics[width=.8\textwidth]{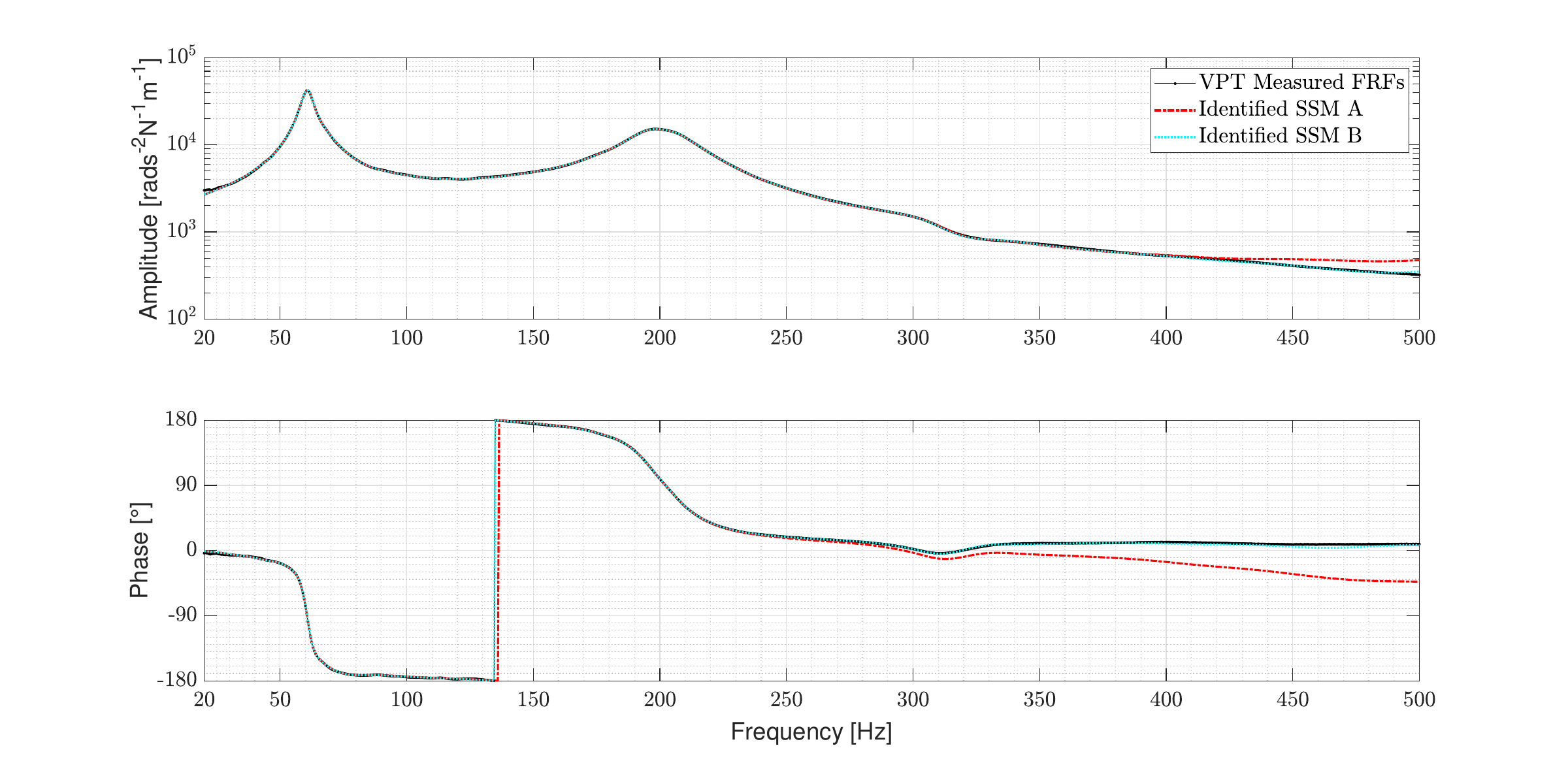}
    \caption{}
     \label{fig:Identified_VPT_Cross_Aluminum_A_comparison_AL_RD_10_4}
  \end{subfigure}
\caption{Comparison of \textcolor{black}{some} interface FRFs obtained by applying VPT on the measured FRFs of assembly A with the same FRFs of two estimated state-space models. State-space model A, which was set up by imposing \textcolor{black}{Newton's second law} by following the approach presented in \cite{AL_2016} (whose FRFs are represented by a red dashed line) and state-space model B, which was constructed by imposing \textcolor{black}{Newton's second law} by exploiting the approach proposed in section \ref{Imposing the second law of Newton} (whose FRFs are represented by a cyan dotted line): a) FRF of the assembly A, whose output is $v_{2}^{y}$ and the input is $m_{1}^{y}$; b) FRF of the assembly A, whose output is $v_{2}^{R_{x}}$ and the input is $m_{1}^{R_{x}}$.}
  \label{fig:Comparison_enforcing_Newton_Second_Law}
\end{figure}

By observing Figure \ref{fig:Comparison_enforcing_Newton_Second_Law} it is evident that the match quality between the FRFs obtained by applying VPT on the measured FRFs (i.e. reference FRFs) and the FRFs of the state-space model B outperformed the quality of the match between the reference FRFs and the FRFs of the state-space model A, specially at higher frequencies (see equations \eqref{eq:CB_AL} and \eqref{eq:CB_RD}). This was indeed expected, hence we conclude that as demonstrated in section \ref{Imposing the second law of Newton}, the approach here proposed to impose \textcolor{black}{Newton's second law} does not require the construction of RCMs by assuming a natural frequency as higher as the one required by the method proposed in \cite{AL_2016}.

\subsection{Coupling results}\label{Coupling results}

In this section, we aim at computing a coupled state-space model representative of assembly B, which will be used in section \ref{Stabilization of the coupled state-space model} to demonstrate the accuracy of the approaches discussed in section \ref{Enforcing_Stability} to enforce stability on coupled state-space models. To compute the intended coupled state-space model by using the state-space models identified in section \ref{State-space models identification}, we must perform two different DS operations. Firstly, we must decouple the state-space models of both aluminum crosses from the model of assembly A to obtain a state-space model representative of the rubber mount. Afterwards, we have to couple the state-space models of both steel crosses with the model representative of the rubber mount to obtain a coupled state-space model of assembly B. To perform the mentioned DS operations, we will exploit the LM-SSS \textcolor{black}{(Lagrange Multiplier State-Space Substructuring) technique (see \cite{BK_2020},\cite{RD_2021},\cite{RD_MSSP_2022})}. Moreover, the state-space models involved in the mentioned operations will be previously transformed into UCF. \textcolor{black}{In this way, by exploiting the post-processing procedure presented in \cite{RD_2021}, which involves the use of a state Boolean localization matrix, the redundant states originated from both decoupling and coupling operations can be eliminated, leading to the computation of a minimal-order coupled state-space model representative of assembly B (see \cite{RD_2021},\cite{RD_MSSP_2022}).} Note that all the DS operations must be performed in accordance with expressions presented in \cite{RD_MSSP_2022} to compute displacement state-space models, otherwise, the procedures outlined in section \ref{Enforcing_Stability} to impose stability on the coupled state-space model cannot be applied.

\begin{figure}
  \begin{subfigure}[t]{1\textwidth}
    \centering
    \includegraphics[width=.8\textwidth]{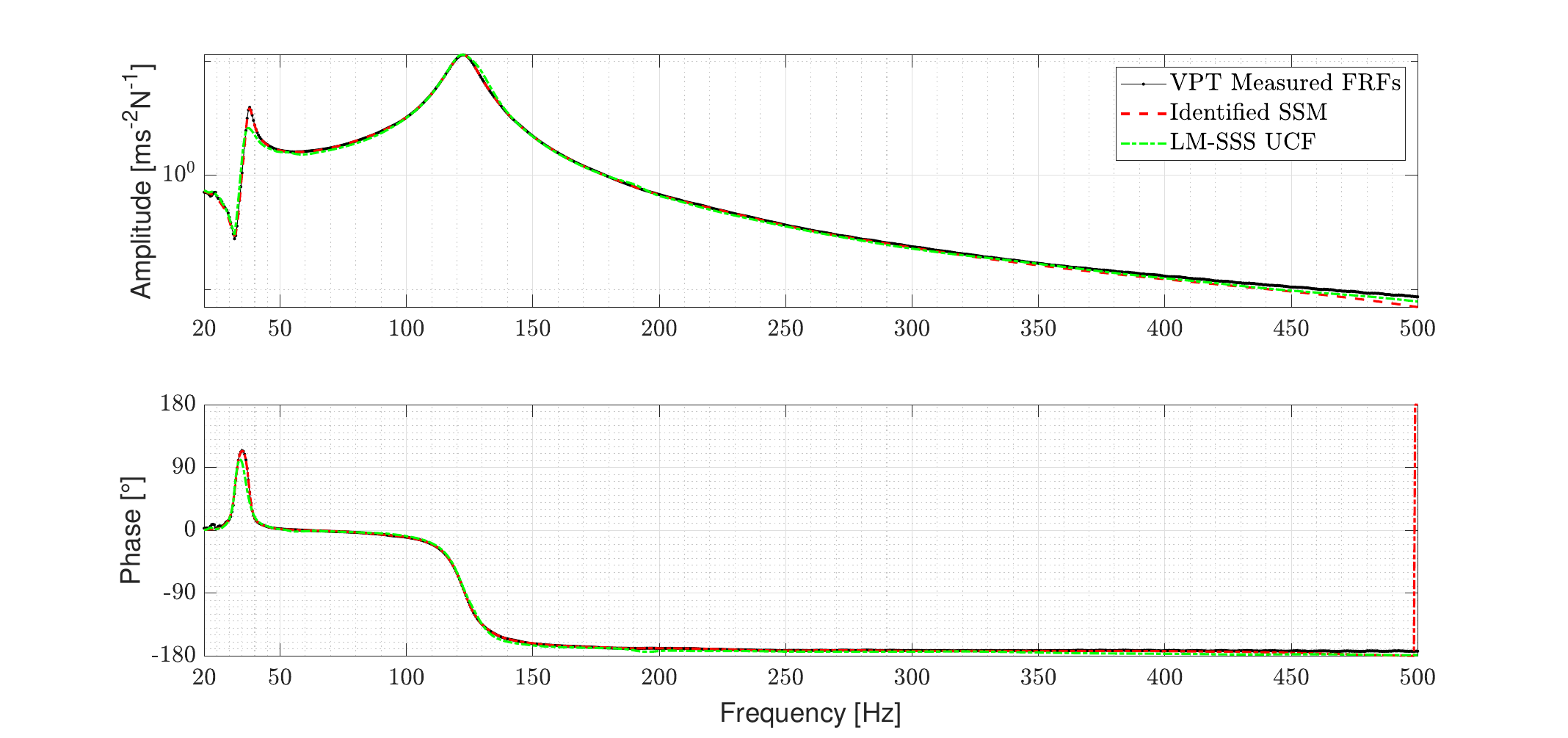}
    \caption{}
     \label{fig:Coupled_Assembly_B_7_1}
  \end{subfigure}
  %\hfill
  \\
  \medskip
  \begin{subfigure}[t]{1\textwidth}
    \centering
    \includegraphics[width=.8\textwidth]{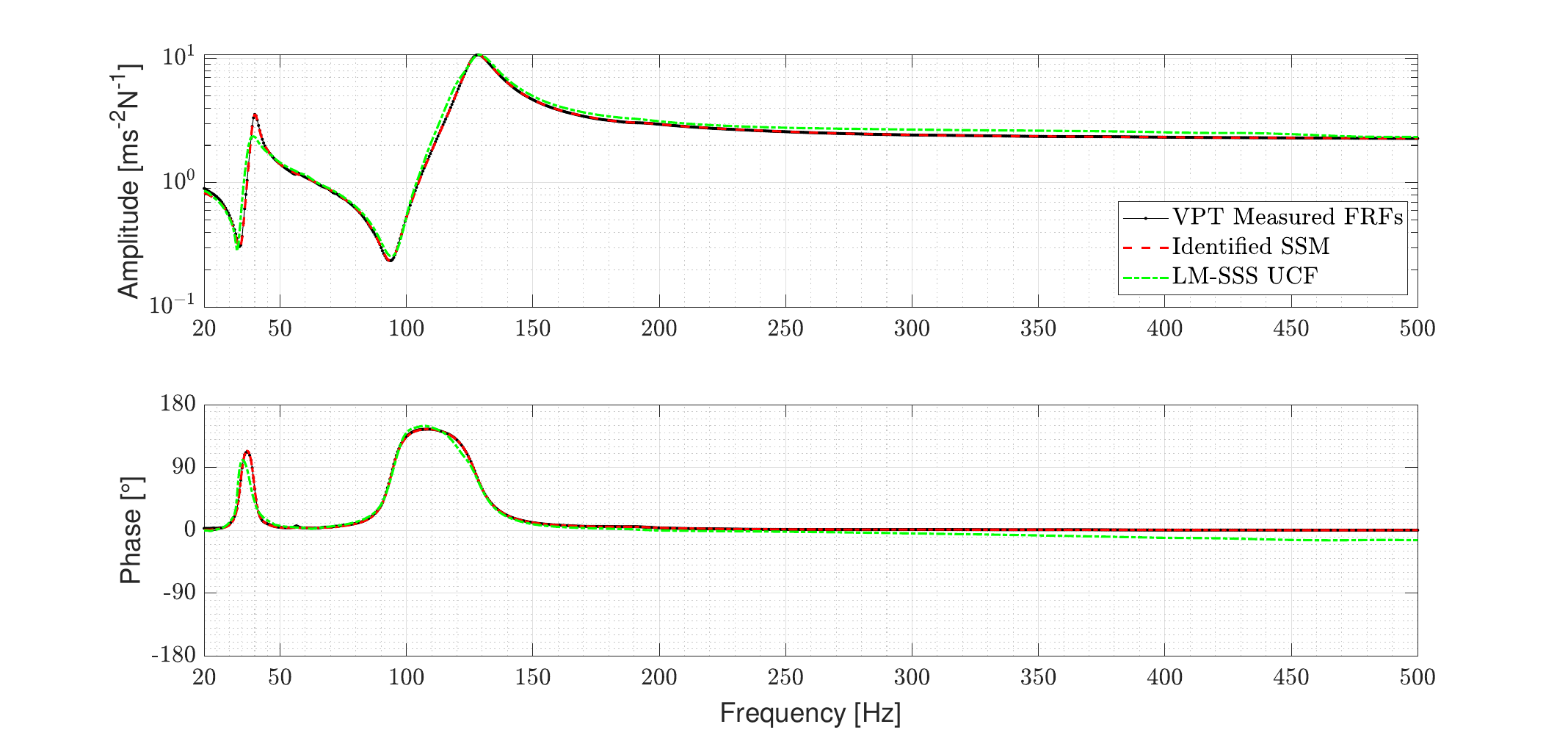}
    \caption{}
     \label{fig:Coupled_Assembly_B_12_12}
  \end{subfigure}
\caption{\textcolor{black}{Comparison of some FRFs obtained by applying VPT on the measured FRFs of assembly B, with the same FRFs of the identified state-space model of assembly B and with the same FRFs of the computed minimal-order coupled state-space model representative of assembly B:} a) FRF of the assembly B, whose output is $v_{2}^{x}$ and the input is $m_{1}^{x}$; b) FRF of the assembly B, whose output is \textcolor{black}{$v_{1}^{y}$ and the input is $m_{1}^{y}$.}}
  \label{fig:FRF_reference_vs_FRF_Coupled_Assembly_B_SSM}
\end{figure}

\textcolor{black}{By following the same strategy used to identify the state-space models of the crosses and of assembly A (see section \ref{State-space models identification}), a state-space model representative of assembly B was computed. It is worth mentioning that, the RCMs responsible for modelling the contribution of the out-of-band modes were constructed by assuming $\omega_{LR}=$0.1 Hz, $\omega_{UR}=1.5 \times 10^{4}$ Hz and $\xi_{LR}=\xi_{UR}=$0.1. While, the RCMs responsible for imposing \textcolor{black}{Newton's second law} were constructed by assuming $\omega_{CB}=1.5 \times 10^{4}$ Hz and $\xi_{CB}=$0.1. Figure \ref{fig:FRF_reference_vs_FRF_Coupled_Assembly_B_SSM} shows the comparison of the FRFs obtained by applying VPT on the measured FRFs of assembly B, with the FRFs of the identified state-space model of assembly B and with the FRFs of the computed  minimal-order acceleration coupled state-space model (obtained by double differentiating the computed displacement coupled model (see \cite{FL_1988})) representative of assembly B. By observing this figure, we can conclude that the FRFs of the identified model and the FRFs of the obtained minimal-order coupled state-space model are very well-matching the FRFs obtained by applying VPT on the measured FRFs of assembly B. Thus, it is evident that the state-space model representative of assembly B was successfully identified. Moreover, we can also conclude that the computed minimal-order coupled state-space model accurately represents the dynamics of assembly B.} However, by checking the poles of the calculated minimal-order coupled model, we conclude that it presents $1024$ poles, being $65$ of them unstable. Hence, this coupled model cannot be used for time-domain simulations as numerical instabilities will for sure be found.

\subsection{Imposing stability on the coupled state-space model}\label{Stabilization of the coupled state-space model}

To force the coupled state-space model computed in section \ref{Coupling results} to be stable, the approach described in section \ref{Enforcing_Stability} was  applied. We divided the original coupled state-space model into two models, one composed by the stable poles and the other by the unstable ones. Then, the state-space model composed by the unstable poles was forced to be stable by following the procedure described in section \ref{Enforcing_Stability}, leading to the computation of a stabilized model. The computed stabilized model was, then, divided into two additional models, one composed by the pairs of complex conjugate poles and the other by the real poles. 

\begin{figure}
  \begin{subfigure}[t]{1\textwidth}
    \centering
    \includegraphics[width=.8\textwidth]{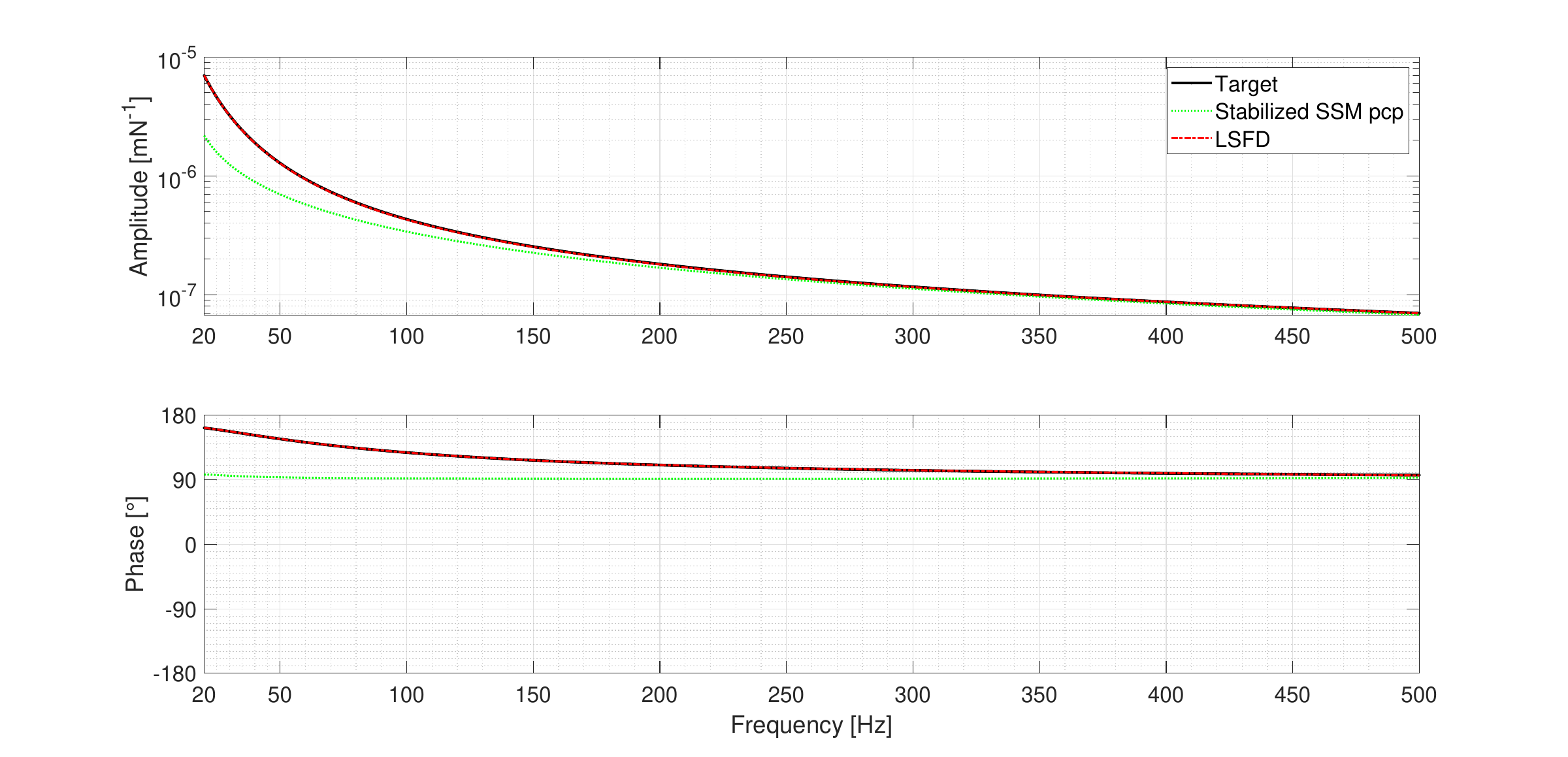}
    \caption{}
     \label{fig:Target_FRF_1_7}
  \end{subfigure}
  %\hfill
  \\
  \medskip
  \begin{subfigure}[t]{1\textwidth}
    \centering
    \includegraphics[width=.8\textwidth]{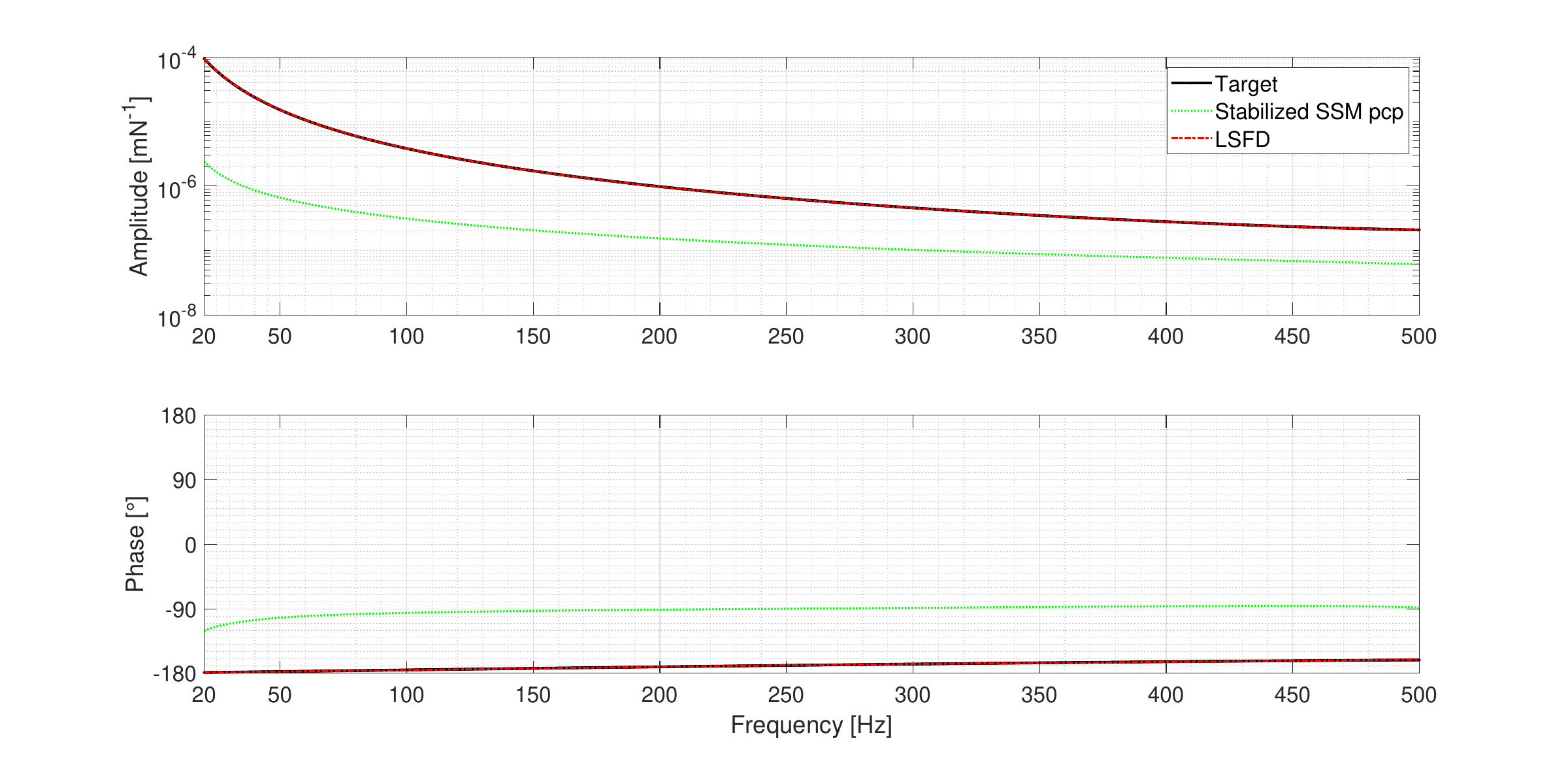}
    \caption{}
     \label{fig:Target_FRF_3_3}
  \end{subfigure}
\caption{Comparison of some displacement target FRFs with: i) the displacement FRFs of the stabilized state-space model composed by the pairs of complex conjugate poles and ii) the displacement FRFs of the modal model computed by LSFD: a) target FRF, whose output is $v_{1}^{x}$ and the input is $m_{2}^{x}$; b) target FRF, whose output is $v_{1}^{z}$ and the input is $m_{1}^{z}$.}
  \label{fig:FRF_Target_vs_MMs_1}
\end{figure}

\begin{figure}
  \begin{subfigure}[t]{1\textwidth}
    \centering
    \includegraphics[width=.8\textwidth]{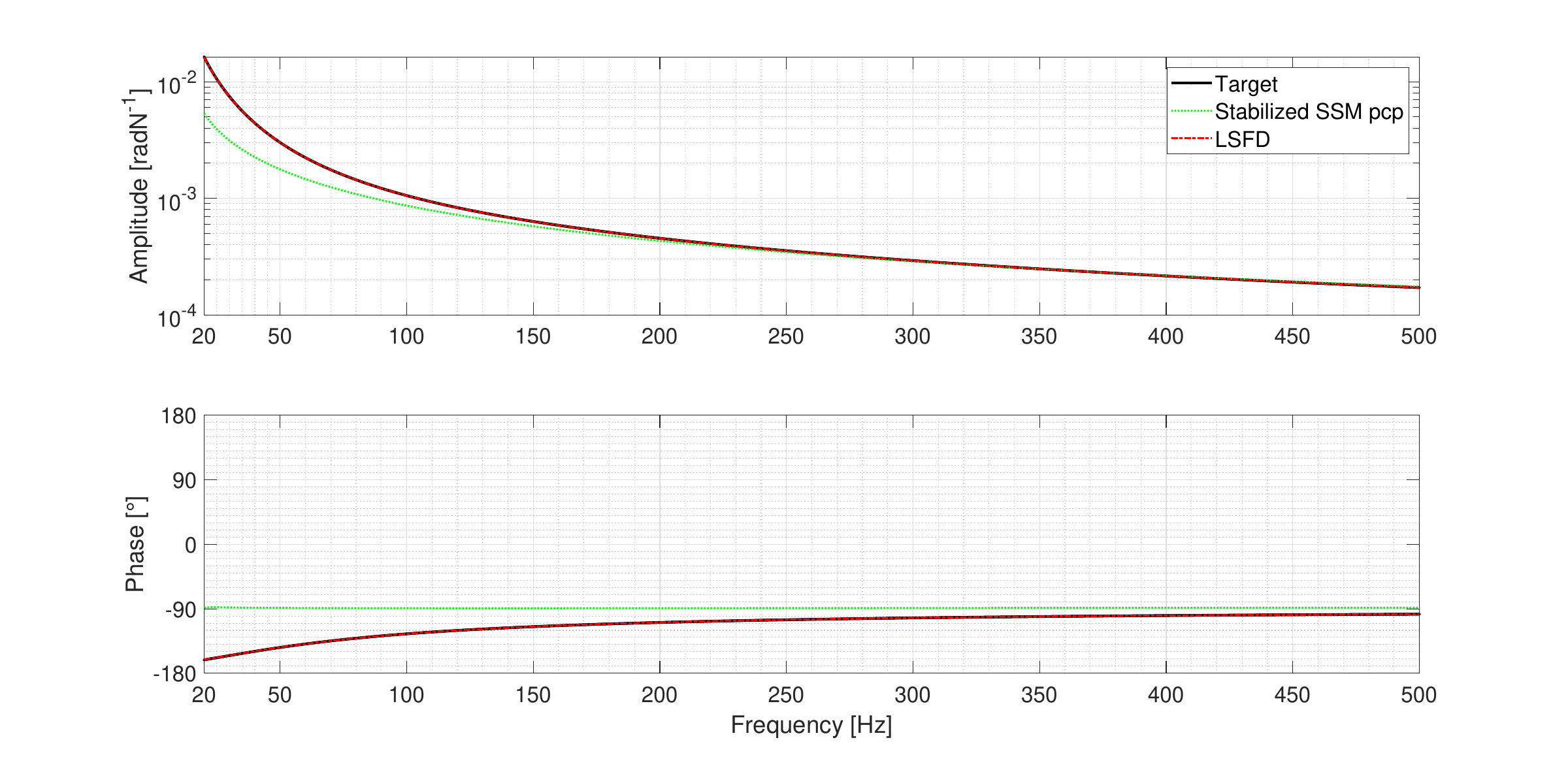}
    \caption{}
     \label{fig:Target_FRF_12_6}
  \end{subfigure}
  %\hfill
  \\
  \medskip
  \begin{subfigure}[t]{1\textwidth}
    \centering
    \includegraphics[width=.8\textwidth]{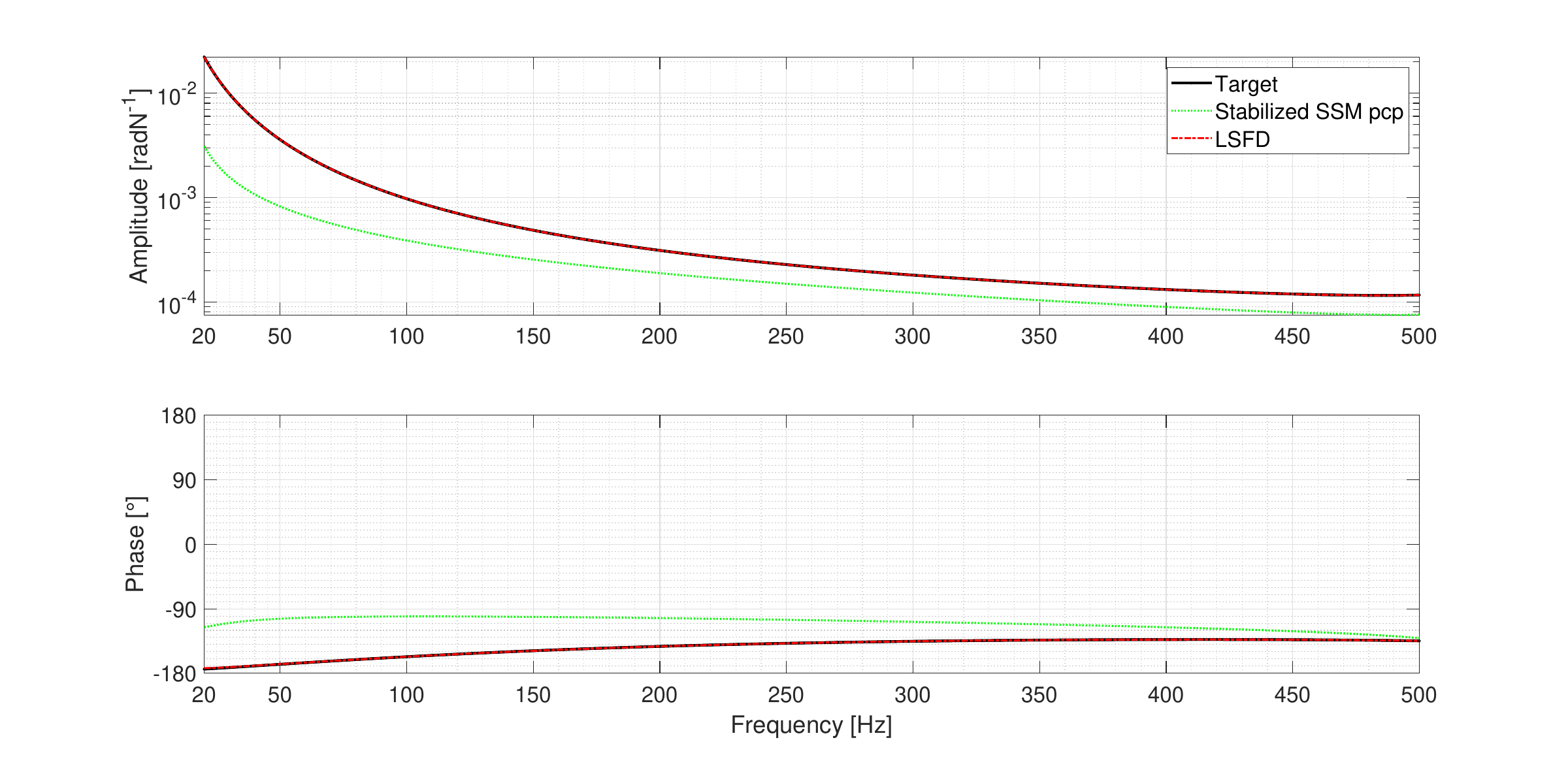}
    \caption{}
     \label{fig:Target_FRF_11_11}
  \end{subfigure}
\caption{Comparison of some displacement target FRFs with: i) the displacement FRFs of the stabilized state-space model composed by the pairs of complex conjugate poles and ii) the displacement FRFs of the modal model computed by LSFD: a) target FRF, whose output is $v_{2}^{R_{z}}$ and the input is $m_{1}^{R_{z}}$; b) target FRF, whose output is $v_{2}^{R_{y}}$ and the input is $m_{2}^{R_{y}}$.}
  \label{fig:FRF_Target_vs_MMs_2}
\end{figure}

Afterwards, the LSFD estimator was exploited to refine the modal parameters of the stabilized state-space model composed by the pairs of complex conjugate poles (see section \ref{Stabilization of the coupled state-space model}). In Figures \ref{fig:FRF_Target_vs_MMs_1} and \ref{fig:FRF_Target_vs_MMs_2}, it is shown the comparison of some displacement target FRFs with: i) the displacement FRFs of the stabilized state-space model composed by the pairs of complex conjugate poles and ii) the displacement FRFs of the modal model computed from the modal parameters refined by LSFD.

By observing Figures \ref{fig:FRF_Target_vs_MMs_1} and \ref{fig:FRF_Target_vs_MMs_2}, \textcolor{black}{we may conclude that, as expected, the match between the FRFs of the stabilized state-space model composed by the pairs of complex conjugate poles and the target FRFs is poor.} Hence, this model will not be used to compute a stable version of the original unstable coupled state-space model. 

In contrast, the FRFs of the modal model estimated by LSFD match well with the correspondent target FRFs. It is worth mentioning that to optimize the modal parameters of the stabilized state-space model composed by the pais of complex conjugate poles, we have exploited LSFD with the acceleration target FRFs, i.e. the FRFs obtained by subtracting the acceleration FRFs of the stabilized state-space model composed by the real poles from the acceleration FRFs of the state-space model composed by the unstable poles (see section \ref{Stabilization of the coupled state-space model}). As the LSFD estimator relies on a linear least-squares formulation, the FRFs of the modal model obtained from the re-estimated modal parameters will better match the reference FRFs at frequencies corresponding to higher amplitudes. In this way, as the amplitude of the displacement target FRFs is substantially lower at higher frequencies when compared to the amplitudes observed at lower frequencies (see Figures \ref{fig:FRF_Target_vs_MMs_1} and \ref{fig:FRF_Target_vs_MMs_2}), if LSFD was exploited with these FRFs, the FRFs of the obtained displacement modal model would not accurately represent the displacement target FRFs at higher frequencies. Conversely, as less significant changes on the amplitude of the acceleration target FRFs in the frequency range of interest are verified, the FRFs of the acceleration modal model obtained by using LSFD with the acceleration target FRFs accurately represent these FRFs. Hence, the correspondent displacement modal model is also representative of the displacement target FRFs (see Figures \ref{fig:FRF_Target_vs_MMs_1} and \ref{fig:FRF_Target_vs_MMs_2}).   

At this point, a stable coupled state-space model representative of the displacement original unstable coupled state-space model can be established as presented in section \ref{Enforcing_Stability}. The stable coupled state-space model is computed by using the stable model identified from the original unstable coupled model, the state-space model composed by the real poles and the models constructed from the modal model obtained by LSFD (from now on this model will be tagged as SSM LSFD). To establish the SSM LSFD model, the RCMs responsible for \textcolor{black}{including the contribution of the out-of-band modes of the modal model obtained by LSFD} and to impose \textcolor{black}{Newton's second law} were computed by selecting $\omega_{LR}=1 \times 10^{-2}$ Hz, $\omega_{UR}=7.5 \times 10^{4}$ Hz, $\omega_{CB}=7.5 \times 10^{4}$ Hz, $\xi_{LR}=0.1$, $\xi_{UR}=0.1$ and $\xi_{CB}=0.1$.

\begin{figure}
  \begin{subfigure}[t]{1\textwidth}
    \centering
    \includegraphics[width=.8\textwidth]{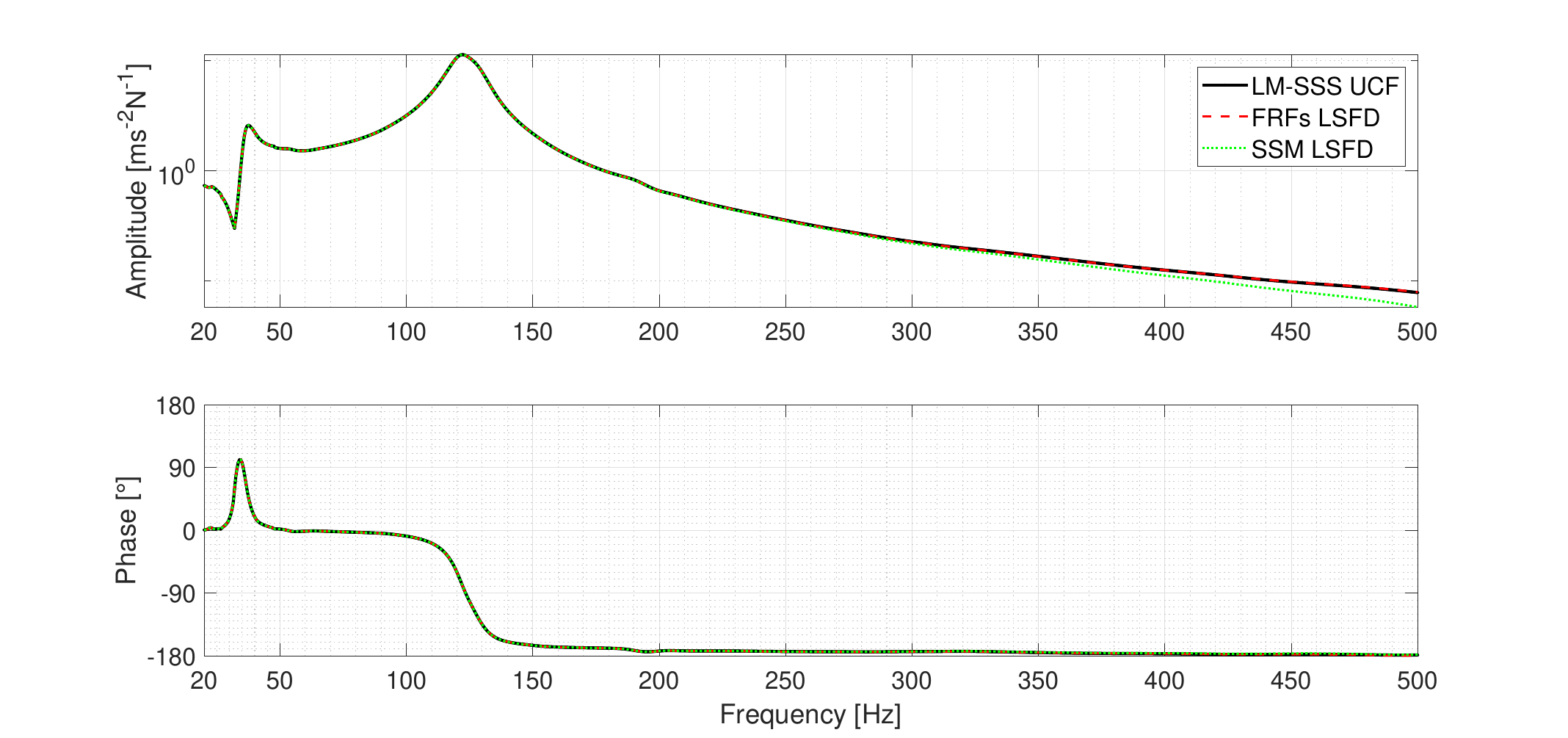}
    \caption{}
     \label{fig:Original_vs_stable_7_1}
  \end{subfigure}
  %\hfill
  \\
  \medskip
  \begin{subfigure}[t]{1\textwidth}
    \centering
    \includegraphics[width=.8\textwidth]{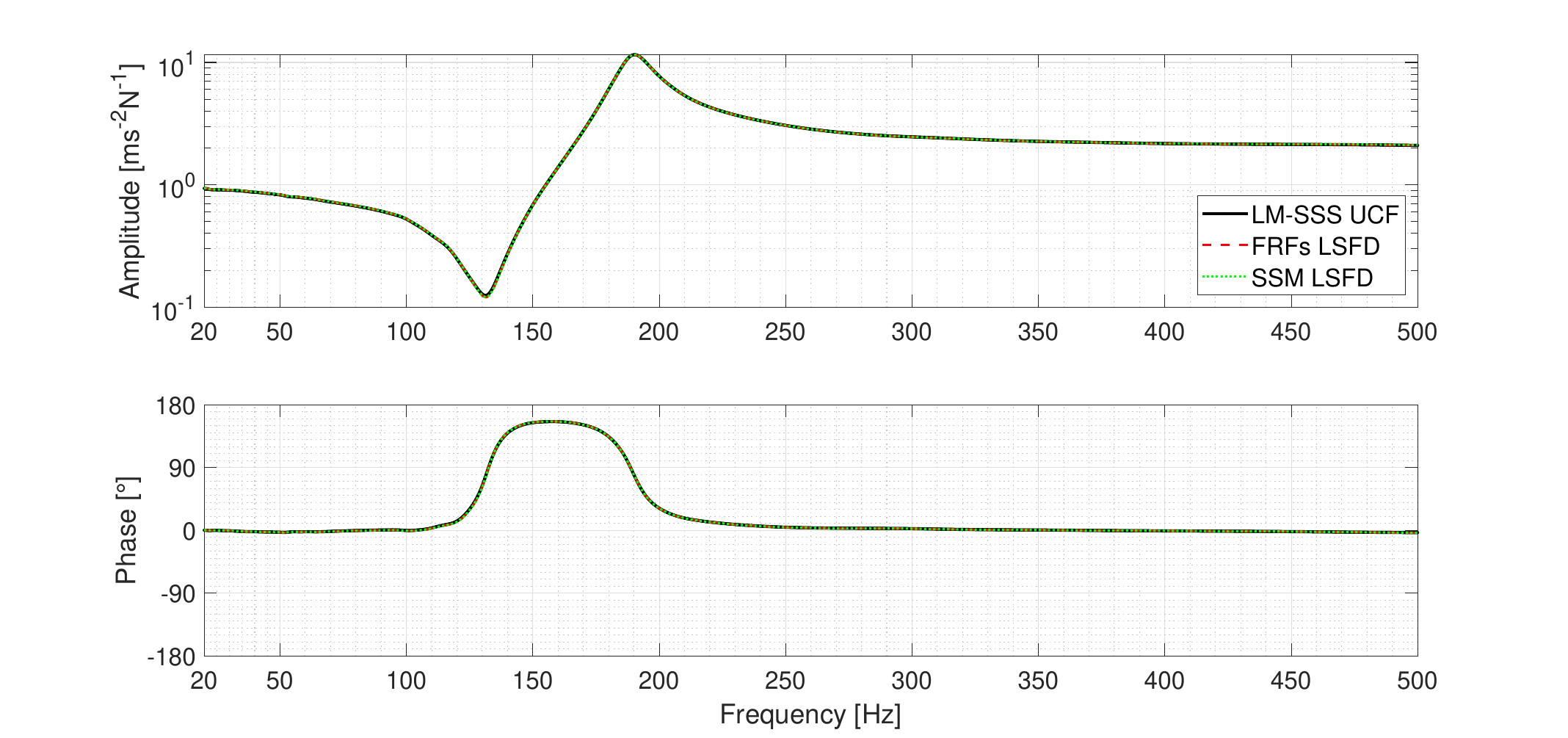}
    \caption{}
     \label{fig:Original_vs_stable_3_3}
  \end{subfigure}
\caption{Comparison of some FRFs of the acceleration original coupled model computed in section \ref{Coupling results} with the same accelerance FRFs of the set FRFs LSFD and the same \textcolor{black}{accelerance} FRFs of the state-space model SSM LSFD: a) FRF of assembly B, whose output is $v_{2}^{x}$ and the input is $m_{1}^{x}$; b) FRF of assembly B, whose output is $v_{2}^{z}$ and the input is $m_{2}^{z}$.}
  \label{fig:Original_coupled_model_vs_stable_versions}
\end{figure}

\begin{figure}
  \begin{subfigure}[t]{1\textwidth}
    \centering
    \includegraphics[width=.8\textwidth]{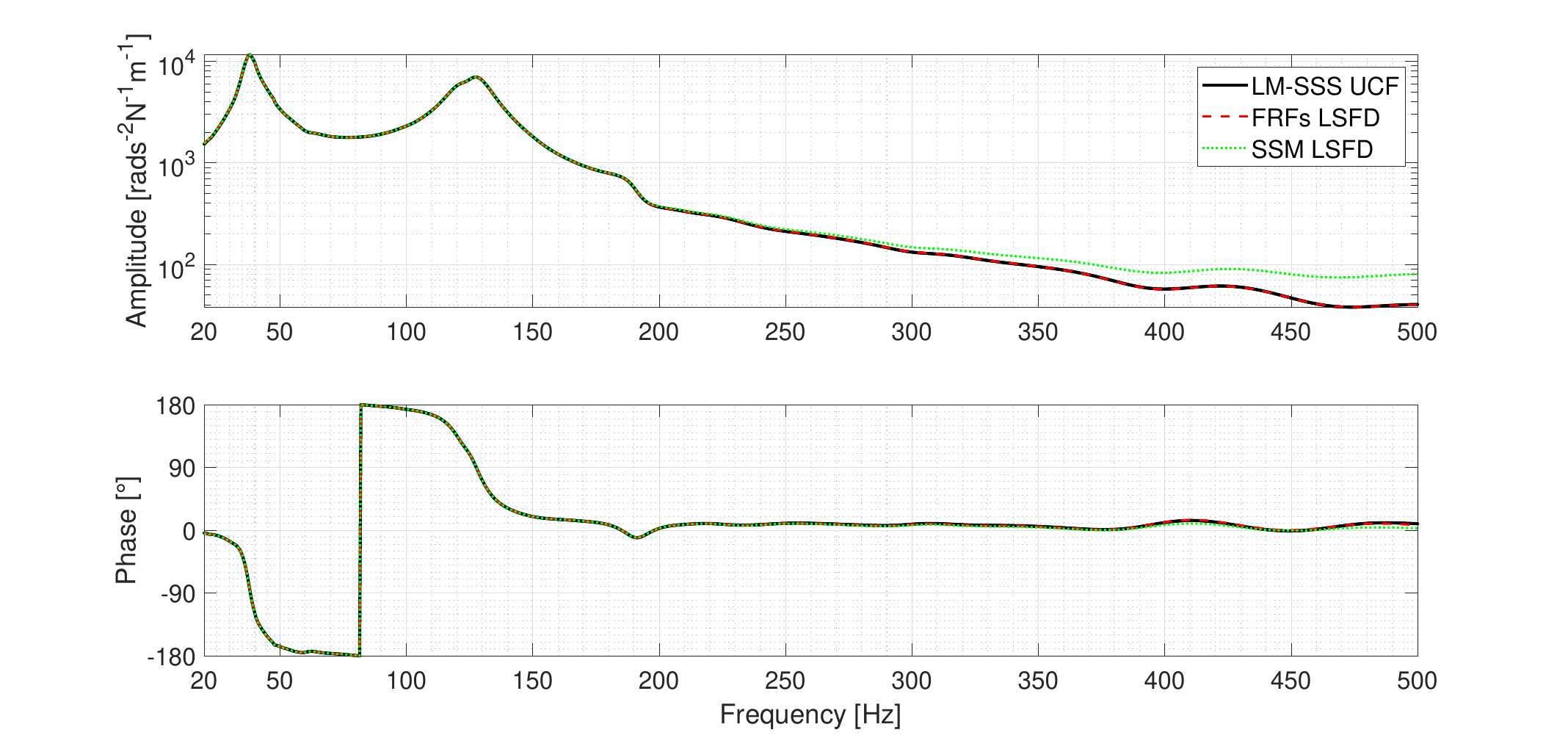}
    \caption{}
     \label{fig:Original_vs_stable_10_4}
  \end{subfigure}
  %\hfill
  \\
  \medskip
  \begin{subfigure}[t]{1\textwidth}
    \centering
    \includegraphics[width=.8\textwidth]{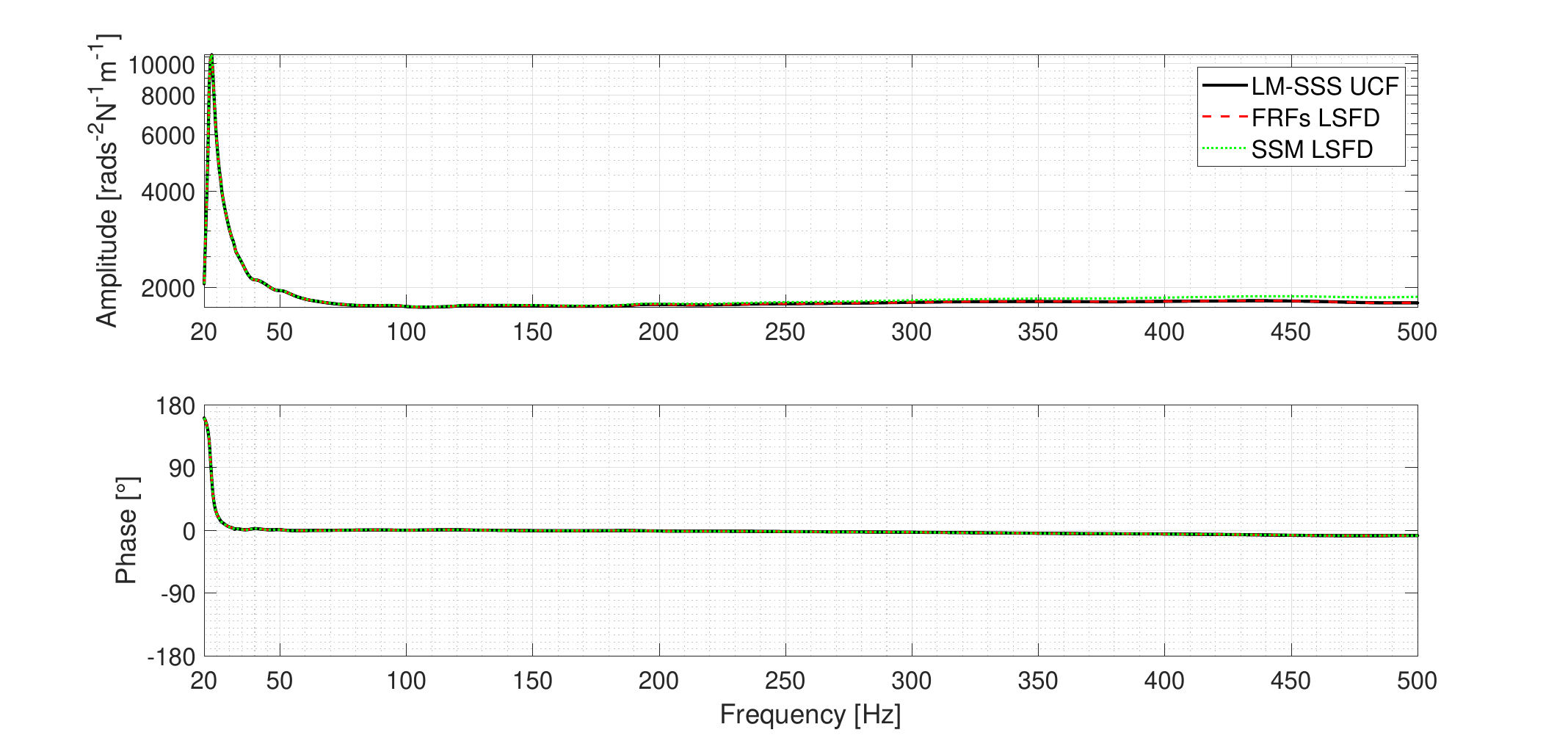}
    \caption{}
     \label{fig:Original_vs_stable_6_6}
  \end{subfigure}
\caption{Comparison of some FRFs of the acceleration original coupled model computed in section \ref{Coupling results} with the same accelerance FRFs of the set FRFs LSFD and the same \textcolor{black}{accelerance} FRFs of the state-space model SSM LSFD: a) FRF of assembly B, whose output is $v_{1}^{R_{x}}$ and the input is $m_{2}^{R_{x}}$; b) FRF of assembly B, whose output is $v_{1}^{R_{z}}$ and the input is $m_{1}^{R_{z}}$.}
  \label{fig:Original_coupled_model_vs_stable_versions_2}
\end{figure}

Furthermore, to evaluate the accuracy of the defined RCMs, a set of FRFs was defined by summing the FRFs of the stable model identified from the original coupled model, the FRFs of the stabilized model composed by the real poles and the FRFs of the modal model obtained by LSFD (from now on the resultant set of FRFs will be denominated as FRFs LSFD). In \textcolor{black}{Figures \ref{fig:Original_coupled_model_vs_stable_versions} and \ref{fig:Original_coupled_model_vs_stable_versions_2}}, some FRFs of the original acceleration unstable coupled model computed in section \ref{Coupling results} are compared with the same accelerance FRFs of the set FRFs LSFD (calculated by multiplying FRFs LSFD by $-\omega^{2}$) and with the accelerance FRFs of the state-space model SSM LSFD (obtained by double-differentiating SSM LSFD). 

\textcolor{black}{By observing Figures \ref{fig:Original_coupled_model_vs_stable_versions} and \ref{fig:Original_coupled_model_vs_stable_versions_2}, we may conclude that the FRFs of SSM LSFD well-match the FRFs of the original unstable coupled model. This suggests accuracy in properly representing the FRFs of the original unstable coupled state-space model. However, we may admit that the FRFs LSFD and the FRFs of SSM LSFD do not perfectly match in the higher frequency range. The mismatches at higher frequencies are justified by two different aspects. On one hand, the FRFs of the state-space model computed from the RCMs used to include the contribution of the upper out-of-band modes (see expression \eqref{eq:ss_RCMs_UR}) of the modal model computed by LSFD do not match so closely the elements of the upper residual matrix of the modal model obtained by LSFD at higher frequencies (see section \ref{Constructing state-space models}). On the other hand, the FRFs of the velocity state-space model computed from the RCMs used to impose \textcolor{black}{Newton's second law} (see expression \eqref{eq:ss_RCMs_CB}) on the stable coupled state-space model are not so close to be null at higher frequencies (see section \ref{Imposing the second law of Newton}). Moreover, the more pronounced deviations verified at the higher frequencies of the FRFs shown in Figures \ref{fig:Original_vs_stable_7_1} and \ref{fig:Original_vs_stable_10_4} are justified by the low amplitudes of these FRFs at those frequencies. Thus, slight inaccuracies on the performance of the computed sets of RCMs will lead to a more important bias on those FRFs than in the FRFs shown in Figures \ref{fig:Original_vs_stable_3_3} and \ref{fig:Original_vs_stable_6_6}, whose amplitude at higher frequencies is still important.} In fact, it would be possible to construct a SSM LSFD model, whose FRFs almost completely match the FRFs LSFD. Nevertheless, we would need to compute the RCMs responsible for modelling the contribution of the upper residual matrix and to impose \textcolor{black}{Newton's second law} by selecting much higher natural frequencies or/and by assuming lower damping ratios (see section \ref{Constructing state-space models}). As we are interested on computing state-space models suitable for time-domain simulations, the natural frequencies of the RCMs must not be extremely high, because it would require a greater sampling frequency to properly discretize them (the sampling frequency should be at least two times higher than the highest natural frequency of the modes included in the model \cite{ME_2022}). On the other hand, as the performance of time-domain simulations with state-space models composed by undamped poles might introduce numerical instabilities, the RCMs must not be constructed by assuming a null damping ratio \cite{ME_2022}. Therefore, to balance the accuracy of the computed \textcolor{black}{stable state-space model and its suitability to be exploited in time-domain simulations, we have computed the RCMs included in SSM LSFD as described above.}

As final analysis, we discuss the advantage of computing a stable coupled state-space model than an unstable coupled model. \textcolor{black}{A time-domain synthesis using the state-space model representative of assembly B (computed in section \ref{Coupling results}), the double-differentiated unstable coupled state-space model computed in section \ref{Coupling results},  and the double-differentiated stable coupled state-space model SSM LSFD is performed.} To implement this time-domain simulation, \textcolor{black}{the three} state-space models were discretized by exploiting the first-order-hold (foh) method (see \cite{JH_04}). A sampling frequency of $1 \times 10^{6}$ Hz was used. This frequency is higher than the highest natural frequency of the modes included in the \textcolor{black}{three simulated} state-space models (which is $7.5 \times 10^{4}$ Hz). In fact, it was found that the highest natural frequency of the modes included in these models was associated with the RCMs computed to model the contribution of the upper out-of-band-modes of the optimized stabilized modal model composed by the pairs of complex conjugate poles and to impose \textcolor{black}{Newton's second law} on SSM LSFD. A faded in/out sine sweep signal (1s length) ranging from 20 Hz to 500 Hz was applied to the input $m_{2}^{z}$ of the \textcolor{black}{three} state-space models in order to excite all the frequencies of interest in that range.

%\begin{equation}\label{eq:input_signal}
%f(t)=\sin \left(20\times2\pi t+2\pi \left(\left(500-20\right)\times \frac{t^2}{2}\right)\right) 
%\end{equation}

\begin{figure}
  \begin{subfigure}[t]{1\textwidth}
    \centering
    \includegraphics[width=.8\textwidth]{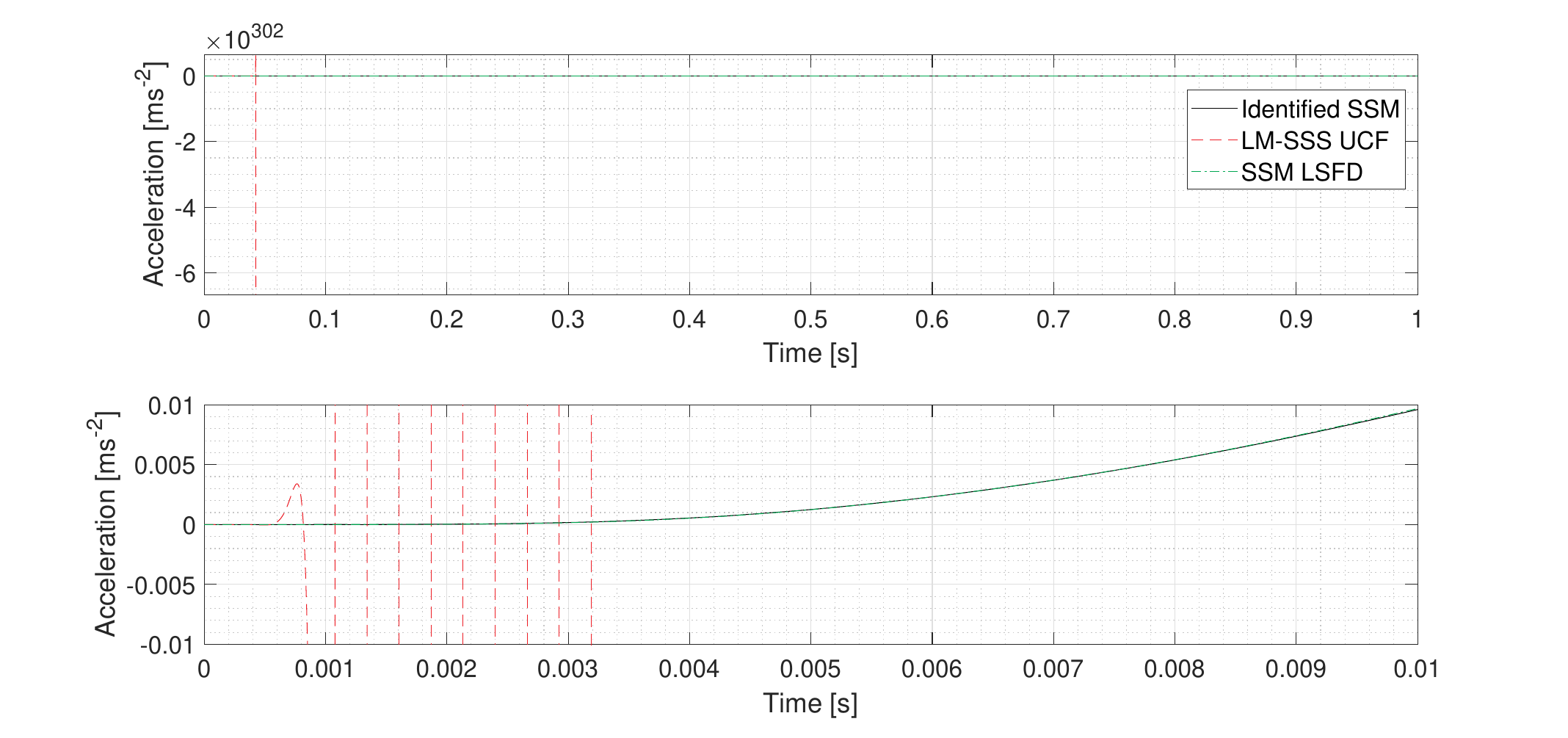}
    \caption{}
     \label{fig:Unstable_time_domain_simulation_1}
  \end{subfigure}
  %\hfill
  \\
  \medskip
  \begin{subfigure}[t]{1\textwidth}
    \centering
    \includegraphics[width=.8\textwidth]{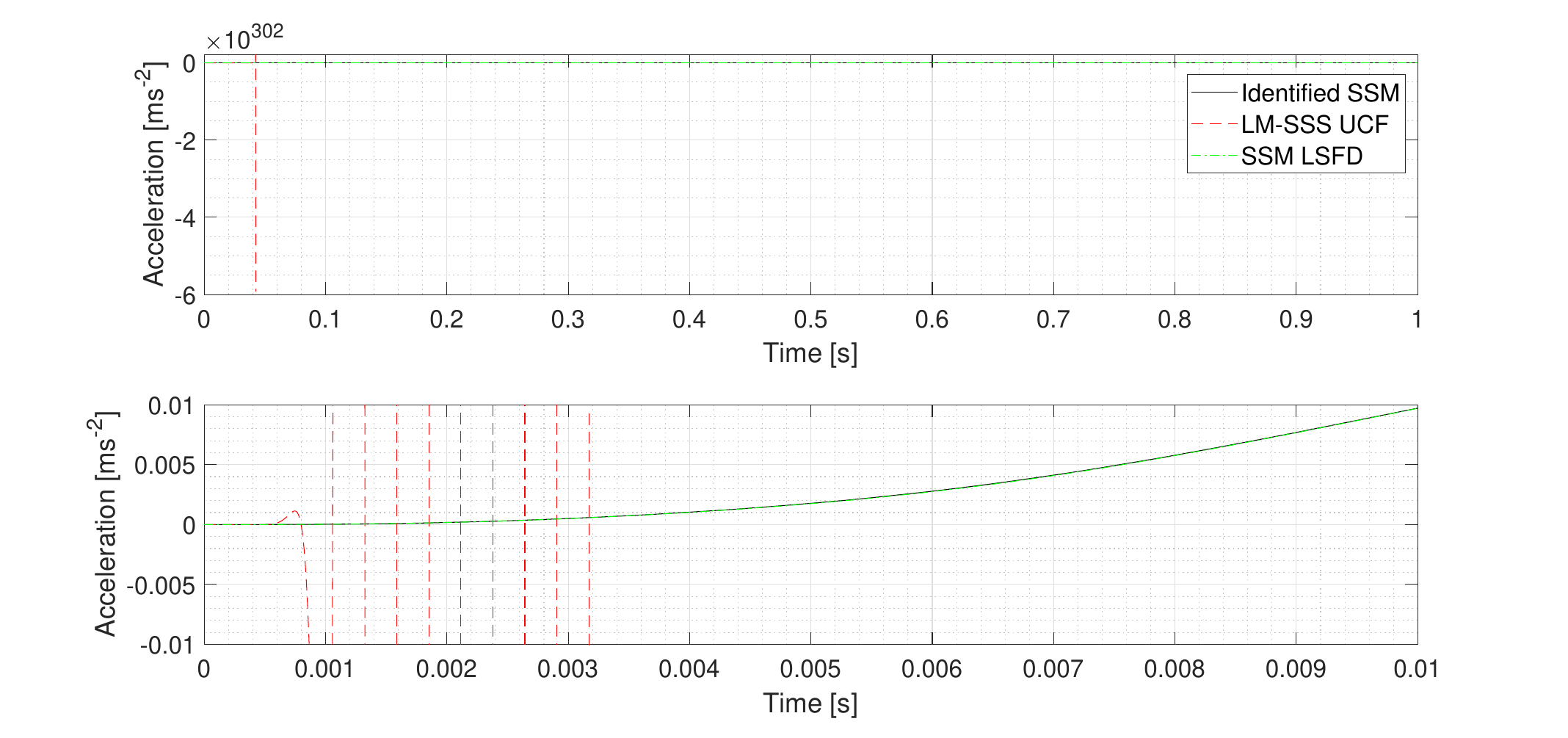}
    \caption{}
     \label{fig:Unstable_time_domain_simulation_12}
  \end{subfigure}
\caption{\textcolor{black}{Comparison of the outputs $v_{1}^{z}$ (Figure \ref{fig:Unstable_time_domain_simulation_1}) and $v_{2}^{z}$ (Figure \ref{fig:Unstable_time_domain_simulation_12}) simulated by using the identified state-space model representative of assembly B computed in section \ref{Coupling results}, the unstable coupled state-space model computed in section \ref{Coupling results} and SSM LSFD.}}
  \label{fig:Unstable_time_domain_simulation}
\end{figure}

Figure \ref{fig:Unstable_time_domain_simulation} show the comparison of the outputs \textcolor{black}{$v_{1}^{z}$} and $v_{2}^{z}$ simulated by using the \textcolor{black}{three} mentioned state-space models. 

\begin{figure}
  \begin{subfigure}[t]{1\textwidth}
    \centering
    \includegraphics[width=.8\textwidth]{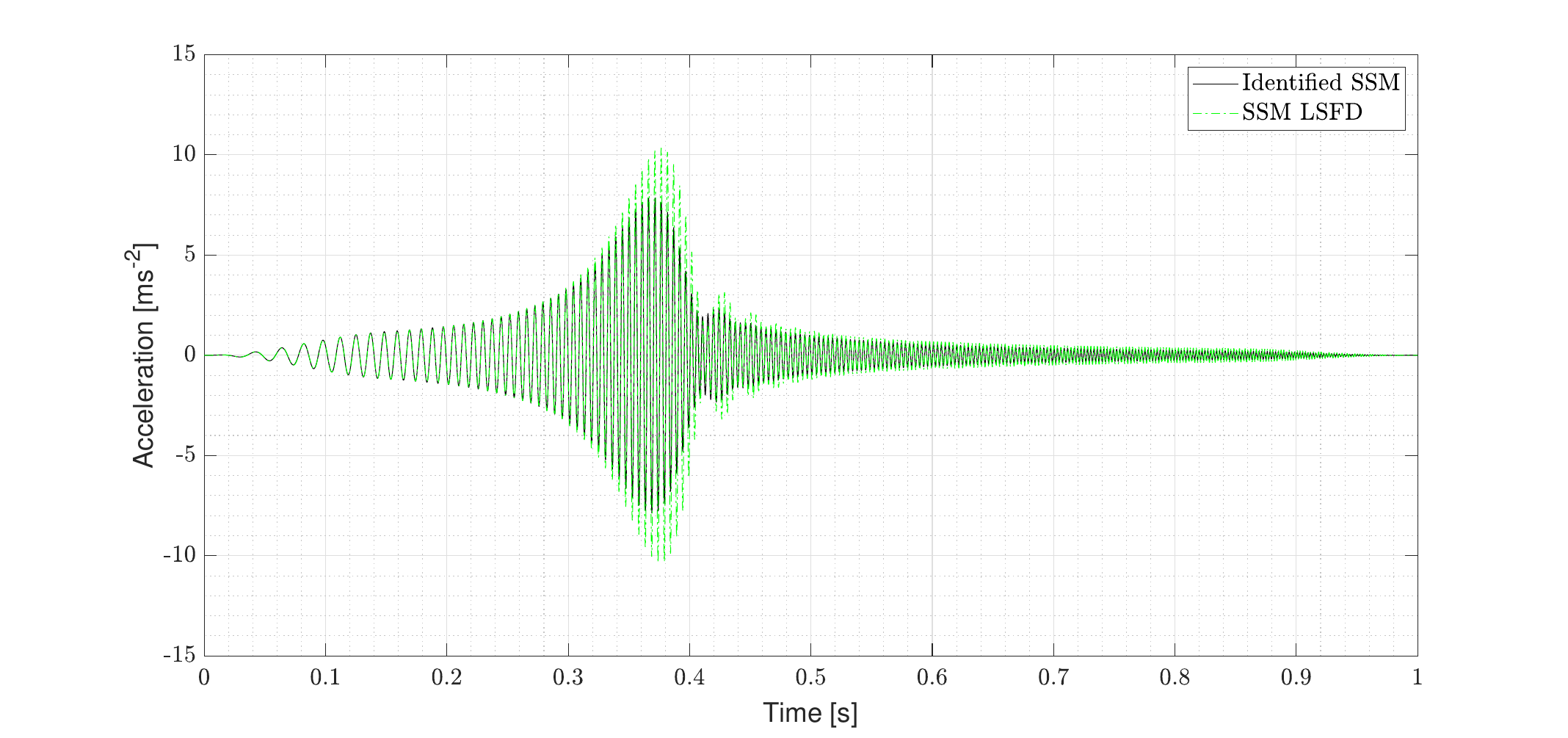}
    \caption{}
     \label{fig:Time_domain_simulation_3}
  \end{subfigure}
  %\hfill
  \\
  \medskip
  \begin{subfigure}[t]{1\textwidth}
    \centering
    \includegraphics[width=.8\textwidth]{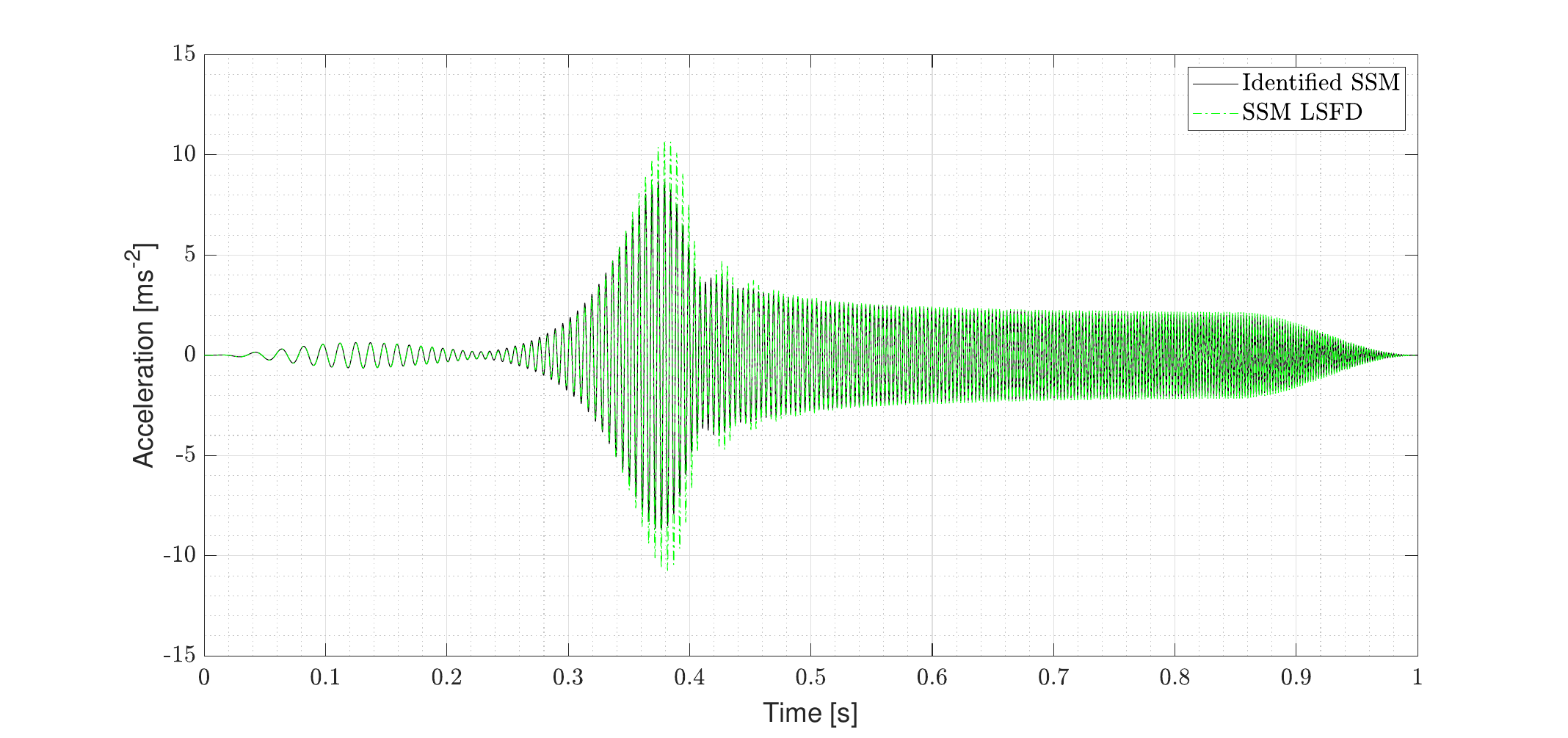}
    \caption{}
     \label{fig:Time_domain_simulation_9}
  \end{subfigure}
\caption{\textcolor{black}{Comparison of the outputs $v_{1}^{z}$ (Figure \ref{fig:Time_domain_simulation_3}) and $v_{2}^{z}$ (Figure \ref{fig:Time_domain_simulation_9}) simulated by using the identified state-space model representative of assembly B computed in section \ref{Coupling results} and by exploiting SSM LSFD.}}
  \label{fig:Stable_time_domain_simulation}
\end{figure}

It is evident that the unstable coupled state-space model is not suitable to be used in time-domain simulations. \textcolor{black}{To better compare the responses obtained by simulating the identified model representative of assembly B with the responses obtained from the time-domain simulation with the double-differentiated SSM LSFD model, in Figure \ref{fig:Stable_time_domain_simulation} these responses are shown without plotting the responses calculated with the unstable coupled state-space model.}

\begin{figure}
    \centering
    \includegraphics[width=.8\textwidth]{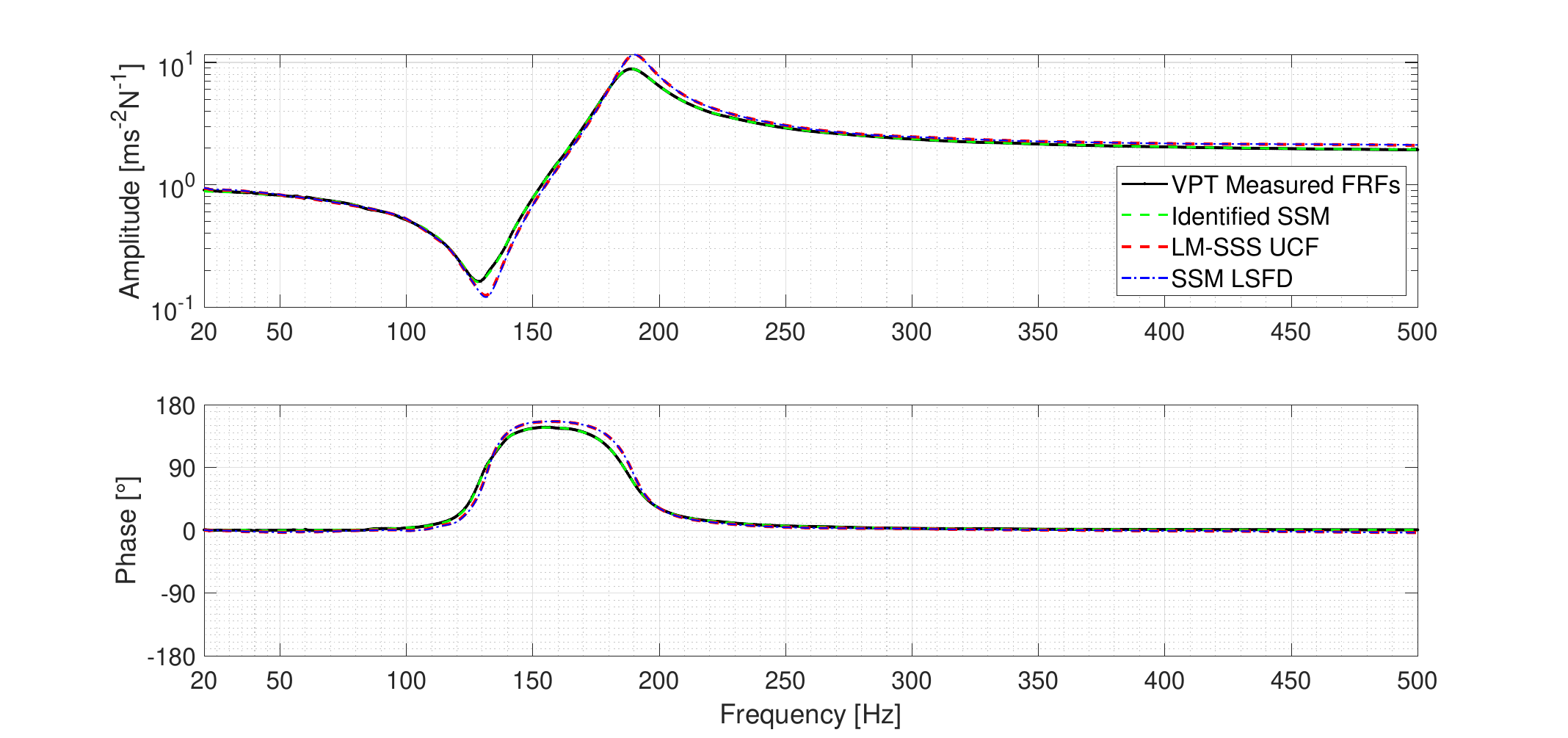}
\caption{\textcolor{black}{Comparison of the accelerance interface FRF of assembly B, whose output is \textcolor{black}{$v_{2}^{z}$} and the input is \textcolor{black}{$m_{2}^{z}$}, computed by applying VPT on the measured FRFs of assembly B with the same accelerance FRF of the identified state-space model representative of assembly B (see section \ref{Coupling results}), with the same accelerance FRFs of the unstable coupled state-space model computed in section \ref{Coupling results} and with the same accelerance FRF of the SSM LSFD model.}}
  \label{fig:Original_coupled_model_vs_unst_model_vs_SSM_LSFD}
\end{figure}

By observing Figure \ref{fig:Stable_time_domain_simulation}, \textcolor{black}{it is evident that the identified model representative of assembly B and the model SSM LSFD are stable and suitable to perform time-domain simulations. Unfortunately, the reference time data was not available to evaluate the accuracy of the results obtained from the time-domain simulation with SSM LSFD. For this reason, the responses obtained from the identified model of assembly B will be taken as reference. By comparing the time domain responses obtained by simulating the identified model (reference responses) with the ones obtained by the double differentiated SSM LSFD model, it is straightforward that they are well-matching. Nevertheless, for the time interval between 0.35 s and 0.4 s the SSM LSFD model is overestimating the time-domain responses. To better understand why this overestimation is happening, in Figure \ref{fig:Original_coupled_model_vs_unst_model_vs_SSM_LSFD} the interface FRF obtained by applying VPT on the measured FRFs of assembly B, whose output is $v_{2}^{z}$ and the input is $m_{2}^{z}$, is compared with the same FRF of the identified model representative of assembly B (see section \ref{Coupling results}), with the same FRF of the unstable coupled model and with the same FRF of the SSM LSFD model.}

\textcolor{black}{By analyzing Figure \ref{fig:Original_coupled_model_vs_unst_model_vs_SSM_LSFD}, we conclude that the FRF of both coupled models very well-match the interface FRF of assembly B. However, around the resonance of this FRF, corresponding to frequencies between 180 Hz and 210 Hz, the amplitude of the FRFs of the coupled models showed to overestimate the amplitude of the reference one, the same is not verified when comparing the FRF of the identified model representative of assembly B with the reference FRF. This frequency interval corresponds to the excited frequencies during the time interval for which the responses simulated with the coupled model SSM LSFD showed to overestimate the reference ones (see Figure \ref{fig:Time_domain_simulation_9}). Thus, we conclude that the observed overestimation of the time-domain responses is, mainly, due to mismatches between the FRFs of the unstable coupled model and the reference FRFs. Thereby, we can claim that the strategy here presented is reliable to compute accurate stable coupled models from unstable coupled state-space models, even when this unstable models are obtained by several DS operations involving state-space models identified from experimentally acquired FRFs.}

\textcolor{black}{It is now worth reflecting on} the number of states of both the computed unstable \textcolor{black}{coupled} state-space model and \textcolor{black}{of} the SSM LSFD model. The number of sates of the original unstable coupled model is $1024$, while SSM LSFD \textcolor{black}{is composed by} $1096$ states. Hence, SSM LSFD presents  $72$ states more than the original unstable coupled model, which represents an increment of, approximately, $7 \%$ of the number of states. This analysis further confirms what was pointed out in section \ref{Enforcing_Stability}, as $n_{m}>> min(n_{o},n_{i})$ is usually verified for state-space models computed from experimental data, the proposed approach to impose stability on the unstable coupled models does not, substantially, increases the number of states of the coupled model.

\color{black}

A comparison between the accuracy of the stable coupled state-space models obtained by coupling passive models and by following the strategy proposed in section \ref{Enforcing_Stability} could not be performed, because it was not possible to estimate accurate passive state-space models from the models identified in section \ref{Constructing state-space models}. Nevertheless, we can still make a justified prediction of the accuracy of the stable coupled state-space model (computed from passive models) that would be obtained if accurate passive models were computed from the identified models in section \ref{Constructing state-space models}. When forcing models to be passive, we usually take action over velocity state-space models in order to force them to verify passivity criteria associated with velocity models (for example, to verify the Positive Real Lemma (see \cite{SB_1994})) or associated with the velocity FRFs (for example, to present positive real part over the entire frequency axis (see \cite{AL_14})). Moreover, at same time, we generally constrain the problem to find a passive model, whose velocity FRFs are as close as possible to the velocity FRFs of the identified models. Thus, in the best case scenario the FRFs of the passive models will perfectly match the FRFs of the originally identified non-passive models. However, this is generally not the case in practice, because the FRFs of the passive model will always be at least slightly different from the FRFs of the identified model. Thus, the accuracy of the passive models will be lower than the one of the identified models.

On top of this, the techniques used to impose passivity on state-space models do not introduce a constraint to make sure that the computed passive state-space models obey \textcolor{black}{Newton's second law}. For this reason, it is very likely that the computed passive models do not obey \textcolor{black}{Newton's second law}. Thus, RCMs to impose this physical law must be included on the determined passive models. In general, the inclusion of these RCMs on the passive models will not be enough to make the model order of the stable coupled model (computed by coupling passive models) to be equal to the model order of the stable coupled model obtained by following the strategy proposed in section \ref{Enforcing_Stability}. Hence, this represents a practical advantage of coupling passive models instead of following the approach here suggested. However, depending  on the number of performed SSS operations at the same interface DOFs, on the number of components involved on those operations and on the number of internal DOFs of each model, the coupled models obtained by coupling passive models may be composed by the same or by an higher number of states than the number of states of the stable coupled models computed with the strategy here proposed. For the experimental case study presented in this section, imposing \textcolor{black}{Newton's second law} in the computed passive state-space models of the aluminum and steel crosses would represent the inclusion of 12 extra states in each of these models, while 24 extra states would be included in the computed passive state-space model of assembly A. Hence, the coupled state-space model obtained by coupling the passive models would be composed by the same number of states as the SSM LSFD model computed with the strategy proposed in section \ref{Enforcing_Stability}. 

Furthermore, the coupled model obtained from the passive models is expected to be less accurate than the SSM LSFD model. This is justified by the loss of accuracy due to the forcing of passivity in the identified state-space models and due to the use of a new set of RCMs to re-impose \textcolor{black}{Newton's second law}, which might introduce some additional error on the passive models (see section \ref{Imposing the second law of Newton}). In addiction, it is expected that the performance of DS operations with the computed passive models lead to more important mismatches between the FRFs of the stable coupled model obtained with the passive models and the FRFs of the unstable coupled model than the mismatches verified between the FRFs of the passive models and the FRFs of the correspondent identified models. Hence, it is evident that the computation of stable coupled models by using passive models is associated with three different steps that foster the degradation of the accuracy of the coupling results (i.e. imposing the identified models to be passive, re-impose \textcolor{black}{Newton's second law} and the performance of DS operations). While the loss of accuracy due to re-imposing \textcolor{black}{Newton's second law} can be well controlled and reduced by increasing the natural frequencies of the RCMs, the loss of accuracy coming from the other two sources of error cannot be controlled neither reduced. Conversely, the approach presented in section \ref{Enforcing_Stability} to compute stable coupled state-space models only involves two steps that decrease the accuracy of the coupling results. These steps are the use of the LSFD estimator to force the FRFs of the stabilized state-space model composed by the pairs of complex conjugate poles to match the target FRFs and the use of RCMs to include the contribution of the out-of-band modes of the modal model computed by LSFD and to impose \textcolor{black}{Newton's second law} on the complete stable coupled state-space model. Moreover, the loss of accuracy coming from these two sources of error can be controlled and reduced. On the one hand, the error associated with using the LSFD estimator to force the stabilized state-space model composed by the pairs of complex conjugate poles to match the target FRFs can be reduced by further refining the modal parameters estimated with LSFD by using the ML-MM method. On the other hand, the accuracy of the RCMs used to set-up the complete stable coupled model can be easily increased by adjusting the value of their natural frequencies (see sections \ref{Constructing state-space models} and \ref{Imposing the second law of Newton}). The advantage of coupling passive models is that the resultant coupled model is, in principal, passive as well (see \cite{AL_14}). Hence, this model would be more consistent with the physical laws than the SSM LSFD model, which is not passive. Nevertheless, in practice we are used to represent the dynamic behaviour of real mechanical systems by using non-passive models. A classical example are the FRFs computed from experimentally acquired data that, due to small inaccuracies on the test set-up are many times not passive. 

For the reasons presented above, we expect the accuracy of the stable coupled models computed by following the strategy proposed in section \ref{Enforcing_Stability}  to be higher than the accuracy of the stable coupled state-space models computed by coupling identified models forced to be passive. Moreover, the strategy here proposed holds many practical advantages over imposing passivity on all the state-space models involved on the DS operations. Firstly, the strategy here proposed showed to be roust in computing stable coupled state-space models resultant from DS operations performed with high order models, without relying on the use of \textcolor{black}{iterative algorithms}. Secondly, this strategy can be directly applied on displacement state-space models, while the approaches used to impose passivity are in general applied on velocity state-space models. Hence, if we intend to compute a displacement coupled model, we have two options. Either we compute the displacement coupled model from the velocity coupled model computed by coupling the passive velocity state-space models of the substructures, or we compute the displacement coupled model from the displacement models of the substructures, which can be computed from the correspondent passive velocity models. Both approaches present the disadvantage of requiring the inversion of the state matrix of the velocity models to compute the correspondent displacement ones (see expression \eqref{eq:CB_disp_ss_matrices}). Lastly, by using the strategy proposed in this article to compute stable coupled models, we only need to take action over the state-space model resultant from all the coupling/decoupling operations performed. This advantage is more relevant as more models are involved in the DS operations.

\color{black}

\section{Conclusion}\label{Conclusion}

The approach developed in this paper to impose \textcolor{black}{Newton's second law} (see section \ref{Imposing the second law of Newton}) on identified state-space models showed to be accurate, even when applied to state-space models estimated from experimentally acquired data (see section \ref{State-space models identification}). Moreover, the procedure here proposed showed to outperform the approach commonly applied in literature, as it makes the use of damped RCMs presenting a lower natural frequency possible. Hence, the approach here presented is advantageous when state-space models suitable for time-domain simulations are intended to be constructed. Furthermore, a novel approach to impose stability on the computed coupled state-space models was proposed in section \ref{Enforcing_Stability}. The presented procedure showed to be simple to apply, leading to accurate results without relying on the use of \textcolor{black}{iterative algorithms}. It turned out that the computed stable coupled state-space model is suitable for time-domain simulations (see section \ref{Stabilization of the coupled state-space model}), demonstrating the benefit of computing stable coupled state-space models.

\section*{Credit authorship contribution statement}

\textbf{R.S.O. Dias}: Conceptualization, Investigation, Methodology, Software, Formal analysis, Validation, Data curation, Writing - original draft, Writing - review \& editing. \textbf{M. Martarelli}: Conceptualization, Methodology, Resources, Funding Acquisition, Writing - review \& editing, Supervision, Project administration. \textbf{P. Chiariotti}: Conceptualization, Methodology, Resources, Funding Acquisition, Writing - review \& editing, Supervision, Project administration.

\section*{Declaration of Competing Interest}

The authors declare that they have no known competing financial interests or personal relationships that could have appeared to influence the work reported in this paper.

\section*{Funding}

This project has received funding from the European Union's Framework Programme for Research and Innovation Horizon 2020 (2014-2020) under the Marie Sklodowska-Curie Grant Agreement nº 858018.

%% appendix sections are then done as normal section

\appendix
%\setcounter{secnumdepth}{-1}
%\setcounter{secnumdepth}{0}
%\setcounter{section}{-1}
%\addcontentsline{toc}{section}{Appendix}

\color{black}

{\section{Suitability of the proposed RCMs to impose \textcolor{black}{Newton's second law}}\label{AppendixA}}

In this section, we will prove that by using the RCMs proposed in section \ref{Imposing the second law of Newton}, we can properly force state-space models to verify \textcolor{black}{Newton's second law}. These RCMs are suitable to impose this physical law, if they guarantee that $[C^{INL}_{full}][B^{INL}_{full}]=[0]$ and if they properly model the contribution of $[C_{ib}][B_{ib}]$ for the complete velocity state-space model. To prove the first requirement, let us develop the product $[C^{INL}_{full}][B^{INL}_{full}]$ (see expression \eqref{eq:Full_ss_matrices_INL}) by having in mind that as discussed in section \ref{Imposing the second law of Newton} $[C_{UR}][B_{UR}]=[C_{LR}][B_{LR}]=[0]$. 

\begin{equation}\label{eq:CB_disp_full_INL_matrix}
[C^{INL}_{full}][B^{INL}_{full}]=[C_{ib}][B_{ib}]+[C_{CB}][B_{CB}]
\end{equation}

By using equations \eqref{eq:CB_ss_matrices_4}, \eqref{eq:CB_poles}, \eqref{eq:CB_mode_shapes} and \eqref{eq:CB_modal_part} and assuming that the state-space model under analysis is of Single Input Single Onput (SISO) type, we may compute the $[A^{vel}_{CB}]$, $[B^{vel}_{CB}]$ and $[C^{vel}_{CB}]$ matrices as follows  

\begin{equation}\label{eq:CB_SISO_A}
[A_{CB}^{vel}]=\left[\begin{matrix}
-\xi_{CB}\omega_{CB}+j\omega_{CB}\sqrt{1-\xi_{CB}^{2}} &\\
& -\xi_{CB}\omega_{CB}-j\omega_{CB}\sqrt{1-\xi_{CB}^{2}}
\end{matrix}
\right]
\end{equation}

\begin{equation}\label{eq:CB_SISO_B}
[B_{CB}^{vel}]=\left[\begin{matrix}
-\frac{j}{2}\sqrt{\sigma_{CB}}V_{CB}^{T}\\
\frac{j}{2}\sqrt{\sigma_{CB}}V_{CB}^{T}
\end{matrix}\right]
\end{equation}

\begin{equation}\label{eq:CB_SISO_C}
[C^{vel}_{CB}]=\left[\begin{matrix}
\frac{\omega_{CB}}{\sqrt{1-\xi_{CB}^2}}\sqrt{\sigma_{CB}}U_{CB} & \frac{\omega_{CB}}{\sqrt{1-\xi_{CB}^2}}\sqrt{\sigma_{CB}}U_{CB} 
\end{matrix}\right].
\end{equation}

From expressions \eqref{eq:CB_SISO_A}, \eqref{eq:CB_SISO_B}, \eqref{eq:CB_SISO_C} and \eqref{eq:CB_disp_ss_matrices}, we know that $[A_{CB}]=[A_{CB}^{vel}]$, $[B_{CB}]=[B_{CB}^{vel}]$ and the matrix $[C_{CB}]$ must be computed as follows.

\begin{equation}\label{eq:CB_ss_matrices_value_CB_disp}
[C_{CB}]=[C^{vel}_{CB}][A_{CB}]^{-1}=\left[\begin{matrix}
\frac{\omega_{CB}\sqrt{\sigma_{CB}}U_{CB}}{\sqrt{1-\xi_{CB}^2}\left(-\xi_{CB}\omega_{CB}+j\omega_{CB}\sqrt{1-\xi_{CB}^{2}}\right)} & \frac{\omega_{CB}\sqrt{\sigma_{CB}}U_{CB}}{\sqrt{1-\xi_{CB}^2}\left(-\xi_{CB}\omega_{CB}-j\omega_{CB}\sqrt{1-\xi_{CB}^{2}}\right)} 
\end{matrix}\right]
\end{equation}

By using equations \eqref{eq:CB_SISO_B} and \eqref{eq:CB_ss_matrices_value_CB_disp} and after performing some mathematical manipulations we may compute the value of $[C_{CB}][B_{CB}]$ as given in \eqref{eq:CB_ss_matrices_value_CB_disp_2}.

\begin{equation}\label{eq:CB_ss_matrices_value_CB_disp_2}
[C_{CB}][B_{CB}]=\left[\begin{matrix}
-U_{CB}\sigma_{CB}V^{T}_{CB}
\end{matrix}\right]
\end{equation}

By using expressions \eqref{eq:CB_SVD} and \eqref{eq:CB_ss_matrices_value_CB_disp_2}, we may rewrite equation \eqref{eq:CB_disp_full_INL_matrix} as given below.

\begin{equation}\label{eq:CB_disp_full_INL_matrix_final}
[C^{INL}_{full}][B^{INL}_{full}]=[U_{CB}\sigma_{CB}V^{T}_{CB}]-[U_{CB}\sigma_{CB}V^{T}_{CB}]=[0]
\end{equation}

Expression \eqref{eq:CB_disp_full_INL_matrix_final} proves that the computed RCMs are suitable to force $[C^{INL}_{full}][B^{INL}_{full}]=[0]$. To demonstrate that the computed RCMs are also able to properly model the contribution of $[C_{ib}][B_{ib}]$ to the complete velocity state-space model, we must compute the FRFs of the velocity state-space model given by expression \eqref{eq:ss_RCMs_CB} as follows.

\begin{equation}\label{eq:Velo_FRF_CB}
    [H^{vel}_{CB}(j\omega)]=[C^{vel}_{CB}](j\omega[I]-[A^{vel}_{CB}])^{-1}[B^{vel}_{CB}]+[C_{CB}][B_{CB}]
\end{equation}

By using expressions \eqref{eq:CB_SISO_A}, \eqref{eq:CB_SISO_B}, \eqref{eq:CB_SISO_C} , \eqref{eq:CB_SVD} and \eqref{eq:CB_ss_matrices_value_CB_disp_2}  and after some mathematical manipulations, equation \eqref{eq:Velo_FRF_CB} may be rewritten as given below.

\begin{equation}\label{eq:Velo_FRF_CB_final}
    [H^{vel}_{CB}(j\omega)]=\frac{\omega_{CB}^{2}[C_{ib}][B_{ib}]}{-\omega^{2}+2j\omega \xi_{CB}\omega_{CB}+\omega_{CB}^{2}}-[C_{ib}][B_{ib}]
\end{equation}

By observing equation \eqref{eq:Velo_FRF_CB_final}, it is straightforward that the first term on the right-hand side of the equation is responsible for modelling the contribution of $[C_{ib}][B_{ib}]$ for the complete velocity state-space model, while the second term is responsible for making sure that $[C^{INL}_{full}][B^{INL}_{full}]=[0]$ is verified. Analyzing the first term of equation \eqref{eq:Velo_FRF_CB_final} we conclude that as the value of the natural frequencies of the RCMs is selected to be higher and as the value of the damping ratios is selected to be lower, more accurate the RCMs will be to model the contribution of $[C_{ib}][B_{ib}]$ for the complete velocity state-space model. In this way, we may conclude that if the values of the natural frequencies and damping ratios of the RCMs are properly selected, the RCMs proposed in section \ref{Imposing the second law of Newton} are suitable for imposing \textcolor{black}{Newton's second law} on estimated state-space models.

Note that, even though we have assumed SISO state-space models to perform the proofs given in this section, the proofs are still valid for Multiple Input Multiple Output (MIMO) state-space models.

\color{black}

{\section{LSFD matrices construction}\label{AppendixB}}

In this section, we will provide a proof for the construction of matrix $[\tilde{A}(L,\lambda,j\omega)]$ (see equation \eqref{eq:A_tilde}). Moreover, we will demonstrate how to construct the same matrix, when exploiting LSFD by using velocity or acceleration reference FRFs. To start, let us consider the \textcolor{black}{modal model} expression for a single mode and frequency line as follows

\begin{equation}\label{eq:modalmodel_sm_sf}
    [H_{ref}(j\omega)]_{1}=\left (\frac{\{\psi_{1}\}\{l_{1}\}}{j\omega_{1}-\lambda_{1}}+\frac{\{\psi_{1}\}^{*}\{l_{1}\}^{*}}{j\omega_{1}-\lambda_{1}^{*}}\right)+\frac{[LR]}{(j\omega_{1})^{2}}+[UR]
\end{equation}
\textcolor{black}{where}, $\left[H_{ref}(j\omega)\right]$ represents the displacement reference FRFs. 

Let us now express $\{\psi_{1}\}$, $\frac{\{l_{1}\}}{j\omega_{1}-\lambda_{1}}$ and $\frac{\{l_{1}\}^{*}}{j\omega_{1}-\lambda_{1}^{*}}$ as given \textcolor{black}{below}.

\begin{equation}\label{eq:expressing_variables_psi}
\{\psi_{1}\}=\{a\}+\{b\}j
\end{equation}

\begin{equation}\label{eq:expressing_variables_term_1}
\frac{\{l_{1}\}}{j\omega_{1}-\lambda_{1}}=\{c\}+\{d\}j
\end{equation}

\begin{equation}\label{eq:expressing_variables_term_2}
\frac{\{l_{1}\}^{*}}{j\omega_{1}-\lambda_{1}^{*}}=\{e\}+\{f\}j
\end{equation}

By using expressions \eqref{eq:expressing_variables_psi}, \eqref{eq:expressing_variables_term_1} and \eqref{eq:expressing_variables_term_2}, we may rewrite the terms $\frac{\{\psi_{1}\}\{l_{1}\}}{j\omega_{1}-\lambda_{1}}$ and $\frac{\{\psi_{1}\}^{*}\{l_{1}\}^{*}}{j\omega_{1}-\lambda_{1}^{*}}$ as given \textcolor{black}{below}.

\begin{equation}\label{eq:expressing_variables_MM_1}
\frac{\{\psi_{1}\}\{l_{1}\}}{j\omega_{1}-\lambda_{1}}=(\{a\}+\{b\}j)(\{c\}+\{d\}j)=\{a\}\{c\}+\{a\}\{d\}j+\{b\}\{c\}j-\{b\}\{d\}
\end{equation}

\begin{equation}\label{eq:expressing_variables_MM_2}
\frac{\{\psi_{1}\}^{*}\{l_{1}\}^{*}}{j\omega_{1}-\lambda_{1}^{*}}=(\{a\}-\{b\}j)(\{e\}+\{f\}j)=\{a\}\{e\}+\{a\}\{f\}j-\{b\}\{e\}j+\{b\}\{f\}
\end{equation}

From expressions \eqref{eq:modalmodel_sm_sf}, \eqref{eq:expressing_variables_MM_1} and \eqref{eq:expressing_variables_MM_2}, we may compute the value of \textcolor{black}{$[H_{ref}(j\omega)]_{1}$} as follows

\begin{equation}\label{eq:H_1}
    [H_{ref}(j\omega)]_{1}=\{a\}\{c\}+\{a\}\{d\}j+\{b\}\{c\}j-\{b\}\{d\}+\{a\}\{e\}+\{a\}\{f\}j-\{b\}\{e\}j+\{b\}\{f\}+\frac{[LR]}{-\omega_{1}^{2}}+[UR]
\end{equation}
\textcolor{black}{hence}, we may also write

\begin{equation}\label{eq:H_Re_Im}
    \left[\begin{matrix}
    \Re([H_{ref}(j\omega)]_{1}) & \Im([H_{ref}(j\omega)]_{1})
    \end{matrix}\right]=\left[\begin{matrix}
    \{a\}\{c\}-\{b\}\{d\}+\{a\}\{e\}+\{b\}\{f\}+\frac{[LR]}{-\omega_{1}^{2}}+[UR] & \{a\}\{d\}+\{b\}\{c\}+\{a\}\{f\}-\{b\}\{e\}
    \end{matrix}\right]
\end{equation}
\textcolor{black}{or}, by expressing the right-hand side of equation \eqref{eq:H_Re_Im} as a product of two matrices, we obtain the expression given \textcolor{black}{below}.

\begin{equation}\label{eq:H_Re_Im_matrix}
    \left[\begin{matrix}
    \Re([H_{ref}(j\omega)]_{1}) & \Im([H_{ref}(j\omega)]_{1})
    \end{matrix}\right]=\left[\begin{matrix}
    \{a\} & \{b\} & [LR] & [UR] 
    \end{matrix}\right]\left[\begin{matrix}
    \{c\}+\{e\} & \{d\}+\{f\}\\
    -\{d\}+\{f\} & \{c\}-\{e\}\\
    \frac{[I]}{-\omega_{1}^{2}} & [0]\\
    [I] & [0]
    \end{matrix}\right]
\end{equation}

By using expression \eqref{eq:LSFD_MM}, we may rewrite expression \eqref{eq:H_Re_Im_matrix} as follows

\begin{equation}\label{eq:LSFD_MM_2}
\left[\tilde{H}_{ref}(j\omega)\right]_{1}=\left[\begin{matrix}
[\Upsilon] & [LR] & [UR]
\end{matrix}\right]\left[\tilde{A}(L,\lambda,j\omega)\right]_{1}
\end{equation}
\textcolor{black}{where}, matrix $\left[\tilde{A}(L,\lambda,j\omega)\right]_{1}$ is given as follows,

\begin{equation}\label{eq:A_matrix_proof}
    \left[\tilde{A}(L,\lambda,j\omega)\right]_{1}=\left[\begin{matrix}
    \left[\tilde{a}_{\Re}(L,\lambda,j\omega)\right]_{1} &\left[\tilde{a}_{\Im}(L,\lambda,j\omega)\right]_{1}\\
    \left[\tilde{b}_{\Re}(\omega)\right]_{1} & [0]
    \end{matrix}\right]
\end{equation}
\textcolor{black}{while}, matrices $\left[\tilde{a}_{\Re}(L,\lambda,j\omega)\right]_{1}$, $\left[\tilde{a}_{\Im}(L,\lambda,j\omega)\right]_{1}$ and $\left[\tilde{b}_{\Re}(\omega)\right]_{1}$ are given below.

\begin{equation}\label{eq:a_Re_a_IM_matrices_proof}
\left[\tilde{a}_{\Re}(L,\lambda,j\omega)\right]_{1}=\left[\begin{matrix}
\Re(\frac{\{l_{1}\}}{j\omega_{1}-\lambda_{1}})+\Re(\frac{\{l^{*}_{1}\}}{j\omega_{1}-\lambda_{1}^{*}})\\
-\Im(\frac{\{l_{1}\}}{j\omega_{1}-\lambda_{1}})+\Im(\frac{\{l^{*}_{1}\}}{j\omega_{1}-\lambda^{*}_{1}})\\
\end{matrix}
\right],\ \ \ 
\left[\tilde{a}_{\Im}(L,\lambda,j\omega)\right]_{1}=\left[\begin{matrix}
\Im(\frac{\{l_{1}\}}{j\omega_{1}-\lambda_{1}})+\Im(\frac{\{l^{*}_{1}\}}{j\omega_{1}-\lambda^{*}_{1}})\\
\Re(\frac{\{l_{1}\}}{j\omega_{1}-\lambda_{1}})-\Re(\frac{\{l^{*}_{1}\}}{j\omega_{1}-\lambda^{*}_{1}})\\
\end{matrix}
\right]
\end{equation}

\begin{equation}\label{eq:b_Re_matrices_proof}
\left[\tilde{b}_{\Re}(\omega)\right]_{1}=\left[\begin{matrix}
\frac{[I]}{-\omega^{2}_{1}}\\
[I]
\end{matrix}
\right]
\end{equation}

Expressions \eqref{eq:LSFD_MM_2} and \eqref{eq:A_matrix_proof} conclude the proof for the $\left[\tilde{A}(L,\lambda,j\omega)\right]$ matrix construction, when implementing LSFD with displacement reference FRFs. In case that velocity reference FRFs are used, expression \eqref{eq:LSFD_MM_2} must be rewritten as follows

\begin{equation}\label{eq:LSFD_MM_2_vel}
j\omega_{1}\left[\tilde{H}_{ref}(j\omega)\right]_{1}=\left[\begin{matrix}
[\Upsilon] & [LR] & [UR]
\end{matrix}\right]\left[\tilde{A}_{vel}(L,\lambda,j\omega)\right]_{1}
\end{equation}
\textcolor{black}{where}, $\left[\tilde{A}_{vel}(L,\lambda,j\omega)\right]_{1}$ is given below.

\begin{equation}\label{eq:A_LSFD_vel}
\left[\tilde{A}_{vel}(L,\lambda,j\omega)\right]_{1}=j\omega_{1}\left[\tilde{A}(L,\lambda,j\omega)\right]_{1}
\end{equation}

Whereas, if LSFD is exploited with acceleration reference FRFs, expression \eqref{eq:LSFD_MM_2} should be rewritten as follows

\begin{equation}\label{eq:LSFD_MM_2_accel}
-\omega^{2}_{1}\left[\tilde{H}_{ref}(j\omega)\right]_{1}=\left[\begin{matrix}
[\Upsilon] & [LR] & [UR]
\end{matrix}\right]\left[\tilde{A}_{accel}(L,\lambda,j\omega)\right]_{1}
\end{equation}
\textcolor{black}{where}, $\left[\tilde{A}_{accel}(L,\lambda,j\omega)\right]_{1}$ is given below.

\begin{equation}\label{eq:A_LSFD_accel}
\left[\tilde{A}_{accel}(L,\lambda,j\omega)\right]_{1}=-\omega^{2}_{1}\left[\tilde{A}(L,\lambda,j\omega)\right]_{1}
\end{equation}

Note that, even thought we have analyzed a modal model presenting a single mode and frequency line to demonstrate how to construct matrices $\left[\tilde{A}(L,\lambda,j\omega)\right]_{1}$, $\left[\tilde{A}_{vel}(L,\lambda,j\omega)\right]_{1}$ and $\left[\tilde{A}_{accel}(L,\lambda,j\omega)\right]_{1}$, the same construction methodology applies for modal models composed by an unlimited number of modes and frequency lines.

\color{black}

 \bibliographystyle{elsarticle-num} 
 \bibliography{elsarticle-template-num}

\color{black}





% \end{thebibliography}
\end{document}